\documentclass[fleqn,12pt,twoside]{article}
\usepackage{mathrsfs, amsfonts, amssymb, amsmath, amsfonts, graphicx, subfigure, epsfig, psfrag}
\usepackage[authoryear]{natbib}
\usepackage[figuresright]{rotating}
 \usepackage{color}
\newcommand{\calL}{\mathcal{L}}

\newcommand{\calM}{\mathcal{M}}

\newcommand{\ISE}{\mathrm{ISE}}
\newcommand{\IMSE}{\mathrm{IMSE}}

\newcommand{\Haus}{\mathrm{Haus}}

\newcommand{\bh}{\mathbf{h}}

\newcommand{\bx}{\mathbf{x}}
\newcommand{\bX}{\mathbf{X}}

\newcommand{\bY}{\mathbf{Y}}

\numberwithin{equation}{section}

\begin{document}

\title{Bandwidth selection for nonparametric modal regression}
\date{}
\author{\small{Haiming Zhou}\\
\small{Division of Statistics, Northern Illinois University, DeKalb, Illinois 60115, U.S.A.}\\
\small{Xianzheng Huang}\\
\small{Department of Statistics, University of South Carolina, Columbia, South Carolina 29208, U.S.A.}}
\maketitle

\noindent {\large (\textit{To appear in Communications in Statistics - Simulation and Computation})}

\begin{abstract}
In the context of estimating local modes of a conditional density based on kernel density estimators, we show that existing bandwidth selection methods developed for kernel density estimation are unsuitable for mode estimation. We propose two methods to select bandwidths tailored for mode estimation in the regression setting. Numerical studies using synthetic data and a real-life data set are carried out to demonstrate the performance of the proposed methods in comparison with several well received bandwidth selection methods for density estimation. 
\end{abstract}


\section{Introduction}
\label{s:intro}
In a regression problem, it is often of interest to infer some typical value(s) of a response, $Y$, given a covariate value, $X=x$. One may view the mean or median associated with the conditional probability density function (pdf), $p(y|x)$, as a typical value. But when $p(y|x)$ is skewed or multimodal, modes of the conditional distribution can be better representations of the response. In such scenarios, the conditional modes can reveal important information regarding the association of $Y$ and $X$ that the conditional mean or quantiles cannot provide; and a mode can yield more precise prediction than the mean and median. These advantages of conditional modes have been especially appreciated among researchers in traffic engineering \citep{Einbeck.Tutz2006}, meteorology \citep{Hyndman.etal1996}, astronomy \citep{bamford2008revealing}, and economics \citep{Huang.Yao2012}, for instance. 

Existing methods for nonparametric estimation of conditional modes are based on kernel density estimators of $p(y|x)$. Given a kernel density estimate, $\hat p(y|x)$, the mean shift algorithm is employed to find local modes of the estimated conditional density \citep{Comaniciu2002, Einbeck.Tutz2006, Chen.etal2016}, leading to a mode (set) estimate. With this dependence of mode estimation on a kernel density estimator, one may naturally adopt well justified bandwidth selection method for density estimation in order to estimate modes. Natural as this idea is, we show in this article that bandwidths desirable, or even optimal (in some sense), for density estimation are usually not suitable for mode estimation. 

To overcome the drawback of existing bandwidth selection approaches, we propose two methods to choose bandwidths in the context of modal regression. We review the methodology of nonparametric modal regression in Section~\ref{s:modalreg}. Then we relate four existing bandwidth selection methods in Section~\ref{s:compete}. These serve as the competing methods with which we compare our proposed strategies, which are described in Section~\ref{s:cvmode}. Section~\ref{s:simulation} presents simulation studies to compare the six methods, and apply them to the Old Faithful geyser data set. Section~\ref{s:diss} gives a recap of our findings in this study, where we also point out limitations of the proposed methods, and suggest future research directions for developing improved bandwidth selection methods. 

\section{Modal regression}
\label{s:modalreg}
Denote by $\mathscr{X}$ the support of $X$ and by $\mathscr{Y}$ the support of $Y$. Given $x\in \mathscr{X}$, the mode set of $p(y|x)$ is $M(x)=\{y\in \mathscr{Y}: \, p_y(y|x)=0 \textrm{ and } p_{yy}(y|x)<0\}$, where $p_y(y|x)=(\partial/\partial y)p(y|x)$ and $p_{yy}(y|x)=(\partial^2/\partial y^2) p(y|x)$. The notational convention of using subscripts attaching to a function to refer to partial derivatives of the function is used throughout the article. We are interested in estimating $M(x)$ in this study. Let $\{(X_i, Y_i)\}_{i=1}^n$ be a random sample from the joint distribution of $(X, Y)$, specified by the joint pdf $p(x,y)$. A simple nonparametric estimator of $p(y|x)$ is 
\begin{equation*}
\hat p(y|x)=\frac{\displaystyle{\frac{1}{nh_1h_2}\sum_{i=1}^n K_1\left(\frac{X_i-x}{h_1}\right)K_2\left(\frac{Y_i-y}{h_2}\right)}}{\displaystyle{\frac{1}{nh_1} \sum_{i=1}^n K_1\left(\frac{X_i-x}{h_1}\right)}},
\end{equation*}
where $h_1$ and $h_2$ are bandwidths, and $K_1(t)$ and $K_2(t)$ are kernel functions. It follows that an estimator of $M(x)$ is given by $\hat{M}(x) = \{y\in \mathscr{Y}: \, \hat{p}_y(y|x)=0 \textrm{ and } \hat{p}_{yy}(y|x)<0\}$. 

A computationally efficient algorithm to find $\hat M(x)$ is the so-called mean shift algorithm. The algorithm is developed when $K_2(t)$ is radially symmetric \citep{Comaniciu2002} and its derivative satisfies $K'_2(t)=cK_3(t)t$, where $c$ is some negative constant and $K_3(t)$ is a kernel. The standard normal kernel and the Epanechnikov kernel are instances where a kernel possesses these features. For illustration purposes, we set $K_2(t)$ as the standard normal kernel in the sequel. With this choice of $K_2(t)$, the equation one solves for $y$ to find the local modes of $\hat p(y|x)$, originating from $\hat p_y(y|x)=0$, reduces to 
\begin{equation*}
\sum_{i=1}^n K_1\left(\frac{X_i-x}{h_1}\right)K_2\left(\frac{Y_i-y}{h_2}\right)(Y_i-y)=0.
\end{equation*}
Starting from multiple initial values, the mean shift algorithm entails repeatedly evaluating the following updating formula until convergence, 
\begin{equation}\label{mean-shift}
y^{(k+1)}=\frac{\displaystyle{\sum_{i=1}^n K_1 \left(\frac{X_i-x}{h_1}\right) K_2\left(\frac{Y_i-y^{(k)}}{h_2}\right)Y_i}}{\displaystyle{\sum_{i=1}^n K_1 \left(\frac{X_i-x}{h_1}\right) K_2\left(\frac{Y_i-y^{(k)}}{h_2}\right)}},
\end{equation}
where $y^{(k)}$ and $y^{(k+1)}$ denote generically two adjacent updated values of $y$.

As in most kernel-based estimation, the so-obtained $\hat M(x)$ is sensitive to the choice of bandwidths, $\bh=(h_1,\, h_2)$. One may conjecture that a good choice of $\bh$ for estimating $p(y|x)$ is also good for estimating $M(x)$. Next we review four bandwidth selection methods that have different rationales and have been shown to perform well in existing literature on density estimation. 
 
\section{Bandwidth selection for density estimation}
\label{s:compete}
Most well received approaches for choosing bandwidths in conditional density estimators aim at finding $\bh$ that minimizes a loss function defined as the weighted integrated squared error,
\begin{equation}\label{ISE:d}
\ISE_{\hbox {\tiny $D$}}(\bh) = \int\int \{\hat{p}(y|x)-p(y|x)\}^2p(x) w(x) dxdy, 
\end{equation}
or the corresponding risk function referred to as the weighted integrated mean squared error,
\[ \IMSE_{\hbox {\tiny $D$}}(\bh) = \int\int E\{\hat{p}(y|x)-p(y|x)\}^2p(x) w(x) dxdy, 
\]
where $p(x)$ is the pdf of $X$, and $w(x)$ is a nonnegative weight function with bounded support used to avoid estimating $p(y|x)$ at an $x$ around which data are too sparse. Given the observed data $\bX=(X_1, \ldots, X_n)$, one may let $w(x)=I(x\in [x_{\hbox {\tiny $L$}}, x_{\hbox {\tiny $U$}}])$, where $I(t)$ is the indicator function, and $x_{\hbox {\tiny $L$}}$ and $x_{\hbox {\tiny $U$}}$ are,  for example, the 2.5th and 97.5th percentile of $\bX$, respectively. All integrals in this article integrate over the support of the corresponding variable.

\cite{Bashtannyk.Hyndman2001} compared a variety of bandwidth selection methods based on parametric estimation of $\IMSE_{\hbox {\tiny $D$}}$. We revisit three of these strategies in this section that are motivated by different viewpoints. The fourth strategy revisited in this section is proposed by \cite{Fan.Yim2004} and \cite{Hall.etal2004}, who derived a cross validation (CV) criterion by approximating $\ISE_{\hbox {\tiny $D$}}$. All four methods have been shown to possess good performance in estimating conditional densities in certain scenarios. 

\subsection{Reference rules}
Letting $K_1(t)=K_2(t)=K(t)$ and defining $\mu_k=\int t^k K(t)dt$, $\nu_k=\int t^k K^2(t) dt$, for $k=0,1,\ldots$, \cite{Hyndman.etal1996} showed that 
\begin{equation}\label{IMSE:approx}
\IMSE_{\hbox {\tiny $D$}}(\bh) \approx \frac{c_1}{nh_1h_2} - \frac{c_2}{nh_1} + c_3h_1^4 + c_4h_2^4 + c_5h_1^2h_2^2,
\end{equation}
where 
\begin{align*}
c_1 &= \int \nu_0^2 w(x) dx,\\
c_2 &= \iint \nu_0 p^2(y|x) w(x) dydx,\\
c_3 &= \iint \frac{\mu_2^2p(x)}{4}\left\{ 2\frac{p_x(x)}{p(x)}p_x(y|x) + p_{xx}(y|x) \right\}^2 w(x) dydx,\\
c_4 &= \iint \frac{\mu_2^2p(x)}{4}\left\{ p_{yy}(y|x) \right\}^2 w(x) dydx,\\
c_5 &= \iint \frac{\mu_2^2p(x)}{2}\left\{ 2\frac{p_x(x)}{p(x)}p_x(y|x) + p_{xx}(y|x) \right\} p_{yy}(y|x) w(x) dydx.
\end{align*}
By minimizing (\ref{IMSE:approx}) with respect to (w.r.t.) $\bh$, \cite{Hyndman.etal1996} showed that the approximated optimal bandwidths are given by
\begin{equation}\label{reference:rule}
\hat{h}_1 = c_1^{1/6}\left\{ 4\left(\frac{c_3^5}{c_4}\right)^{1/4} + 2c_5\left(\frac{c_3}{c_4}\right)^{3/4} \right\} n^{-1/6}, \qquad
\hat{h}_2 = \hat{h}_1 \left(\frac{c_3}{c_4}\right)^{1/4}.
\end{equation}
Further assuming that $p(x)$ is a normal pdf, and $p(y|x)$ is a normal pdf with mean and standard deviation both being linear functions of $x$, \cite{Bashtannyk.Hyndman2001} elaborated (\ref{reference:rule}), resulting in the reference rule denoted by $\bh_{\hbox {\tiny $N$}}=(\hat{h}_{1, {\hbox {\tiny $N$}}}, \, \hat{h}_{2, \hbox {\tiny $N$}})$.

\subsection{Regression-based bandwidth selection}
Motivated by the property of the scaled kernel that $E\{ K_{h_2}(Y-y)|X=x \}\approx p(y|x)$ as $h_2\to 0$, where $K_h(t)=K(t/h)/h$, \cite{Fan.etal1996} developed estimators of $p(y|x)$ following local polynomial estimation of the mean function \citep{Fan&Gijbels1996}, viewing $p(y|x)$ as the conditional mean when regressing $K_{h_2}(Y-y)$ on $X$. Following this view, \cite{Bashtannyk.Hyndman2001} proposed to first fix $h_2$ at a reference rule, then select $h_1$ by minimizing the penalized mean squared prediction error defined by
\begin{equation}\label{regression}
Q_{h_2}(h_1) = \frac{\Delta}{n} \sum_{i=1}^{n} \sum_{k=1}^{\calM} \left\{ \hat{p}(y_k|X_i)-K_{h_2}(Y_i-y_k) \right\}^2w(X_i) \Xi\{W_{h_1,i}(X_i)\},
\end{equation}
where $\{y_k\}_{k=1}^{\calM}$ is a sequence of equally spaced grid points over $\mathscr{Y}$, $\Delta$ is the distance between two adjecent grid points, $\Xi(u)$ is a penalty function that has the first order Taylor expansion of the form $\Xi(u)=1+2u+O(u^2)$, and $W_{h_1,i}(x) = K_{h_1}\left(X_i-x\right)/\sum_{j=1}^n K_{h_1}\left(X_j-x\right)$, for $i=1, \ldots, n$. \cite{Koehler.etal2014} provided a review of popular choices of the penalty function. We use the penalty $\Xi(u)=(1+u)/(1-u)$ as in \cite{Akaike1970} in the simulation study in Section~\ref{s:simulation}. Denote by $\bh_{\hbox {\tiny $R$}}=(\hat{h}_{1,\hbox {\tiny $R$}}, \, \hat{h}_{2,\hbox{\tiny $R$}})$ the bandwidths resulting from this method. 

\subsection{Bootstrap bandwidth selection}
Yet another approach considered in \cite{Bashtannyk.Hyndman2001} involves parametric bootstrap, mimicking the idea in \cite{Hall1999}. This method targets at finding $\bh$ that minimizes the following empirical version of $\ISE_{\hbox {\tiny $D$}}(\bh)$,
\begin{equation}
A(\bh; \bX, \bY, \,\hat{p}, \,p) = \frac{\Delta}{n} \sum_{i=1}^{n} \sum_{k=1}^{\calM} \left\{ \hat{p}(y_k|X_i)-p(y_k|X_i) \right\}^2w(X_i),
\label{eq:Aboot}
\end{equation}
where $\hat{p}(y|x)$ is an estimate of $p(y|x)$ based on the observed data $(\bX, \bY)$ given $\bh$. Since (\ref{eq:Aboot}) depends on the unknown $p(y|x)$, \cite{Bashtannyk.Hyndman2001} constructed the following estimate of (\ref{eq:Aboot}) based on $\calL$ bootstrap samples, $\{(\bX,\bY^{(\ell)})\}_{\ell=1}^\calL$, simulated from an estimate of $p(y|x)$ denoted by $\hat p^*(y|x)$, 
\begin{equation}\label{bootstrap}
\hat A(\bh) = \frac{1}{\calL} \sum_{\ell=1}^{\calL} A(\bh; \bX, \bY^{(\ell)}, \,\hat{p}^{(\ell)}, \,\hat p^*),
\end{equation}
where $\hat{p}^{(\ell)}$, referring to $\hat p^{(\ell)}(y|x)$, is the same type of estimate as $\hat{p}(y|x)$ but computed based on the $\ell$th bootstrap sample, $(\bX,\bY^{(\ell)})$, for $\ell=1, \ldots, \calL$; and $\hat p^*$, referring to $\hat p^*(y|x)$, is a parametric estimate of $p(y|x)$. More specifically, $\hat p^*(y|x)$ results from fitting a parametric model (using data $(\bX, \bY)$) of $Y$ given $X$ specified by 
\[ Y_i=\beta_0 +\beta_1X_i + \ldots + \beta_kX_i^k + \sigma\epsilon_i, \textrm{ for $i=1, \ldots, n$},
\]
assuming $\{\epsilon_i\}_{i=1}^n$ independent model errors from $N(0, 1)$, where $k$ is determined by the Akaike information criterion (AIC). Once $\hat{p}^*(y|x)$ is obtained, one uses this model to generate the bootstrap responses, $\bY^{(\ell)}$, given $\bX$, for $\ell=1,\ldots, \calL$. Denote by $\bh_{\hbox {\tiny $B$}}=(\hat{h}_{1, \hbox {\tiny $B$}}, \, \hat{h}_{2, \hbox {\tiny $B$}})$ the bandwidths selected by this approach. 

\subsection{A cross validation method}
\cite{Fan.Yim2004} and \cite{Hall.etal2004} proposed a cross-validation criterion based on an elaboration of $\ISE_{\hbox {\tiny $D$}}(\bh)$ in (\ref{ISE:d}) as follows,  
\begin{equation}\label{ISE:CV}
\begin{aligned}
\ISE_{\hbox {\tiny $D$}}(\bh) = &\iint \hat{p}(y|x)^2 p(x)w(x)dxdy - 2 \iint \hat{p}(y|x) p(x,y)w(x) dxdy\\
 & + \int p(y|x)^2 p(x)w(x)dxdy.
\end{aligned}
\end{equation}
Note that the third term does not depend on $\bh$ and thus can be ignored when one minimizes $\ISE_{\hbox {\tiny $D$}}(\bh)$ w.r.t. $\bh$. Therefore, they proposed the following estimator of the first two terms in (\ref{ISE:CV}) as a CV criterion, 
\begin{equation}\label{cv:density}
\textrm{CV}_{\hbox {\tiny $D$}}(\bh) = \frac{1}{n}\sum_{i=1}^{n}w(X_i)\int\hat{p}_{-i}(y|X_i)^2dy - \frac{2}{n}\sum_{i=1}^{n}w(X_i)\hat{p}_{-i}(Y_i|X_i),
\end{equation} 
where $\hat{p}_{-i}(y|X_i)$ is an estimate of $p(y|x)$ based on data $\{(X_j, Y_j), j\neq i\}$ given $\bh$. With the kernel associated with $Y$ being a standard normal pdf, the integral in (\ref{cv:density}) can be derived explicitly. Denoted by $\bh_{\hbox {\tiny $D$}}=(\hat{h}_{1, \hbox {\tiny $D$}},\,  \hat{h}_{2, \hbox {\tiny $D$}})$ the value of $\bh$ that minimizes (\ref{cv:density}). 

\section{Bandwidth selection for mode estimation}
\label{s:cvmode}
\subsection{Preliminary}
In contrast to mean and quantile regression, the unknown quantity to be estimated in modal regression is not a one-to-one function, but a set, or a one-to-many function, of which the size is also unknown. This makes constructing a sensible proxy of a loss function less straightforward. For a mode estimate $\hat M(x)$, a reasonable loss function is the weighted integrated squared error defined by 
\begin{equation}\label{ISE:m}
\ISE_{\textrm{\hbox {\tiny $M$}}}(\bh) = \int \{\Haus(\hat{M}(x), \, M(x))\}^2 p(x) w(x) dx, 
\end{equation}
where $\Haus(A, B)=\inf \{ r: \, A\subset B\oplus r,\, B \subset A\oplus r\}$ is the Hausdorff distance between two sets, $A$ and $B$, in which 
$A\oplus r=\{ b: \inf_{a\in A} d(a, b) \leq r \}$, and $B\oplus r$ is similarly defined, in which $d(a, b)$ denotes the Euclidean distance between two points, $a$ and $b$. In mean and quantile regression, a proxy of a loss function associated with a mean/quantile estimate is easily obtained using residuals. For example, the weighted mean squared residuals, $n^{-1}\sum_{i=1}^n (\hat Y_i- Y_i)^2w(X_i)$, is a proxy of the weighted integrated squared error of a mean estimate $\hat m(x)$, $\int \{\hat m(x)-m(x)\}^2p(x)w(x) dx$, where $m(x)=E(Y|X=x)$ and $\hat Y_i=\hat m(X_i)$. In quantile regression, the check function evaluated at the residual is used to construct a CV criterion \citep{koenker2005quantile}. When both $X$ and $Y$ are continuous, given an $X_i$, there is typically only one corresponding $Y_i$ observed, and thus it is not clear how to construct a residual associated with a set estimate $\hat M(X_i)$. It certainly should not be $d(\hat M(X_i), \, Y_i,)$, where $d(A, a)$ denotes the distance between a set $A$ and a point $a$, defined as the minimum Euclidean distance between $a$ and $b\in A$. This is because $d(\hat M(X_i), \,Y_i)$ is not guaranteed to represent well $\Haus(\hat{M}(X_i), \, M(X_i))$ when $p(y|X_i)$ is multimodal, even if $Y_i\in M(X_i)$.

If one considers $\hat M(x)$ as a byproduct of density estimation, one may want to use the bandwidths chosen for estimating $p(y|x)$ in modal regression. But, since a smaller $\ISE_{\hbox {\tiny $D$}}$ or $\IMSE_{\hbox {\tiny $D$}}$ does not necessarily imply a smaller $\ISE_{\hbox {\tiny $M$}}$, we conjecture that bandwidth selection methods designed for density estimation are not adequate for modal regression. The intuition is that, in order to infer the modes of a conditional density, one does not have to estimate well features like the tails of the distribution or the height of a mode. Yet most well accepted density estimation methods do strive to capture these features, and the accompanying bandwidth selection methods are often distracted by, for instance, the tail behavior of $p(y|x)$ \citep{hall1992global}. 

\cite{Chen.etal2016} assumed $h_1=h_2=h$ and proposed to use the volume of an estimated prediction set to choose $h$. Following this idea, a loss function of $\hat M(x)$ is defined by $\textrm{Vol}(h)=\hat \epsilon_{1-\alpha, h} \int N(x) dx$, where $\epsilon_{1-\alpha, h}$ is the $(1-\alpha)$ quantile of $\{d(\hat M(X_i), \, Y_i)\}_{i=1}^n$ and $N(x)$ is the size of $\hat M(x)$, i.e., the number of points in $\hat M(x)$. This loss function is constructed to balance between the number of estimated local modes and the distance between the estimated modes and $\bY$. This method has two pitfalls. First, it relies on an extra tuning parameter $\alpha$; and, second, setting $h_1=h_2$ is not well justified, especially in a regression setting where $X$ and $Y$ play very different roles. 

\subsection{Two proposed methods}
\label{s:ours}
Also hoping to account for the size of a mode set estimate while striving for accurate prediction as in \cite{Chen.etal2016}, we propose the following CV criterion, 
\begin{equation}\label{cv:mode}
\textrm{CV}_{\hbox {\tiny $M$}}(\bh) = \frac{1}{n}\sum_{i=1}^{n} d^2(\hat M_{-i}(X_i), \, Y_i) N_{-i}^2(X_i) w(X_i),
\end{equation} 
where $\hat M_{-i}(X_i)$ is an estimate of $M(X_i)$ based on data $\{(X_j, Y_j), j\neq i\}$ given $\bh$, and $N_{-i}(X_i)$ is the size of $\hat M_{-i}(X_i)$, for $i=1, \ldots, n$. Denote by $\bh_{\hbox {\tiny $M$}}=(\hat{h}_{1, \hbox {\tiny $M$}}, \, \hat{h}_{2, \hbox {\tiny $M$}})$ the bandwidths that minimize (\ref{cv:mode}). 

Our second proposal is motivated by estimating $\ISE_{\hbox {\tiny $M$}}(\bh)$ in (\ref{ISE:m}). Given $M(X_i)$, an empirical version of $\ISE_{\hbox {\tiny {$M$}}}(\bh)$ is 
\[ A_{\hbox {\tiny $M$}}(\bh; \bX, \bY, \, \hat{M}, \, M) = \frac{1}{n} \sum_{i=1}^{n} \left\{ \Haus(\hat{M}(X_i), \,{M}(X_i)) \right\}^2w(X_i),
\]
where $\hat{M}(x)$ is the nonparametric estimate of $M(x)$ based on data $(\bX, \bY)$ given $\bh$. Since $A_{\hbox {\tiny $M$}}(\bh; \bX, \bY, \, \hat{M}, \, M)$ is not available in practice due to its dependence on the unknown $M(X_i)$, we use $\calL$ bootstrap samples, $\{(\bX,\, \bY^{(\ell)})\}_{\ell=1}^\calL$, to estimate it via
\begin{equation}\label{bootstrap:m}
\hat A_{\hbox {\tiny $M$}}(\bh) = \frac{1}{\calL} \sum_{\ell=1}^{\calL} A_{\hbox {\tiny $M$}}(\bh; \bX, \bY^{(\ell)}, \, \hat{M}^{(\ell)}, \, \hat M^*),
\end{equation}
where $\hat{M}^{(\ell)}$ refers to the mode estimate, of the same type of estimate as $\hat M(x)$, based on the $\ell$th bootstrap sample, for $\ell=1, \ldots, \calL$, and $\hat M^*$ refers to a mode estimate obtained from a parametric estimate of $p(y|x)$, denoted by $\tilde p^*(y|x)$. In particular, $\tilde p^*(y|x)$ results from fitting a finite mixture model, implemented by R package \texttt{flexmix},
\[ Y|X\sim \sum_{k=1}^{K}\pi_k \phi
\left(\frac{y-\beta_{k0}-\beta_{k1}b_1(X)-\ldots-\beta_{kJ}b_J(X)}{\sigma_k}\right),
\]
where $\beta_{k0}, \ldots, \beta_{kJ}$ and $\sigma_k$ are estimated using data $(\bX, \bY)$, $\{b_1(\cdot), \ldots, b_J(\cdot)\}$ form a B-spline basis, and $J$ is determined by AIC. Once a parametric estimate $\tilde{p}^*(y|x)$ is obtained, we generate bootstrap sample $\bY^{(\ell)}$ given $\bX$, for $\ell=1, \ldots, \calL$. Denote by $\bh^*_{\hbox {\tiny $B$}}=(\hat h^*_{1, \hbox {\tiny $B$}}, \, \hat h^*_{2, \hbox {\tiny $B$}})$ the resultant bandwidths. 

\section{Empirical study}
\label{s:simulation}
\subsection{Simulation design}
We design simulation experiments aiming to, first, demonstrate the bandwidths chosen by different methods, second, compare performance of the density estimator and mode estimator when different bandwidths are used. In the simulation experiment, we consider the following five true model configurations: 
\begin{enumerate}
	\item[(C1)] $Y = m(X)-1+\epsilon$, where $m(x)=x+x^2$, $\epsilon\sim\Gamma(3,2)$, and $X\sim N(0, 1)$. Here $\Gamma(a,b)$ is the gamma distribution with mean $a/b$. In this case, $p(y|x)$ is unimodal with $ M(x) \approx \{m(x)\}$.
	\item[(C2)] $[Y|X=x] \sim 0.5N\left(m_1(x), \, 1\right)+0.5N\left(m_2(x), \, 1\right)$, where $m_1(x)=x+x^2$, $m_2(x)=m_1(x)-6$, and $X\sim N(0,1)$. In this case, $p(y|x)$ is bimodal with $M(x) \approx \{m_1(x), \, m_2(x)\}$.
	\item[(C3)] For $x \le 0$, the model of $[Y|X=x]$ is the same as that under (C1); for $x>0$, the model of $[Y|X=x]$ is the same as that under (C2), where $X\sim N(0,1)$. In this case, $p(y|x)$ is unimodal if $x\leq 0$, and it is bimodal if $x>0$.
	\item[(C4)] $[Y|X=x] \sim 0.5N\left(m_1(x), \, 0.5^2\right)+0.3N\left(m_2(x), \, 0.5^2\right)+0.2N\left(m_3(x), \, 0.5^2\right)$, where $m_1(x)=x+x^2$, $m_2(x)=m_1(x)-3$, $m_3(x)=m_1(x)-6$, and $X\sim N(0,1)$. In this case, $p(y|x)$ is trimodal with $M(x) \approx \{m_1(x), \, m_2(x), \, m_3(x)\}$.
	\item[(C5)] $[Y|X=x] \sim 0.2 \sum_{j=1}^{5} N\left(m_j(x), \, 0.2^2\right)$, where $m_j(x)=x+x^2-1.5(j-1)$, and $X\sim N(0,1)$. In this case, $p(y|x)$ has five modes in $M(x) \approx \{m_j(x), \, j=1,\ldots, 5\}$.
\end{enumerate}

Under each true model configuration, we generate $500$ Monte Carlo (MC) replicates, each of sample size $n=500$, from the true model of $(X, Y)$. Based on each MC replicate, we use the R package \texttt{hdrcde} to obtain the bandwidths $\bh_{\hbox {\tiny $N$}}$, $\bh_{\hbox {\tiny $R$}}$, and $\bh_{\hbox {\tiny $B$}}$, and use the R package \texttt{lpme}, created by the first author, to obtain the bandwidths $\bh_{\hbox {\tiny $D$}}$, $\bh_{\hbox {\tiny $M$}}$, and $\bh^*_{\hbox {\tiny $B$}}$. We then use each of the six pairs of bandwidths to estimate the conditional density, $p(y|x)$, followed by estimating the mode set, $M(x)$. To address the aforementioned second aim, we compute two metrics associated with density estimation and mode estimation, respectively, one is the density-based empirical integrated squared error (EISE),
\[ \textrm{EISE}_{\hbox {\tiny $D$}} = \sum_{j=1}^{\mathcal{M}'} \sum_{k=0}^{\mathcal{M}}\left\{ \hat{p}(y_j|x_k)-p(y_j|x_k) \right\}^2 p(x_k) \Delta\Delta',
\]
the other is the mode-based EISE,
\[ \textrm{EISE}_{\hbox {\tiny $M$}} = \sum_{k=0}^{\mathcal{M}}\left\{{\Haus}(\hat{M}(x_k), {M}(x_k)) \right\}^2 p(x_k) \Delta,
\]
where $\{x_k=x_{\hbox {\tiny $L$}}+k\Delta\}_{k=0}^\calM$, $\Delta$ is the partition resolution, $\mathcal{M}$ is the largest integer no greater than $(x_{\hbox {\tiny $U$}}-x_{\hbox {\tiny $L$}})/\Delta$, and $\{y_j\}_{j=1}^{{\calM}'}$ is a sequence of grid points equally spaced over the observed sample range of $\bY$ with $y_{j+1}-y_j=\Delta'$.  

Finally, to formulate some benchmarks with which we compare the six pairs of bandwidths, we also find the bandwidths that minimize $\textrm{EISE}_{\hbox {\tiny $D$}}$ and the ones that minimize $\textrm{EISE}_{\hbox {\tiny $M$}}$, denoted by $\tilde\bh_{\hbox {\tiny $D$}}=(\tilde h_{1,\hbox {\tiny $D$}}, \, \tilde h_{2,\hbox {\tiny $D$}})$ and $\tilde\bh_{\hbox {\tiny $M$}}=(\tilde h_{1,\hbox {\tiny $M$}}, \, \tilde h_{2,\hbox {\tiny $M$}})$, respectively. Naturally, $\tilde\bh_{\hbox {\tiny $D$}}$ can be viewed as the optimal choice of $\bh$ for the purpose of density estimation, which is practically unattainable due to the dependence of $\textrm{EISE}_{\hbox {\tiny $D$}}$ on the unknown true density. Similarly, $\tilde\bh_{\hbox {\tiny $M$}}$ can be viewed as the (unrealistic) optimal choice of $\bh$ for the purpose of mode estimation. Using these two pairs of optimal (in different senses) bandwidths, we also obtain the corresponding density/mode estimates, which are the benchmark estimates with which we compare the other density/mode estimates.

\subsection{Simulation results}
\begin{figure}[p]
	\centering
	\setlength{\linewidth}{0.2\textwidth}
	\subfigure[]{ \includegraphics[width=\linewidth]{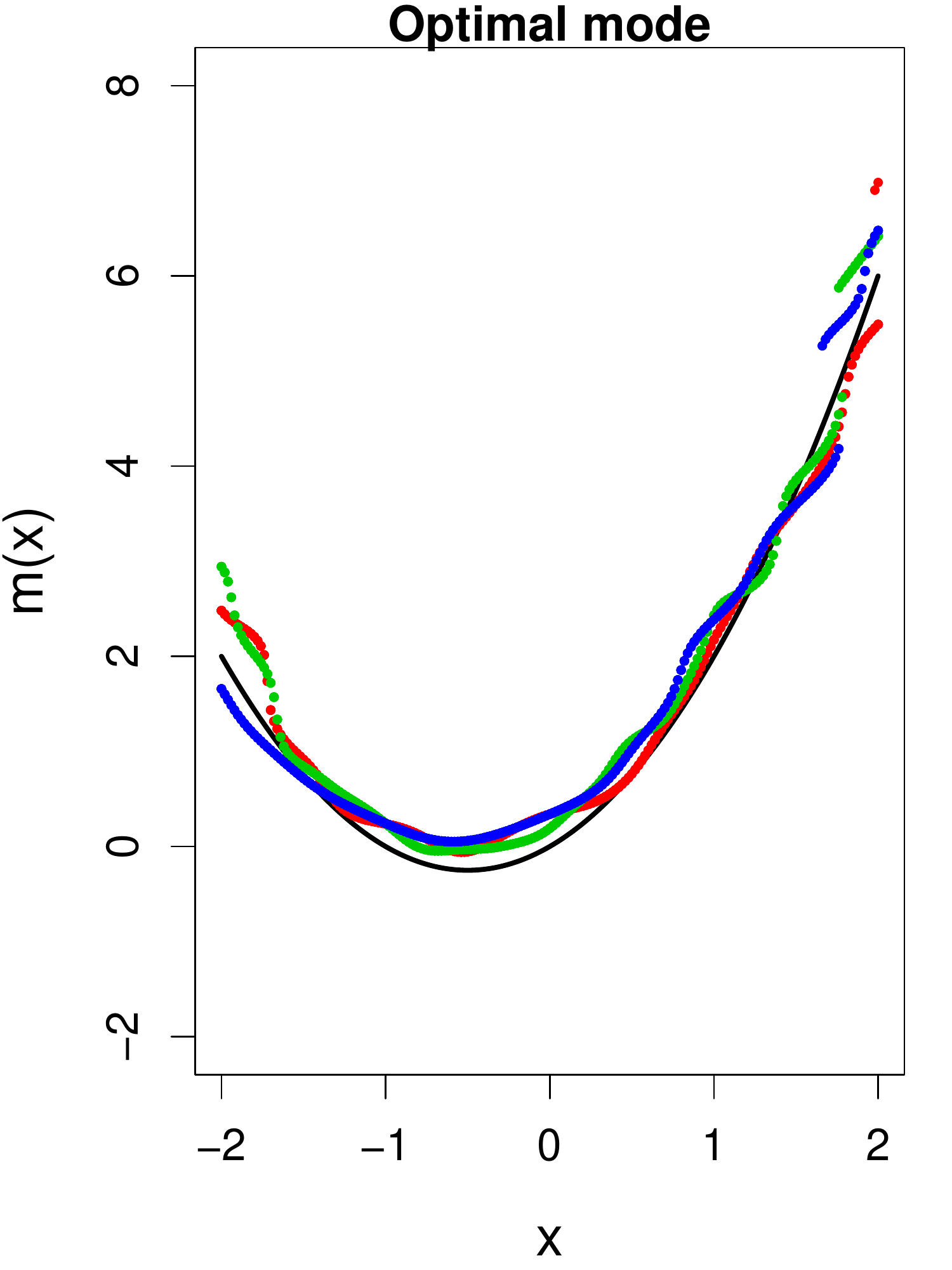} }
	\subfigure[]{ \includegraphics[width=\linewidth]{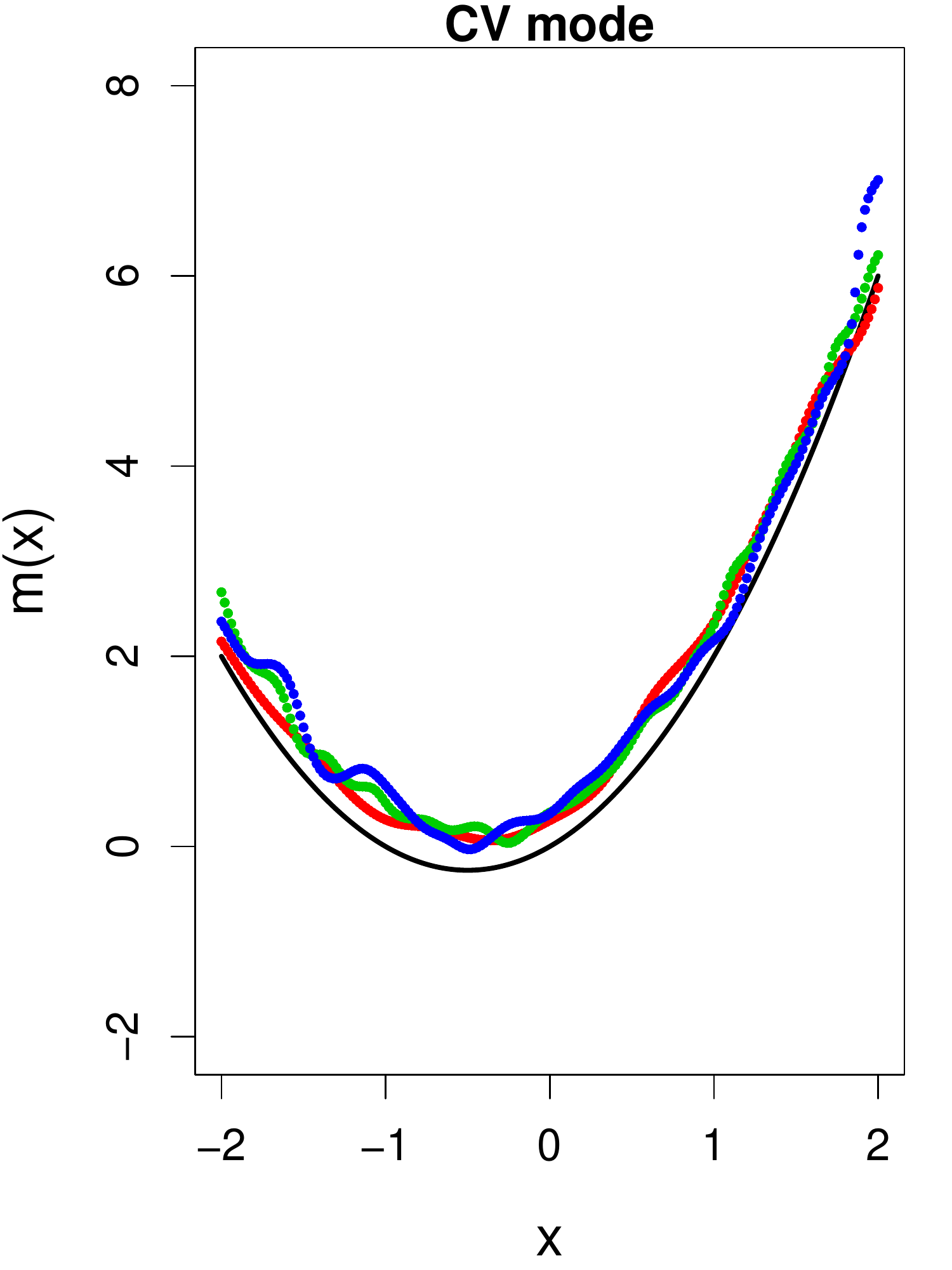} }
	\subfigure[]{ \includegraphics[width=\linewidth]{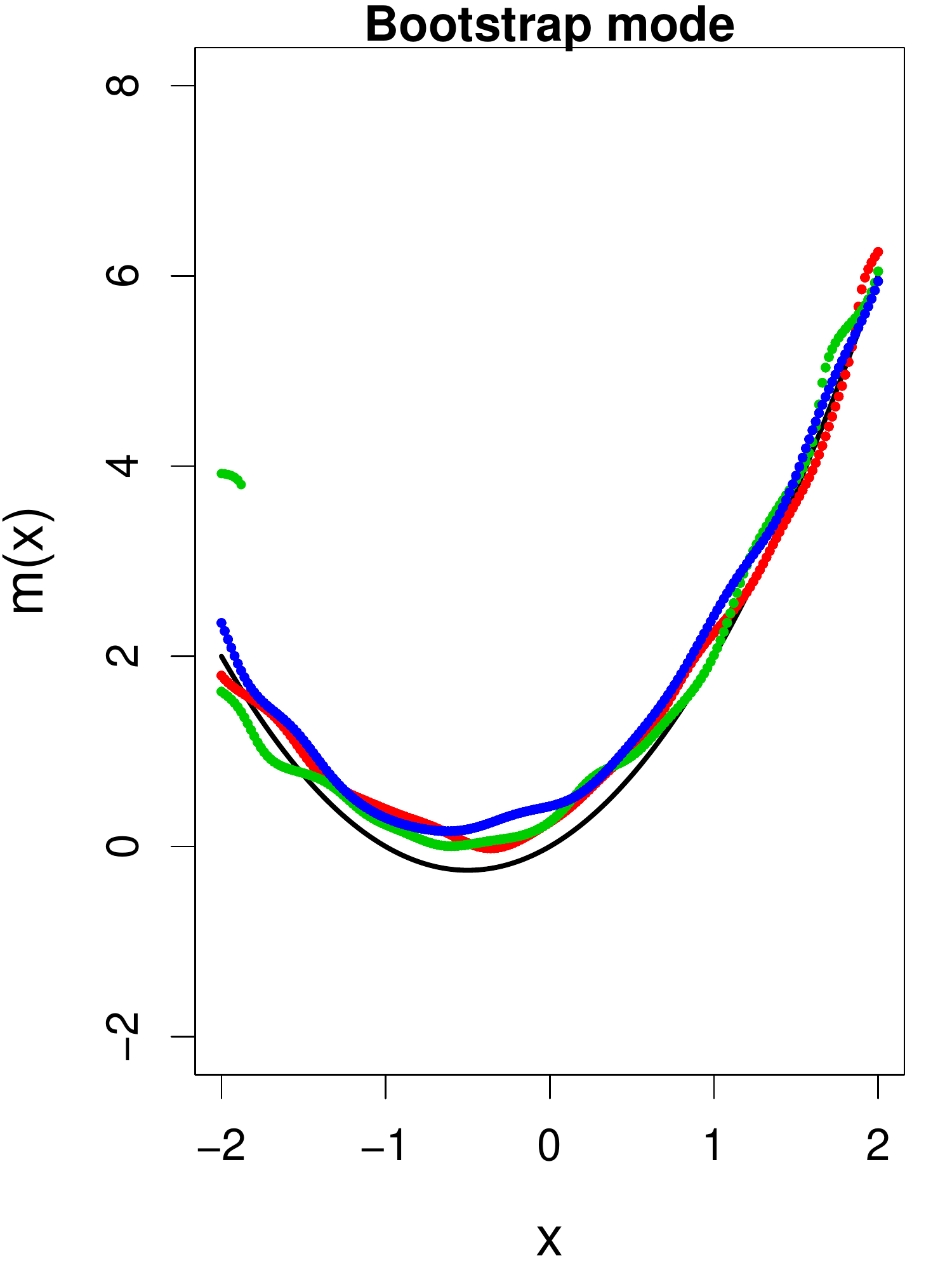} }
	\subfigure[]{ \includegraphics[width=\linewidth]{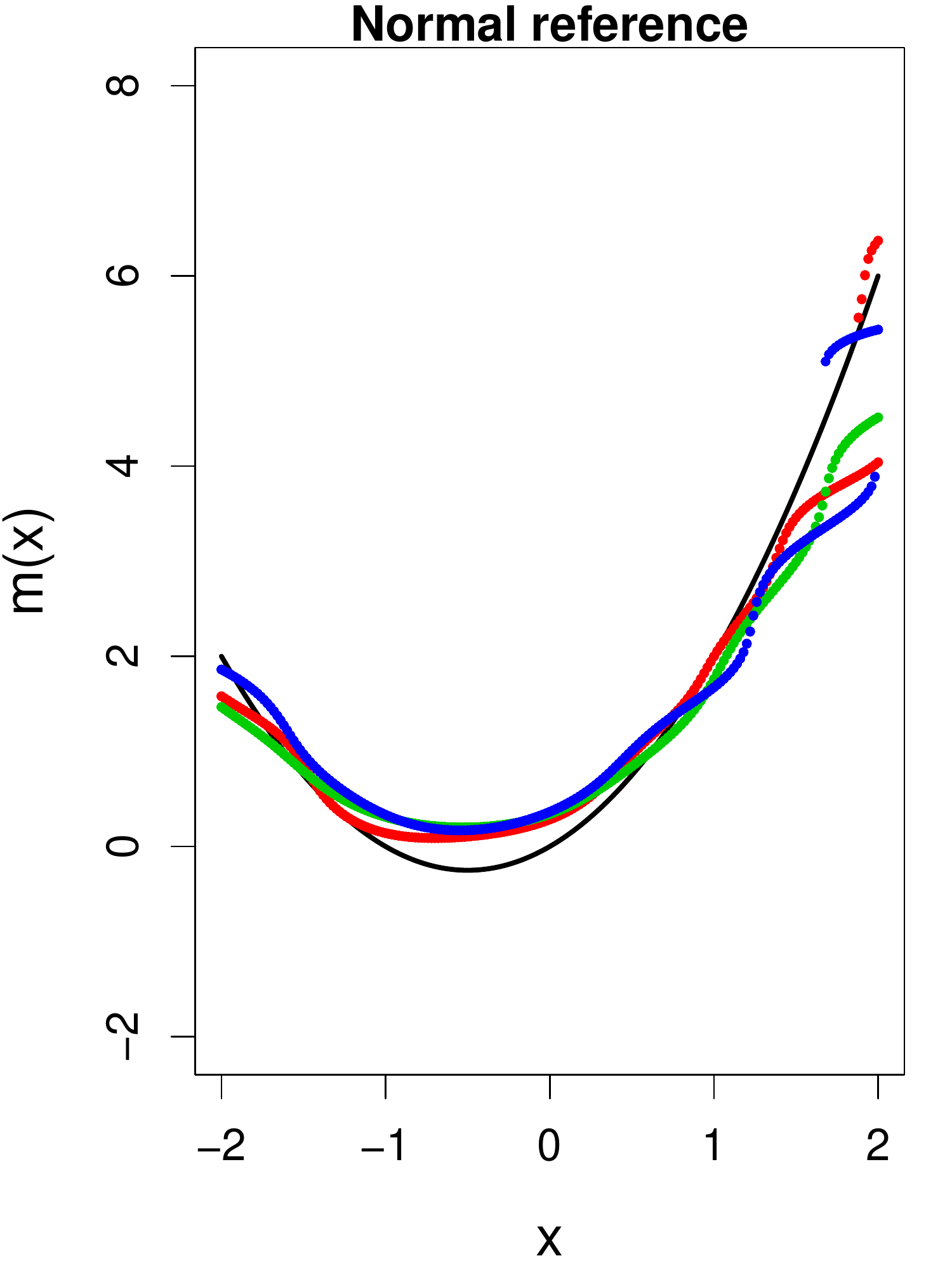} }\\
	\subfigure[]{ \includegraphics[width=\linewidth]{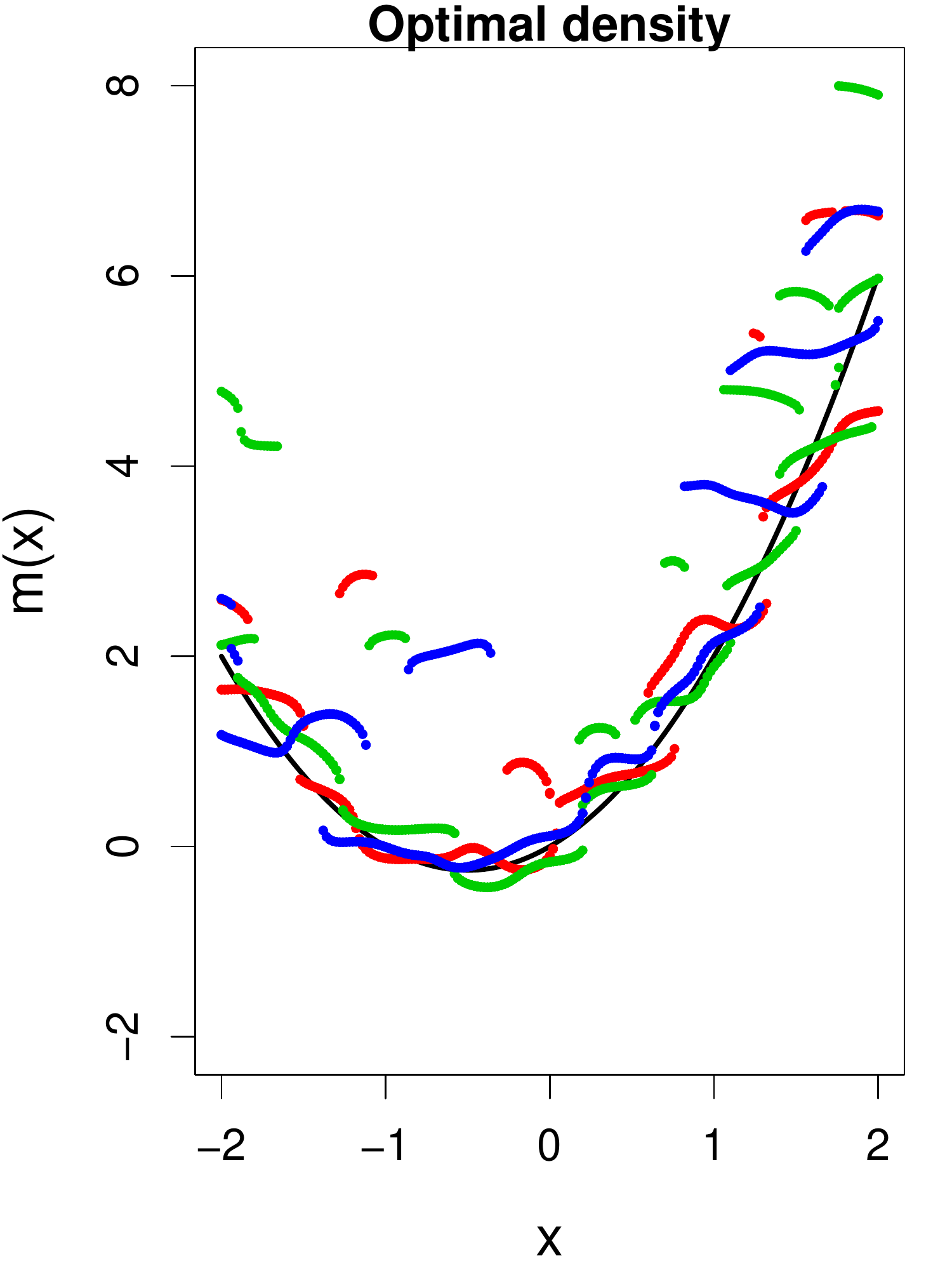} }
	\subfigure[]{ \includegraphics[width=\linewidth]{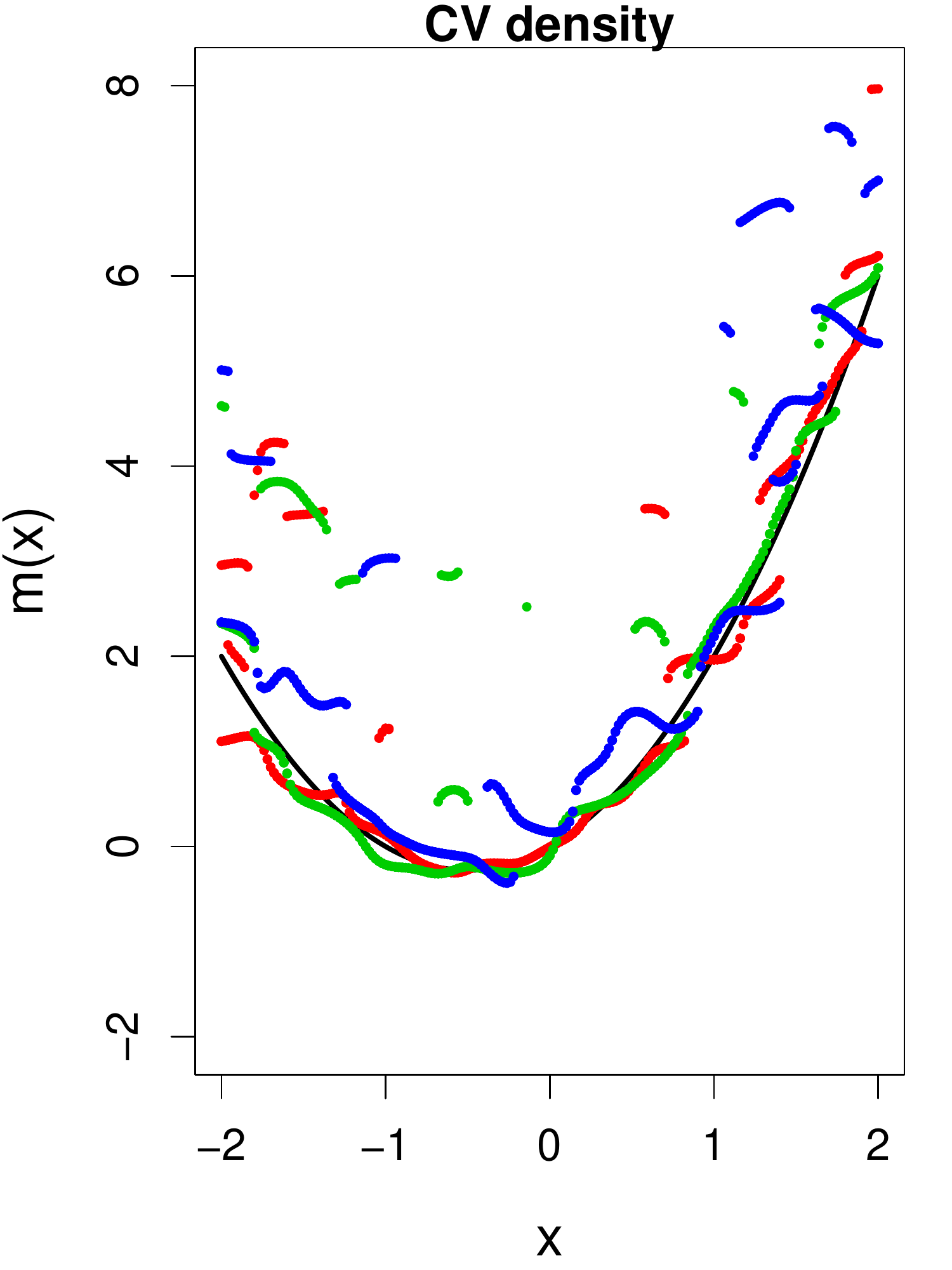} }
	\subfigure[]{ \includegraphics[width=\linewidth]{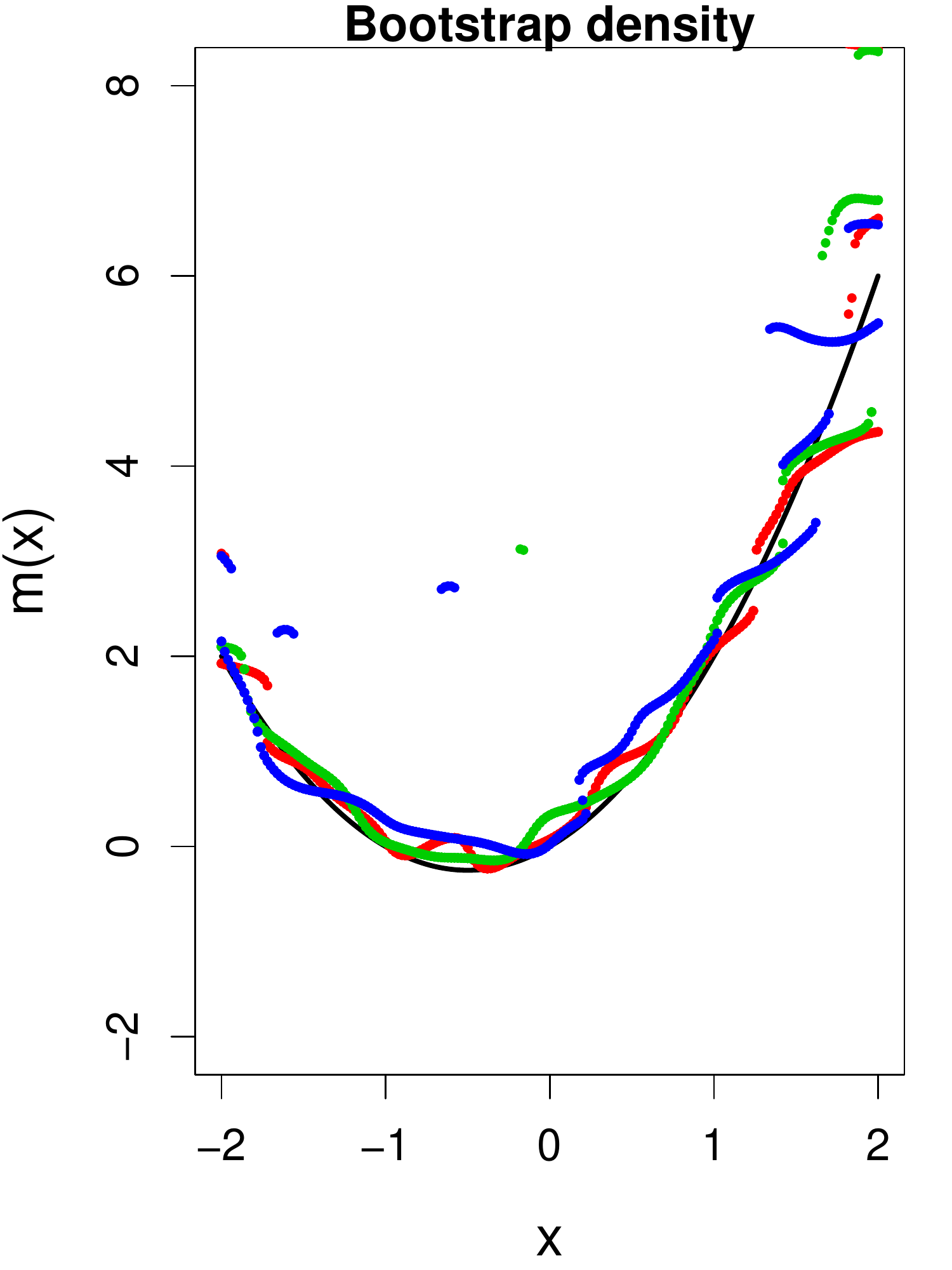} }
	\subfigure[]{ \includegraphics[width=\linewidth]{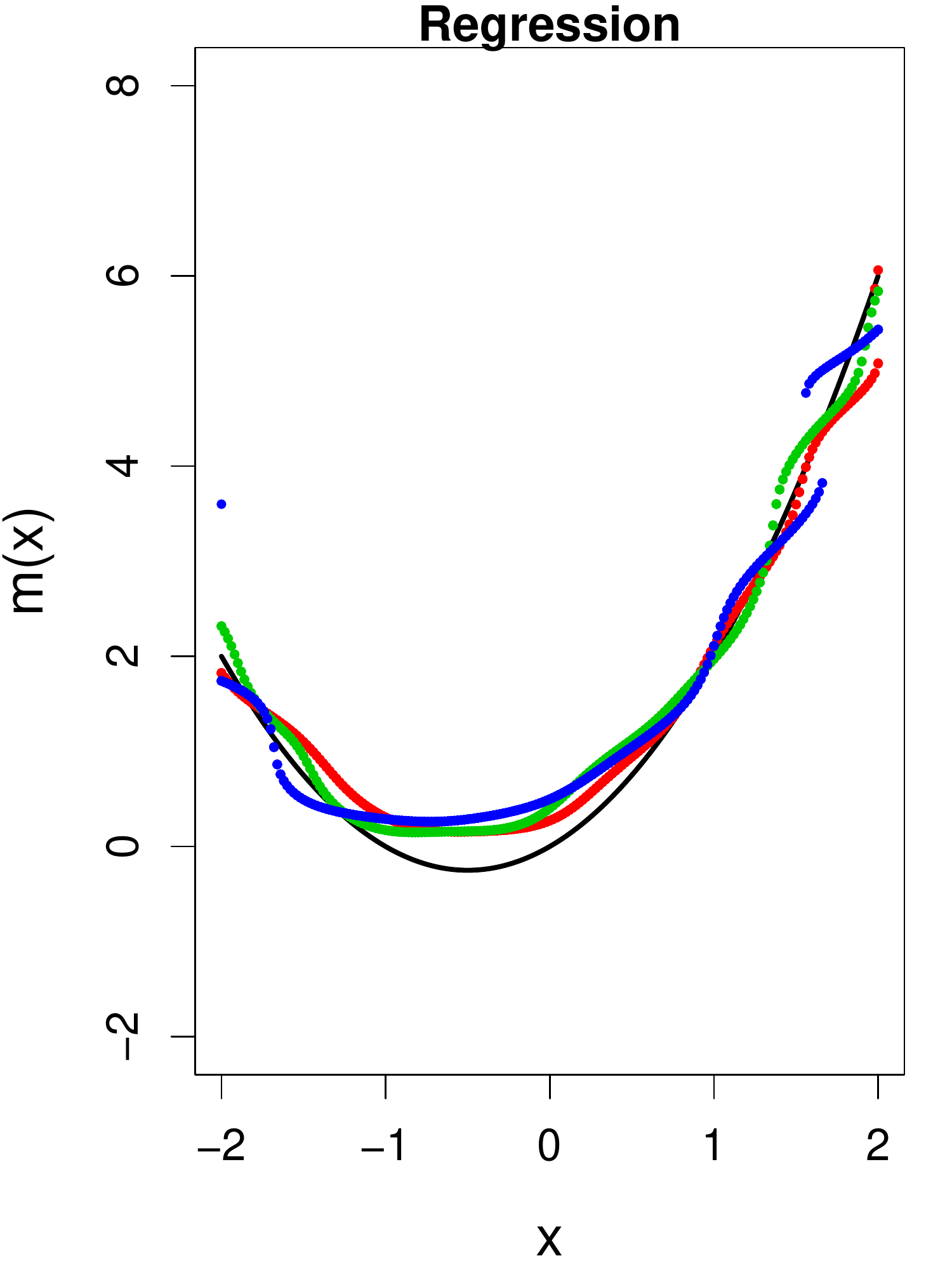} }
	\caption{Estimated mode curves resulting from eight choices of bandwidths $\bh$ under (C1).	These bandwidths are (a) $\tilde\bh_{\hbox {\tiny $M$}}$, the optimal bandwidths that minimize $\textrm{EISE}_{\hbox {\tiny $M$}}$; (b) $\bh_{\hbox {\tiny $M$}}$, the mode-based bandwidths involving CV; (c) $\bh^*_{\hbox {\tiny $B$}}$, the mode-based bandwidths involving bootstrap; (d) $\bh_{\hbox {\tiny $N$}}$, the normal reference; (e) $\tilde\bh_{\hbox {\tiny $D$}}$, the optimal bandwidths that minimize $\textrm{EISE}_{\hbox {\tiny $D$}}$; (f) $\bh_{\hbox {\tiny $D$}}$, the density-based bandwidths involving CV; (g) $\bh_{\hbox {\tiny $B$}}$, the density-based bandwidths involving bootstrap; (h) $\bh_{\hbox {\tiny $R$}}$, resulting from the regression-based method. 	In each panel, the black line depicts the true mode curve, the red, green, and blue lines are three estimated mode curves from the same method that yield $\textrm{EISE}_{\hbox {\tiny $M$}}$ being the first, second, and third quantiles among the 500 $\textrm{EISE}_{\hbox {\tiny $M$}}$'s for that method from the simulation, respectively.}
	\label{Sim1:curves}
\end{figure}

\begin{figure}[p]
	\centering
	\setlength{\linewidth}{0.2\textwidth}
	\subfigure[]{ \includegraphics[width=\linewidth]{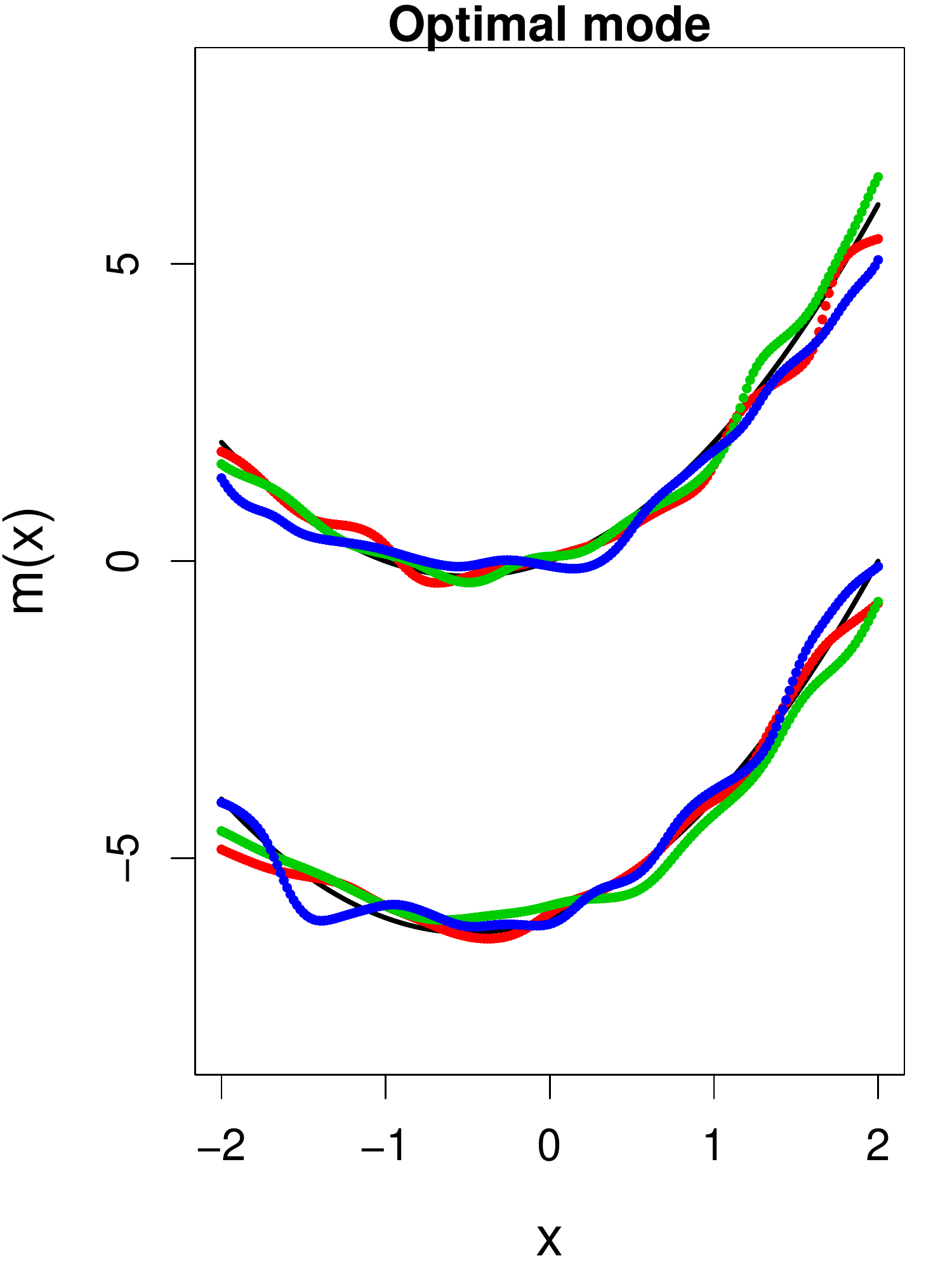} }
	\subfigure[]{ \includegraphics[width=\linewidth]{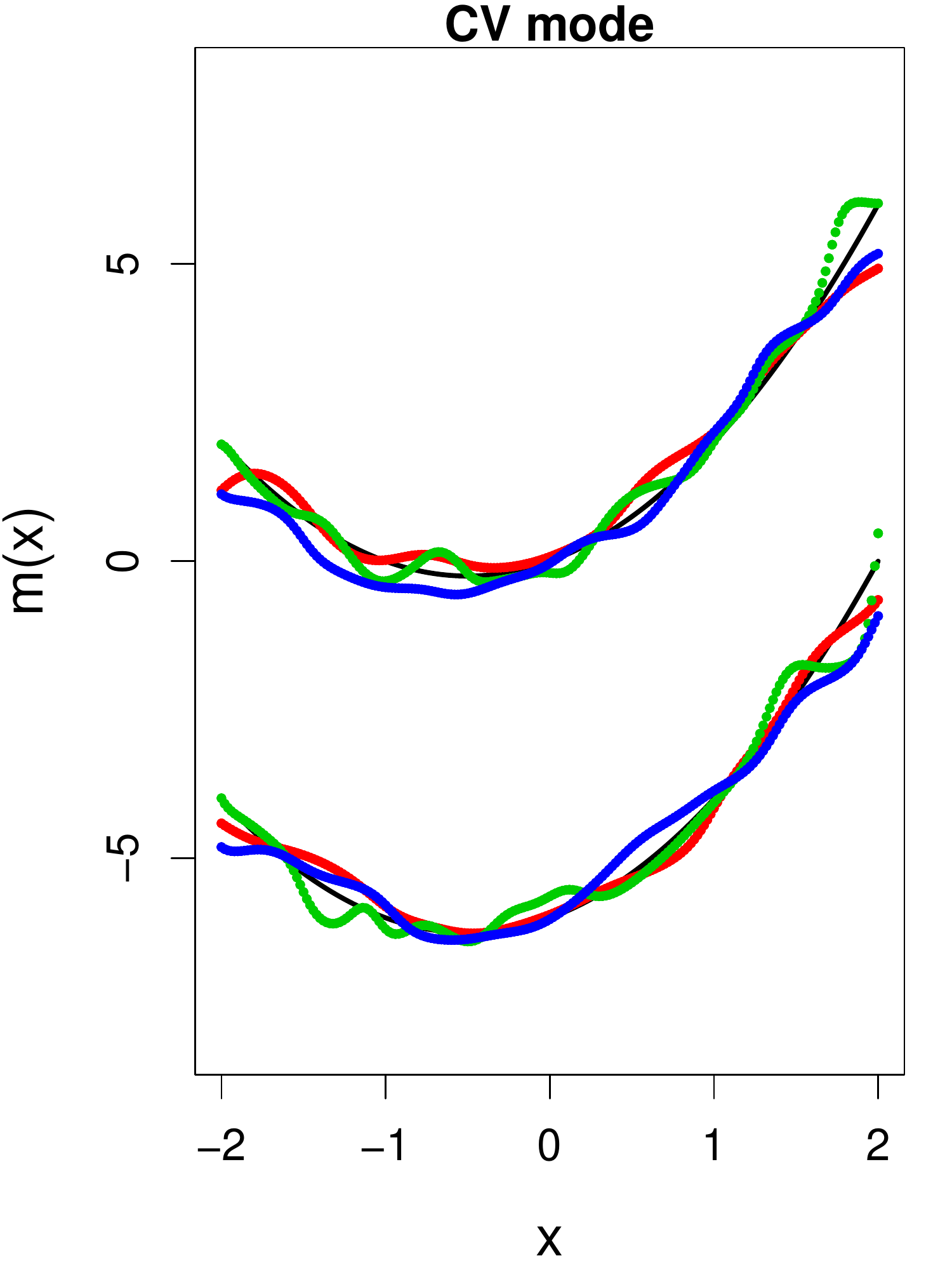} }
	\subfigure[]{ \includegraphics[width=\linewidth]{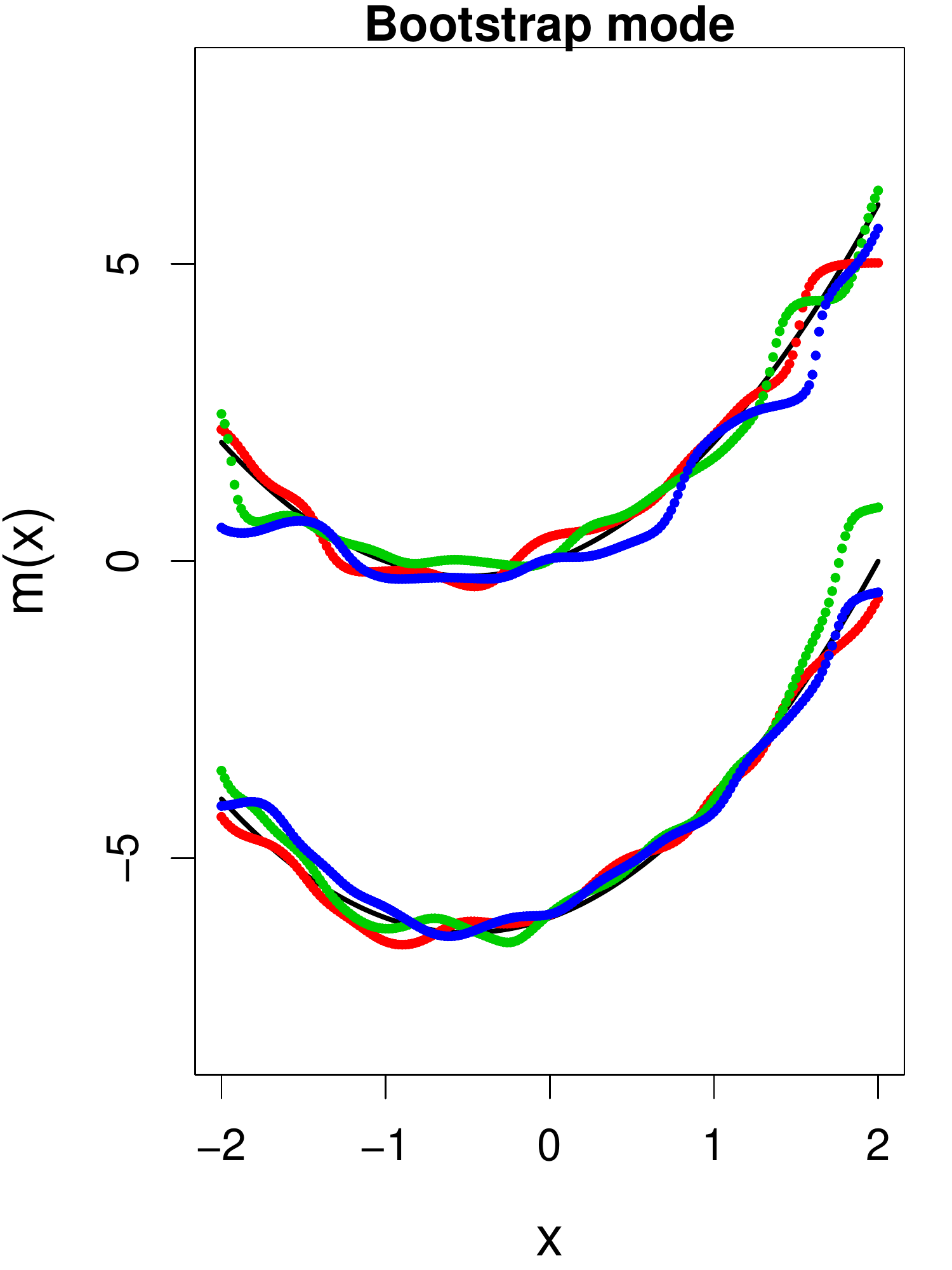} }
	\subfigure[]{ \includegraphics[width=\linewidth]{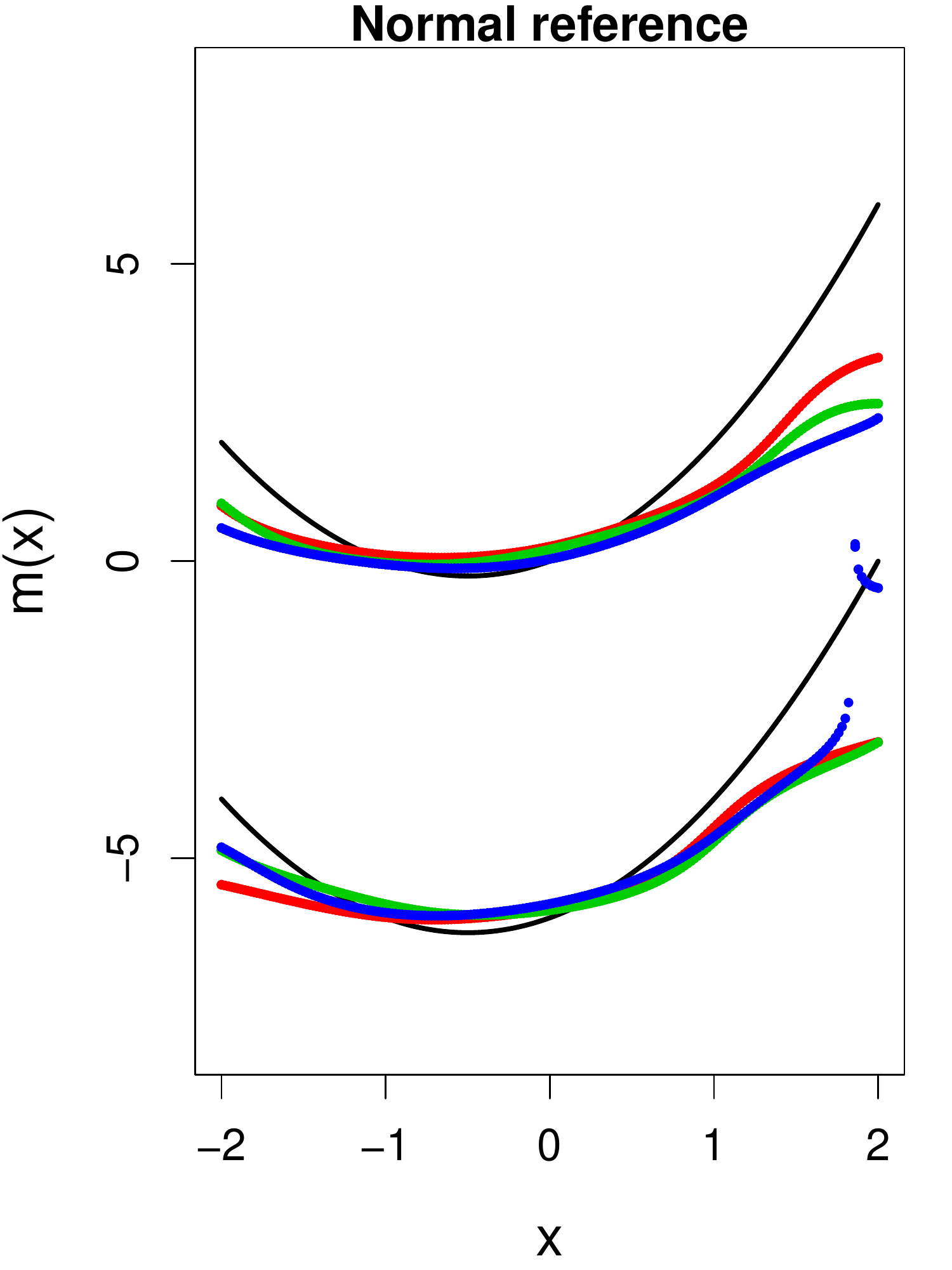} }\\
	\subfigure[]{ \includegraphics[width=\linewidth]{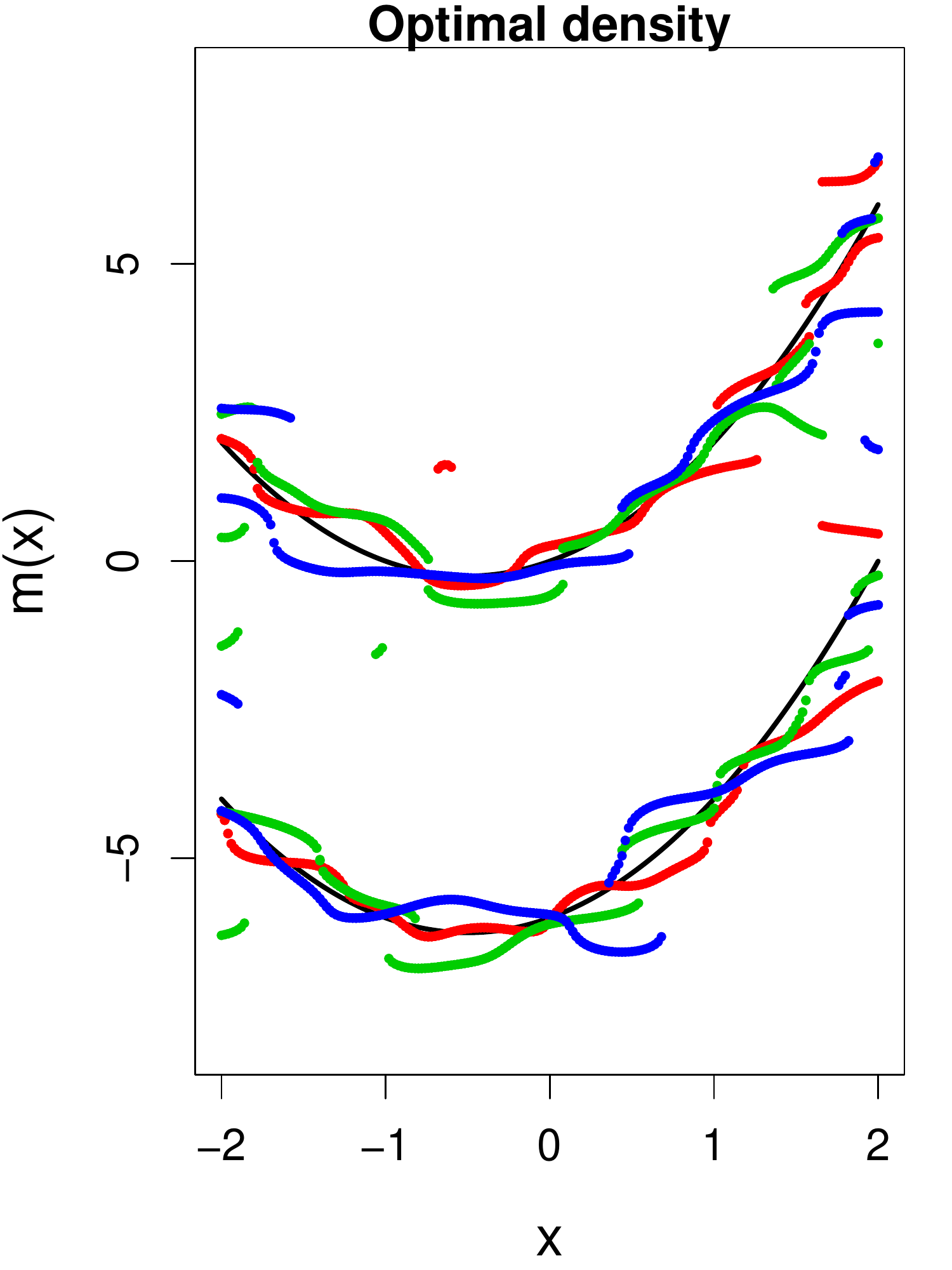} }
	\subfigure[]{ \includegraphics[width=\linewidth]{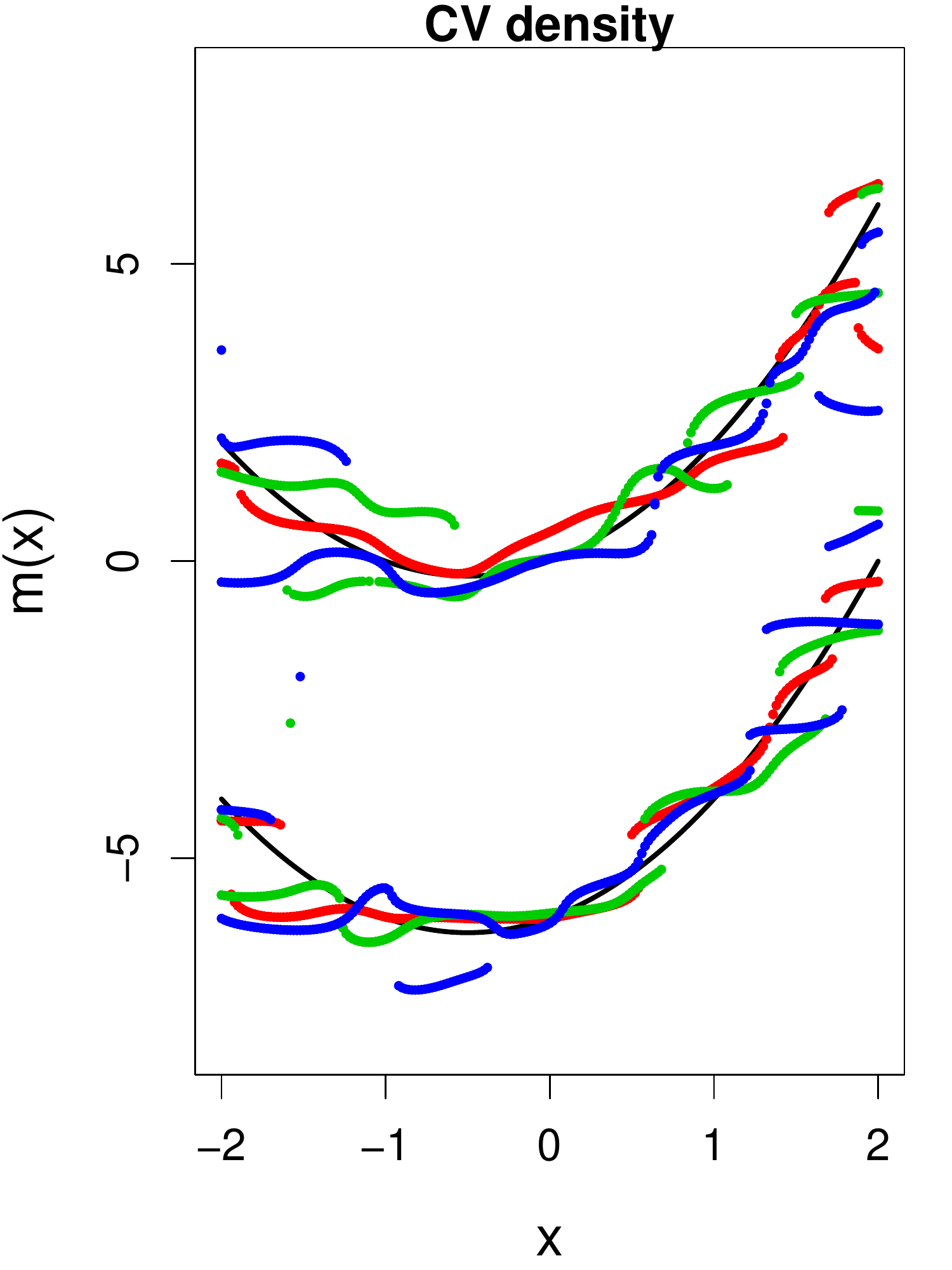} }
	\subfigure[]{ \includegraphics[width=\linewidth]{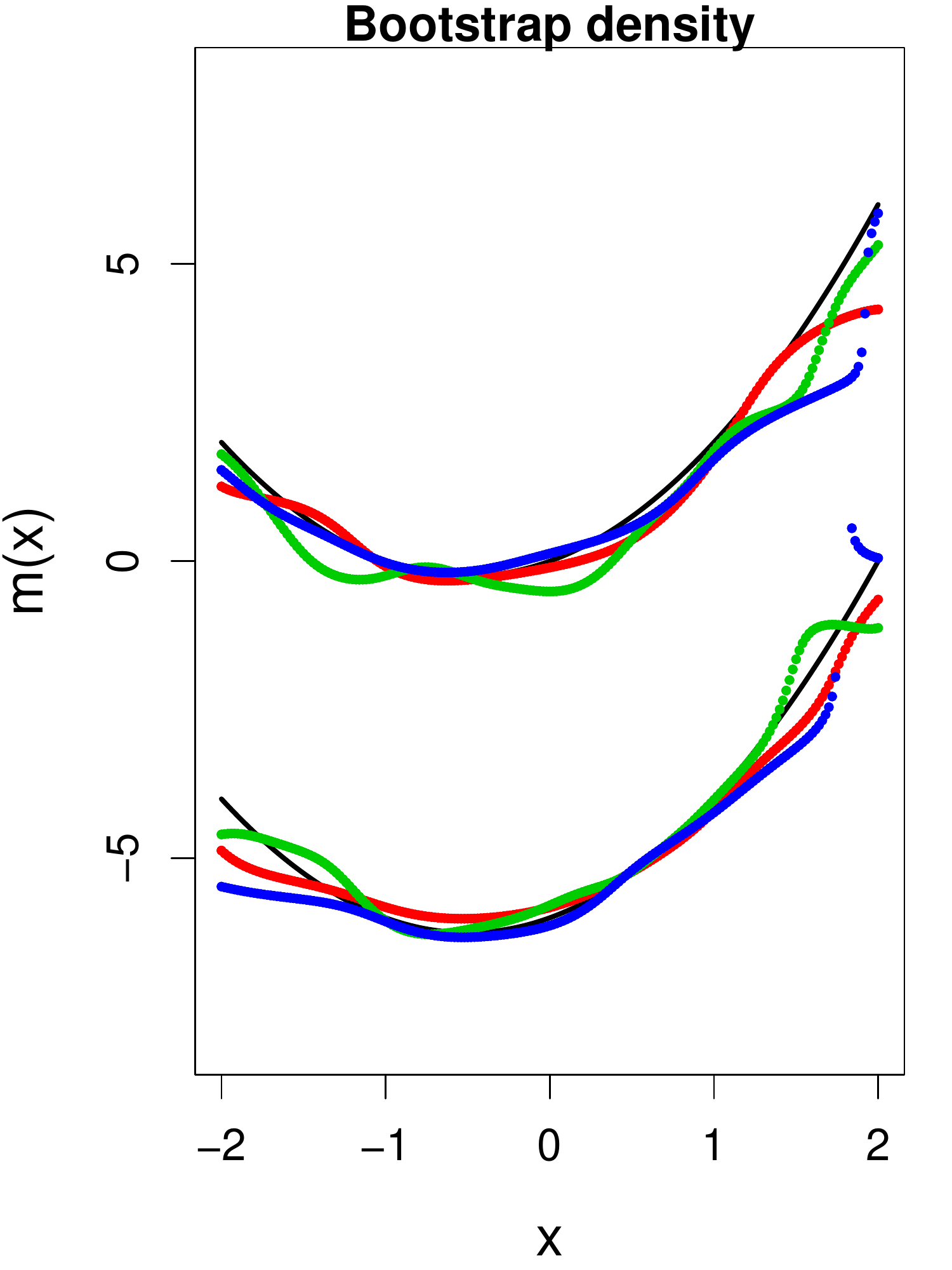} }
	\subfigure[]{ \includegraphics[width=\linewidth]{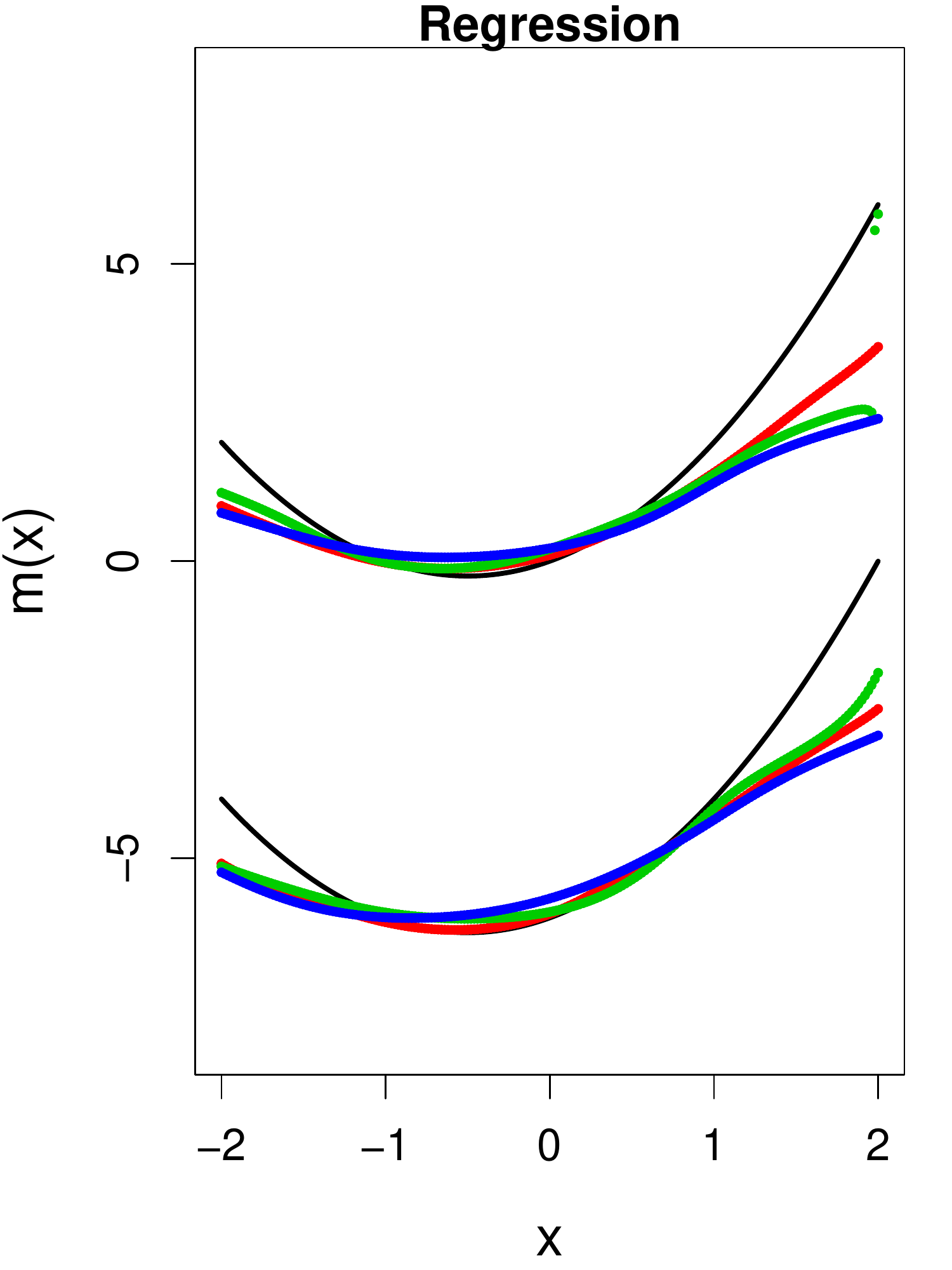} }
	\caption{Estimated mode curves resulting from eight choices of bandwidths $\bh$ under (C2). The correspondence of the eight panels to eight methods and the correspondence of different colors to different lines are the same as those in Figure~\ref{Sim1:curves}.}
	\label{Sim2:curves}
\end{figure}

\begin{figure}[p]
	\centering
	\setlength{\linewidth}{0.2\textwidth}
	\subfigure[]{ \includegraphics[width=\linewidth]{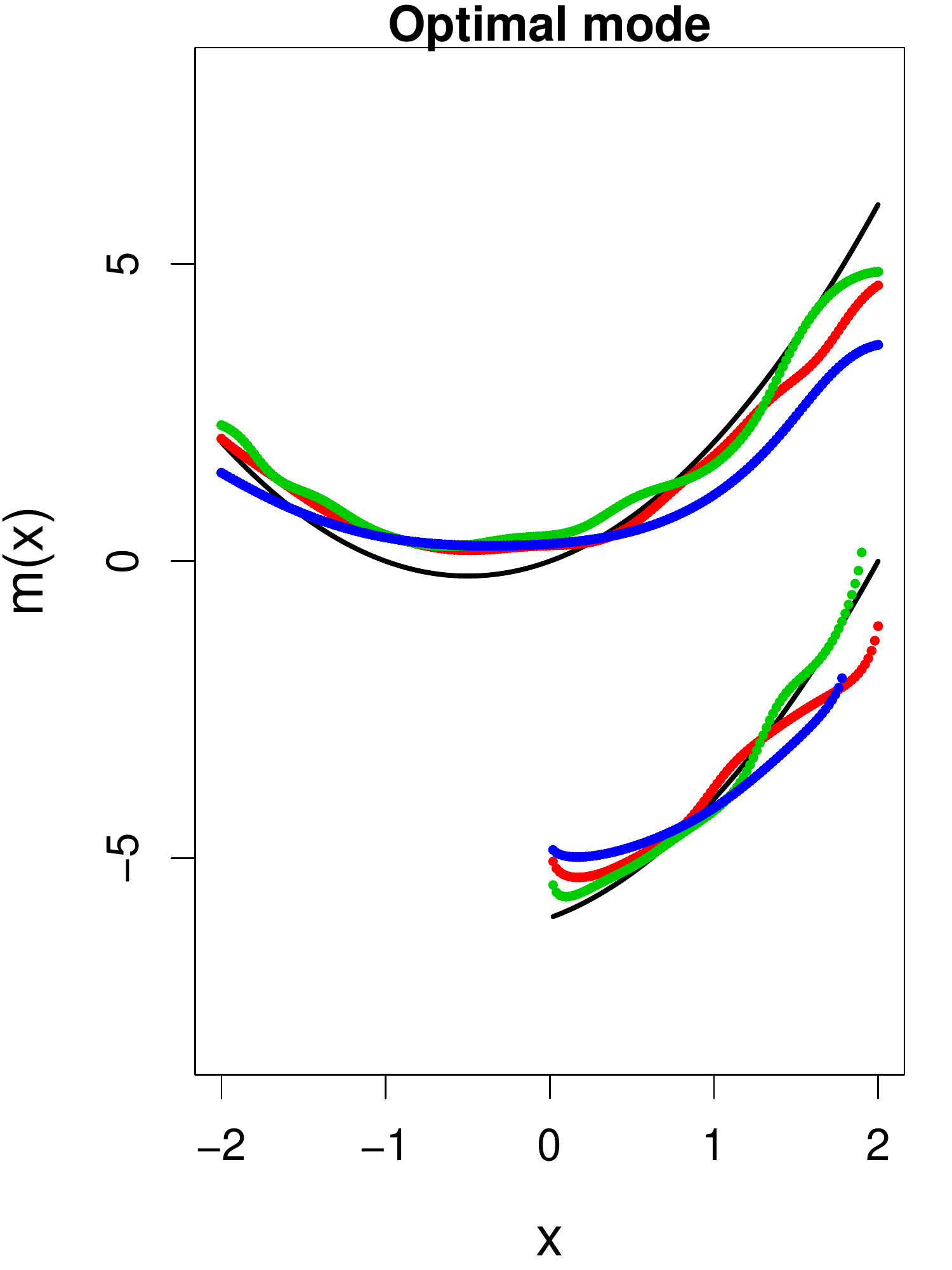} }
	\subfigure[]{ \includegraphics[width=\linewidth]{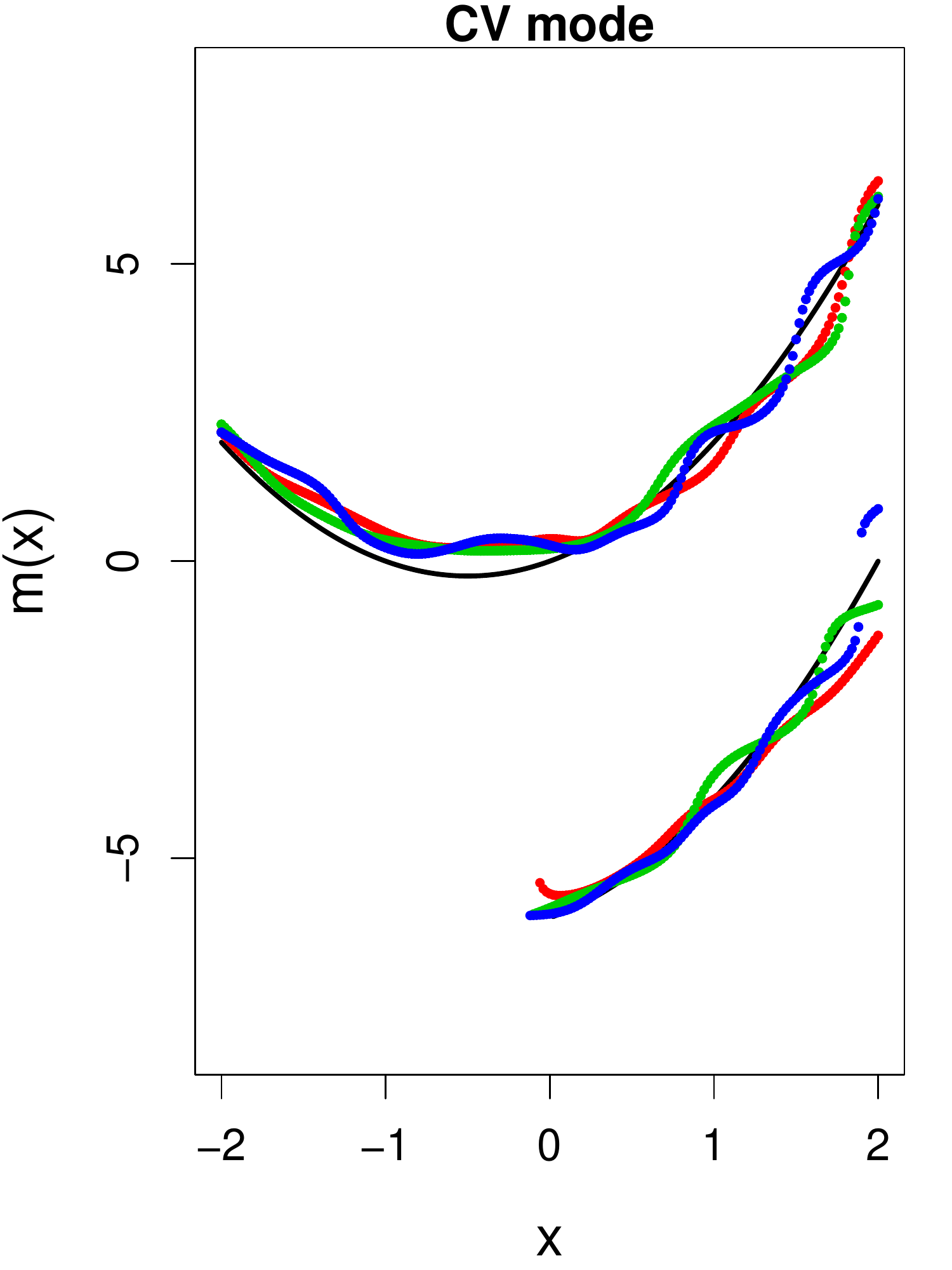} }
	\subfigure[]{ \includegraphics[width=\linewidth]{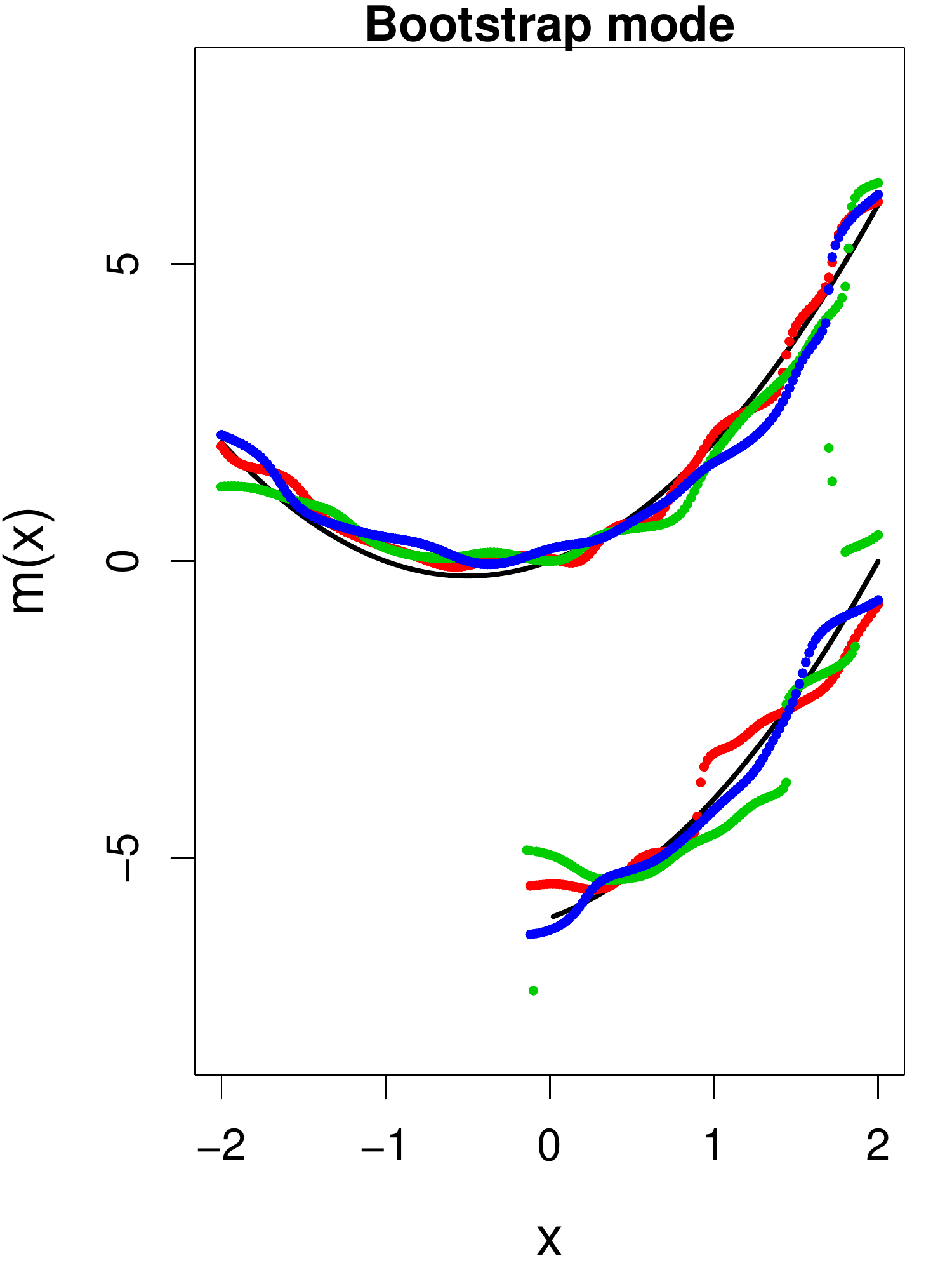} }
	\subfigure[]{ \includegraphics[width=\linewidth]{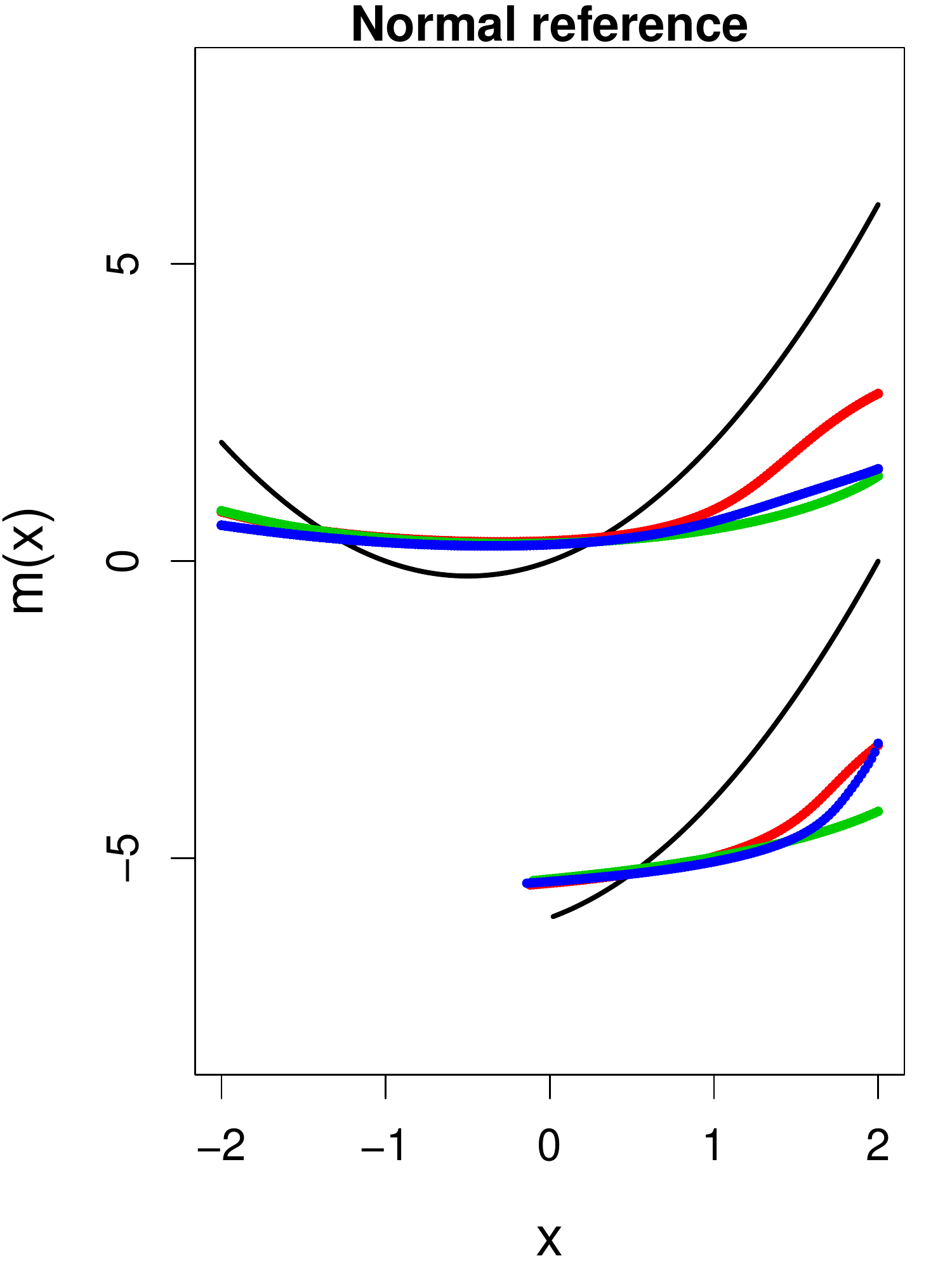} }\\
	\subfigure[]{ \includegraphics[width=\linewidth]{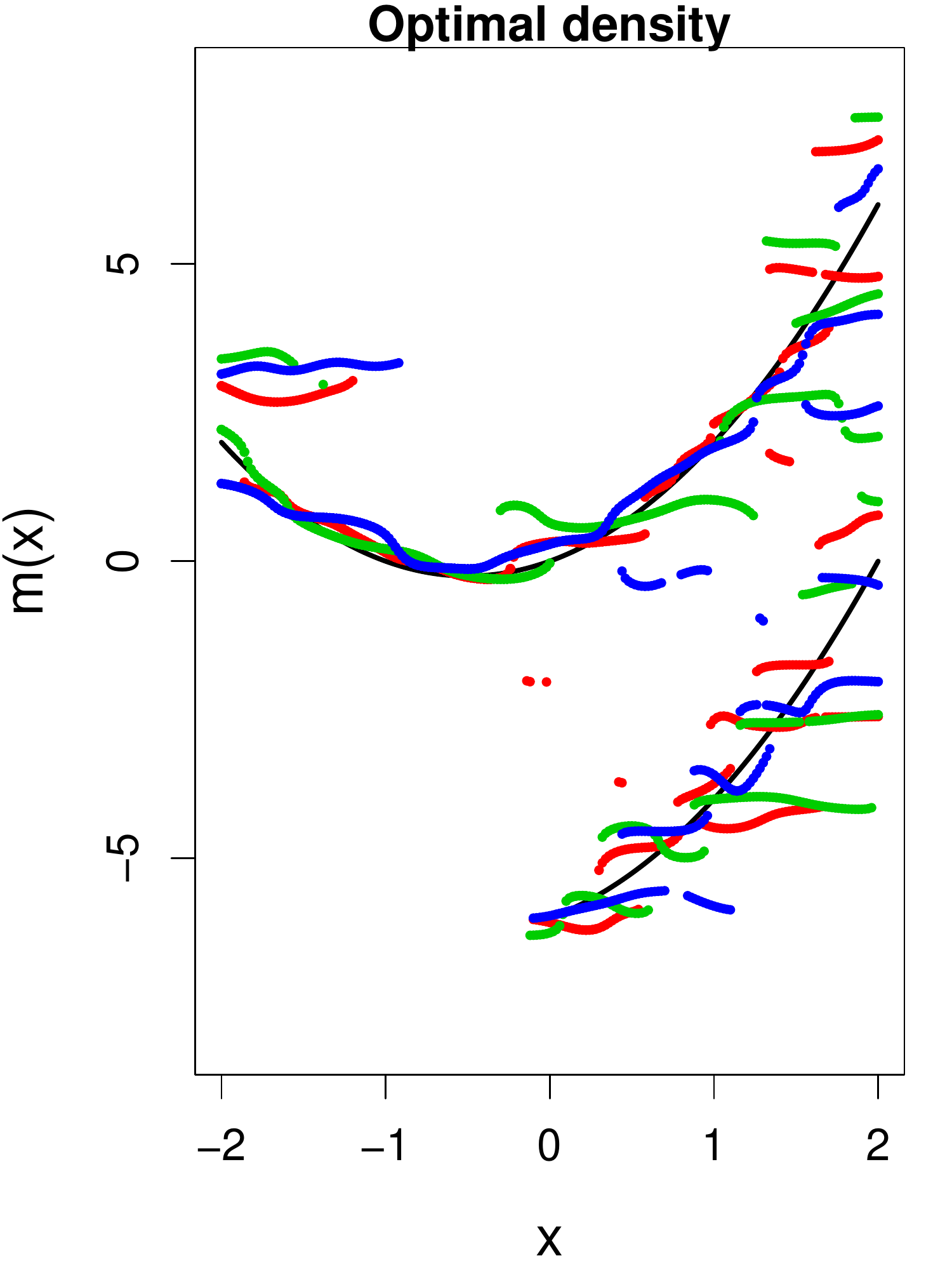} }
	\subfigure[]{ \includegraphics[width=\linewidth]{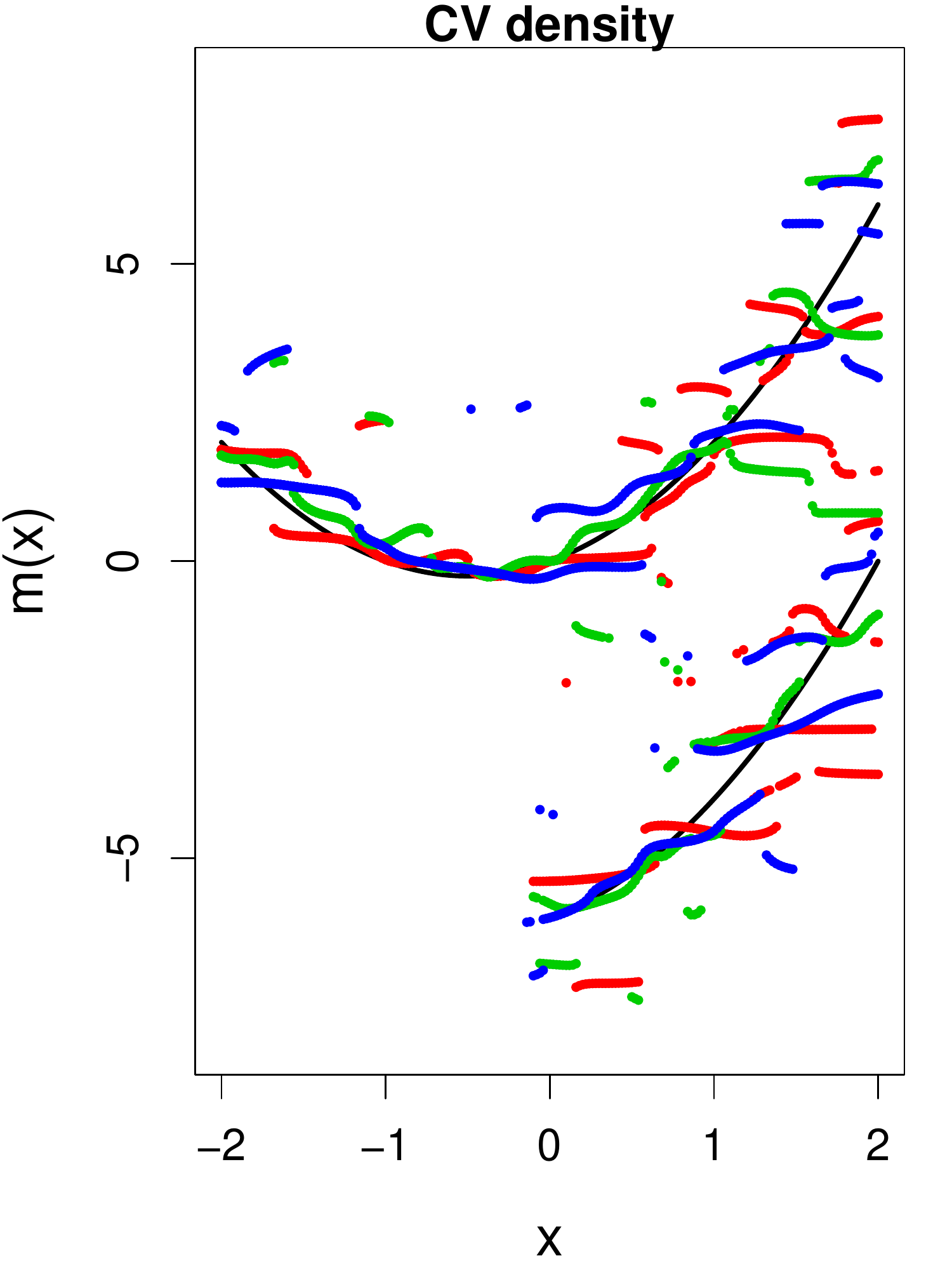} }
	\subfigure[]{ \includegraphics[width=\linewidth]{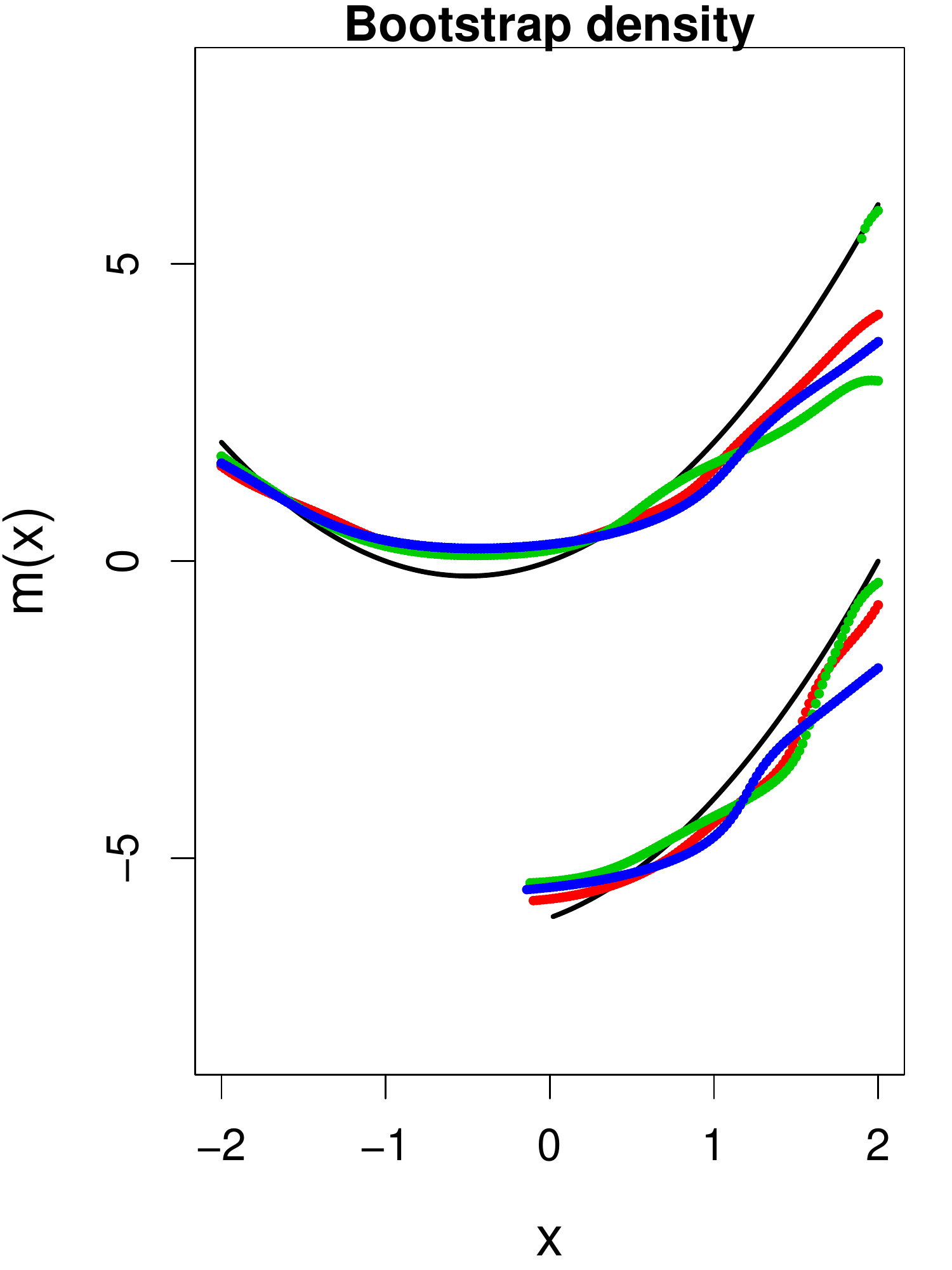} }
	\subfigure[]{ \includegraphics[width=\linewidth]{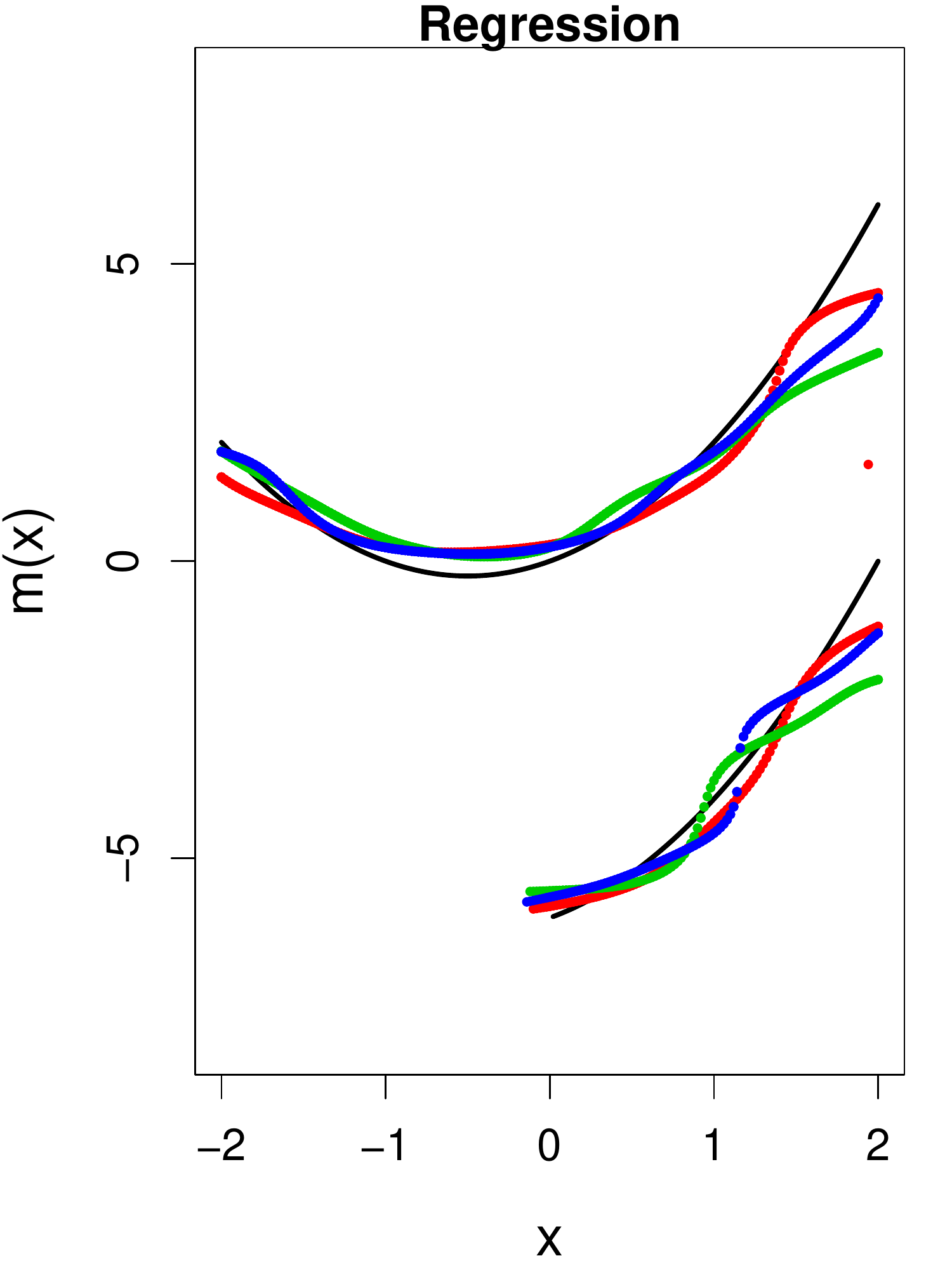} }
	\caption{Estimated mode curves resulting from eight choices of bandwidths $\bh$ under (C3). The correspondence of the eight panels to eight methods and the correspondence of different colors to different lines are the same as those in Figure~\ref{Sim1:curves}.}
	\label{Sim3:curves}
\end{figure}

\begin{figure}
	\centering
	\setlength{\linewidth}{0.2\textwidth}
	\subfigure[]{ \includegraphics[width=\linewidth]{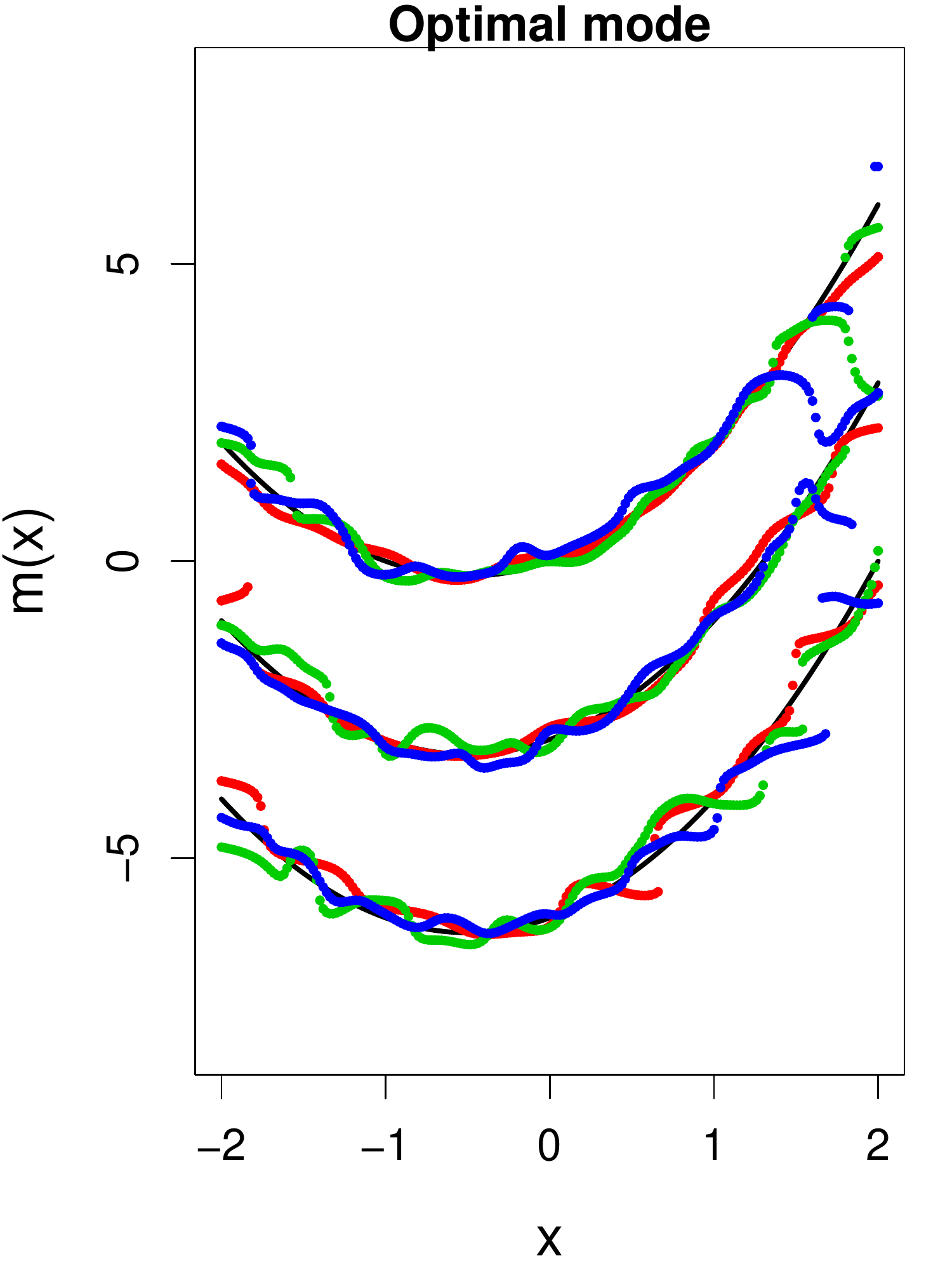} }
	\subfigure[]{ \includegraphics[width=\linewidth]{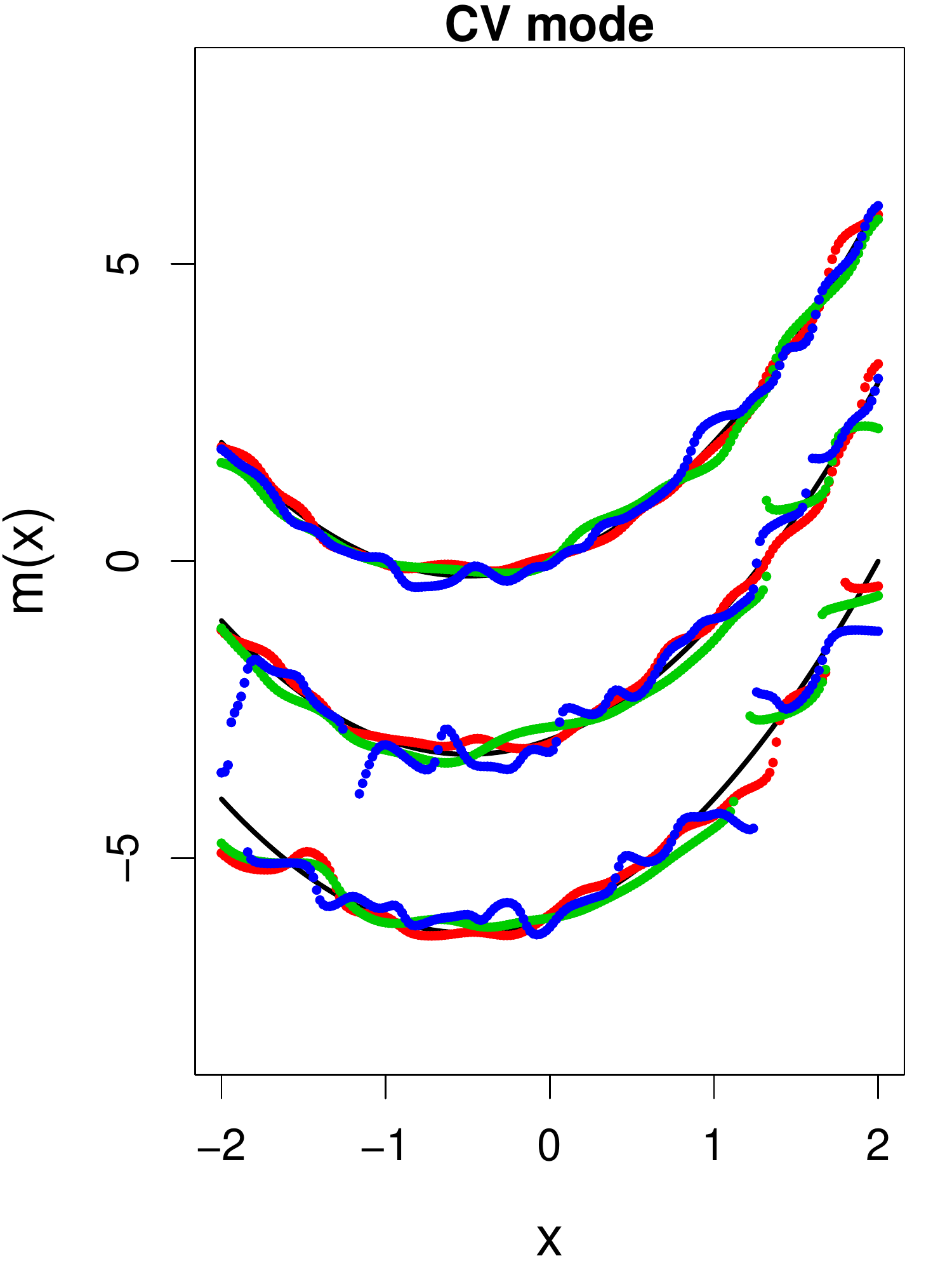} }
	\subfigure[]{ \includegraphics[width=\linewidth]{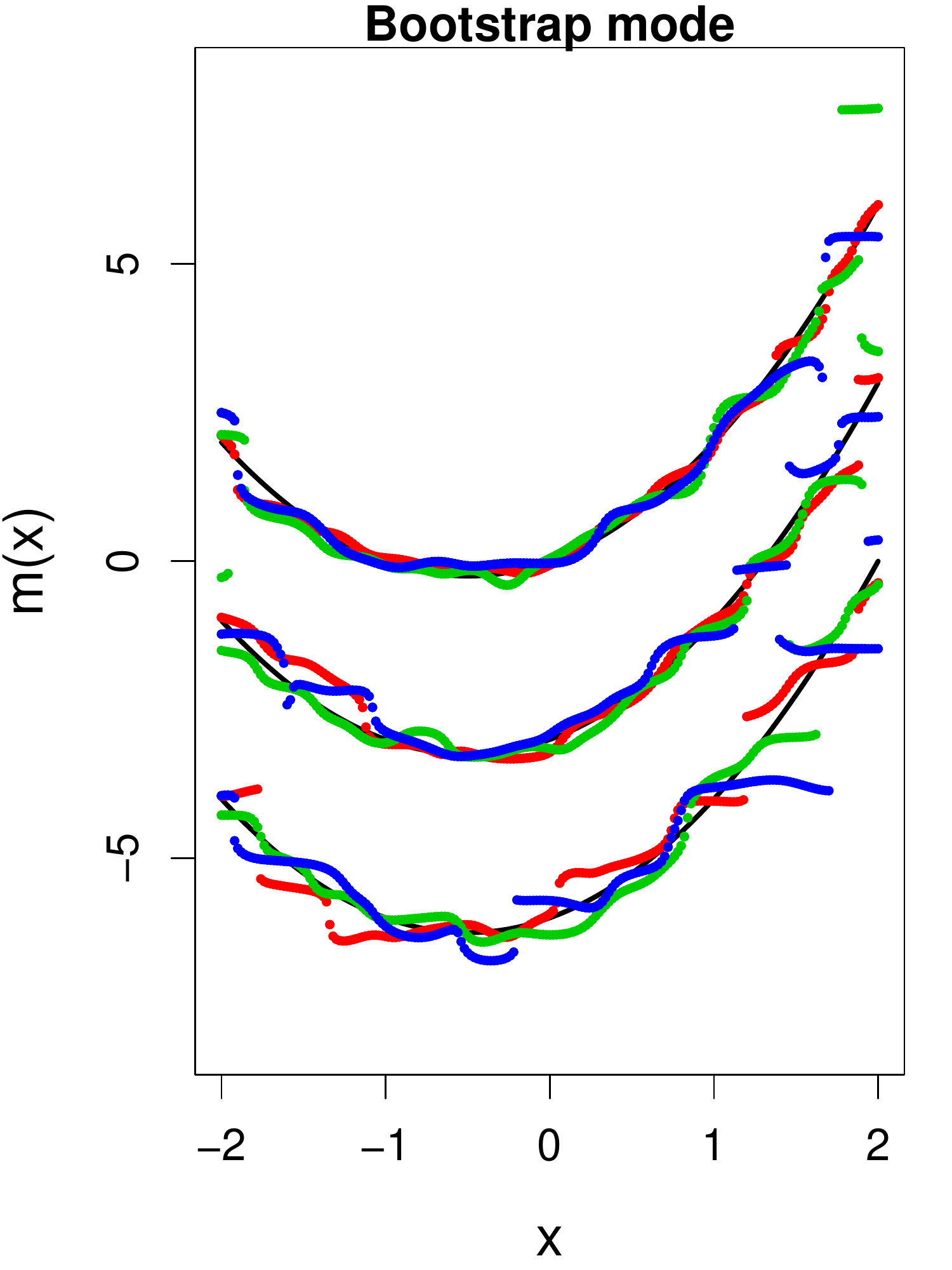} }
	\subfigure[]{ \includegraphics[width=\linewidth]{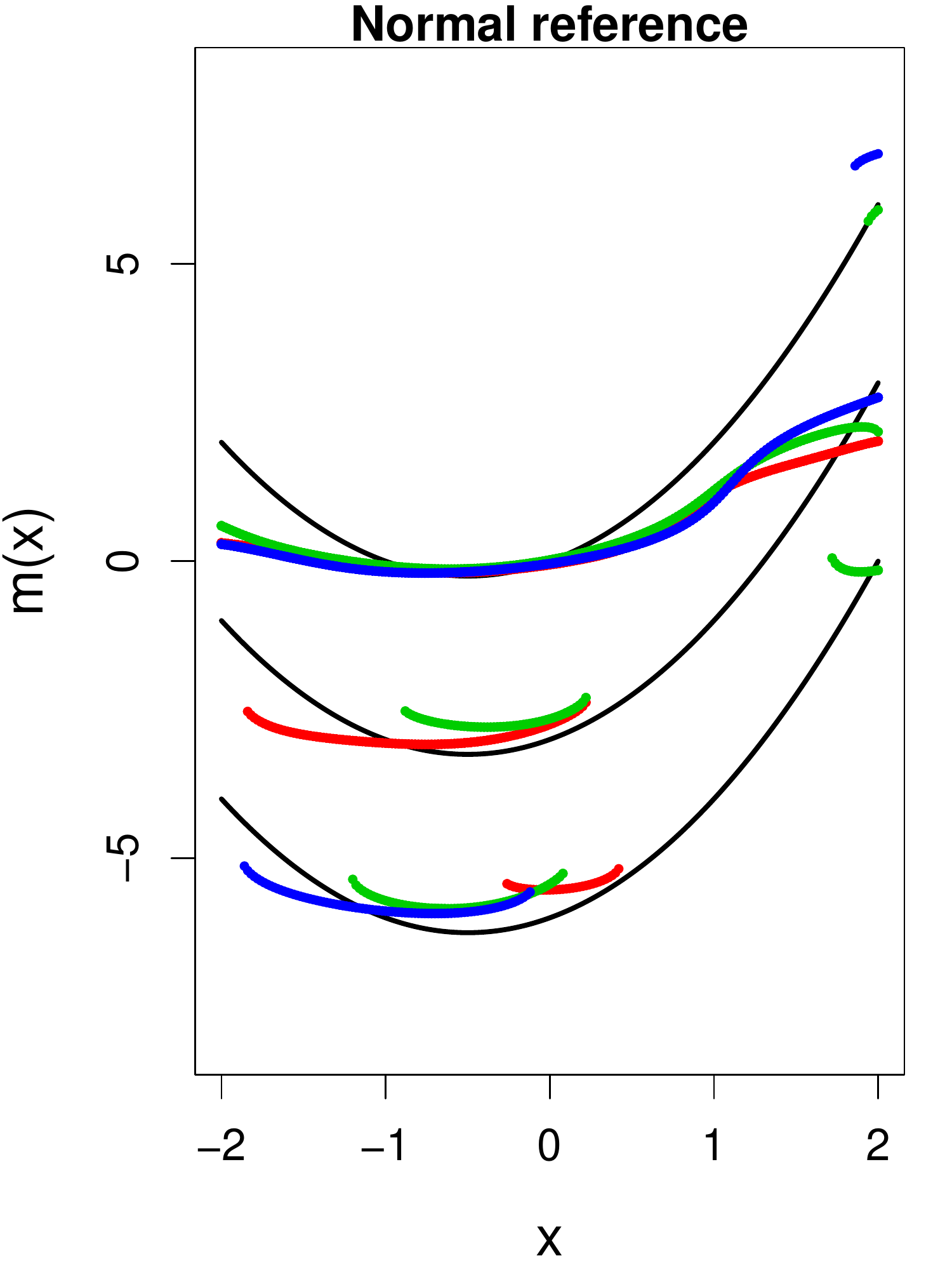} }\\
	\subfigure[]{ \includegraphics[width=\linewidth]{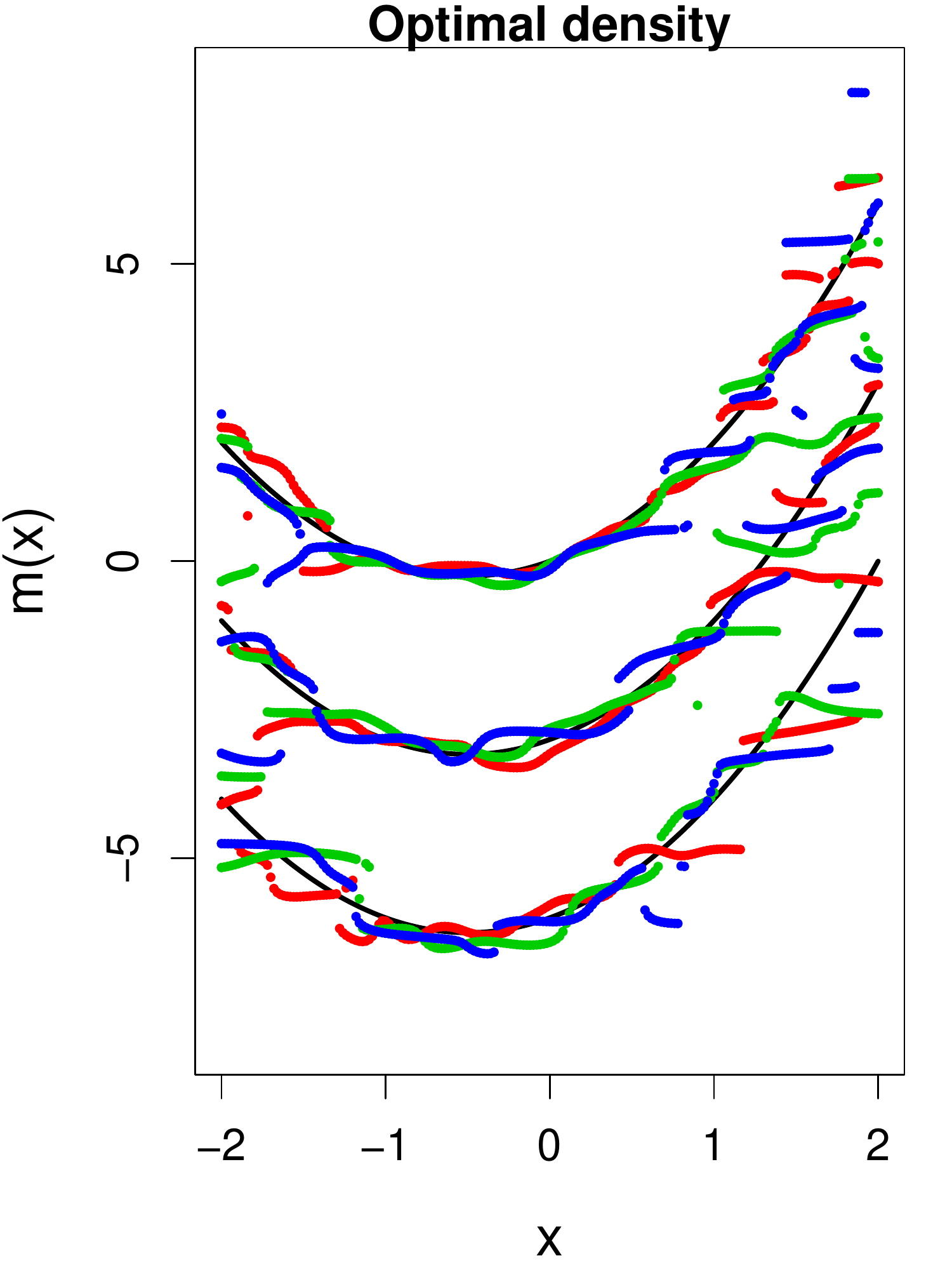} }
	\subfigure[]{ \includegraphics[width=\linewidth]{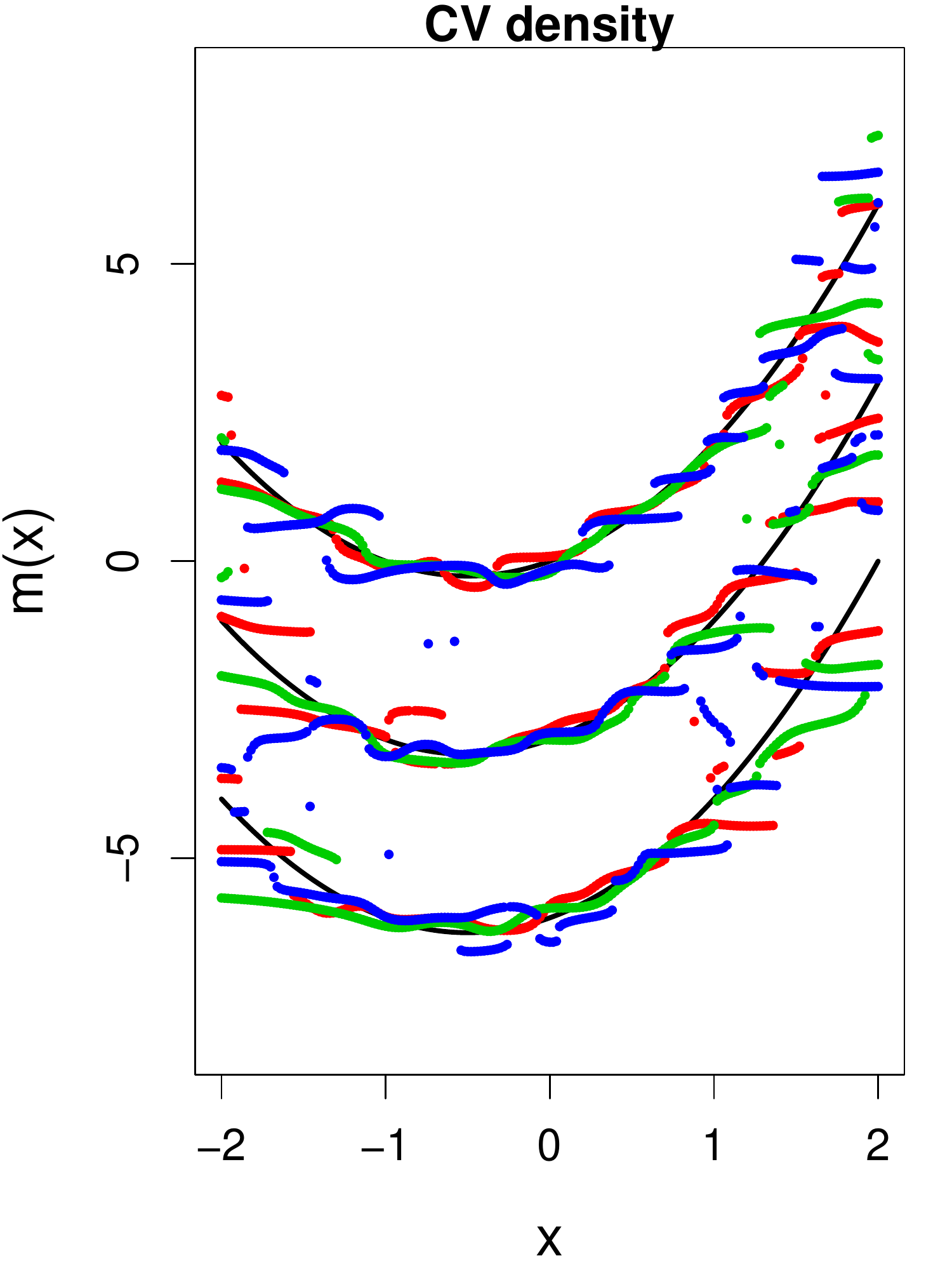} }
	\subfigure[]{ \includegraphics[width=\linewidth]{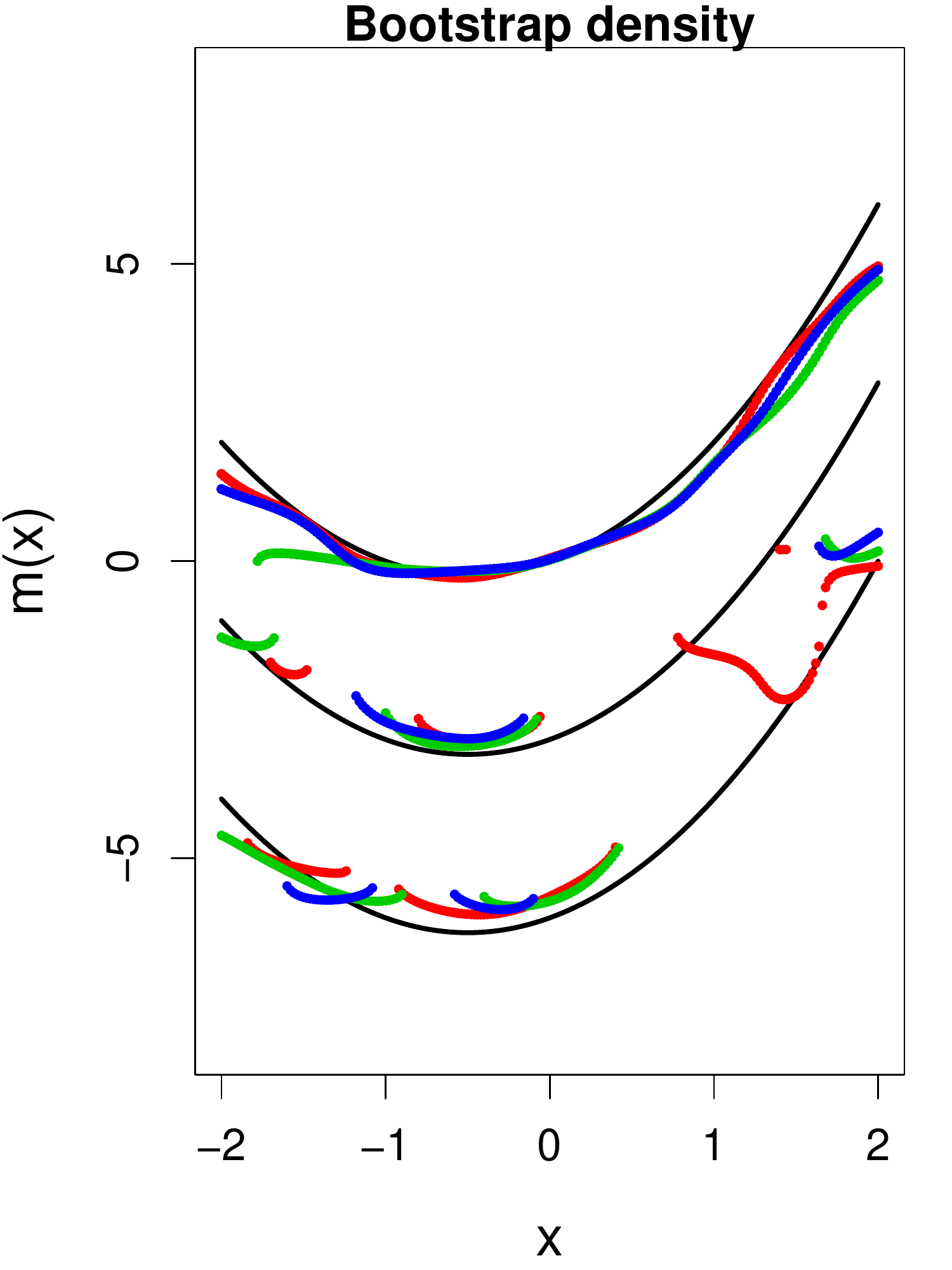} }
	\subfigure[]{ \includegraphics[width=\linewidth]{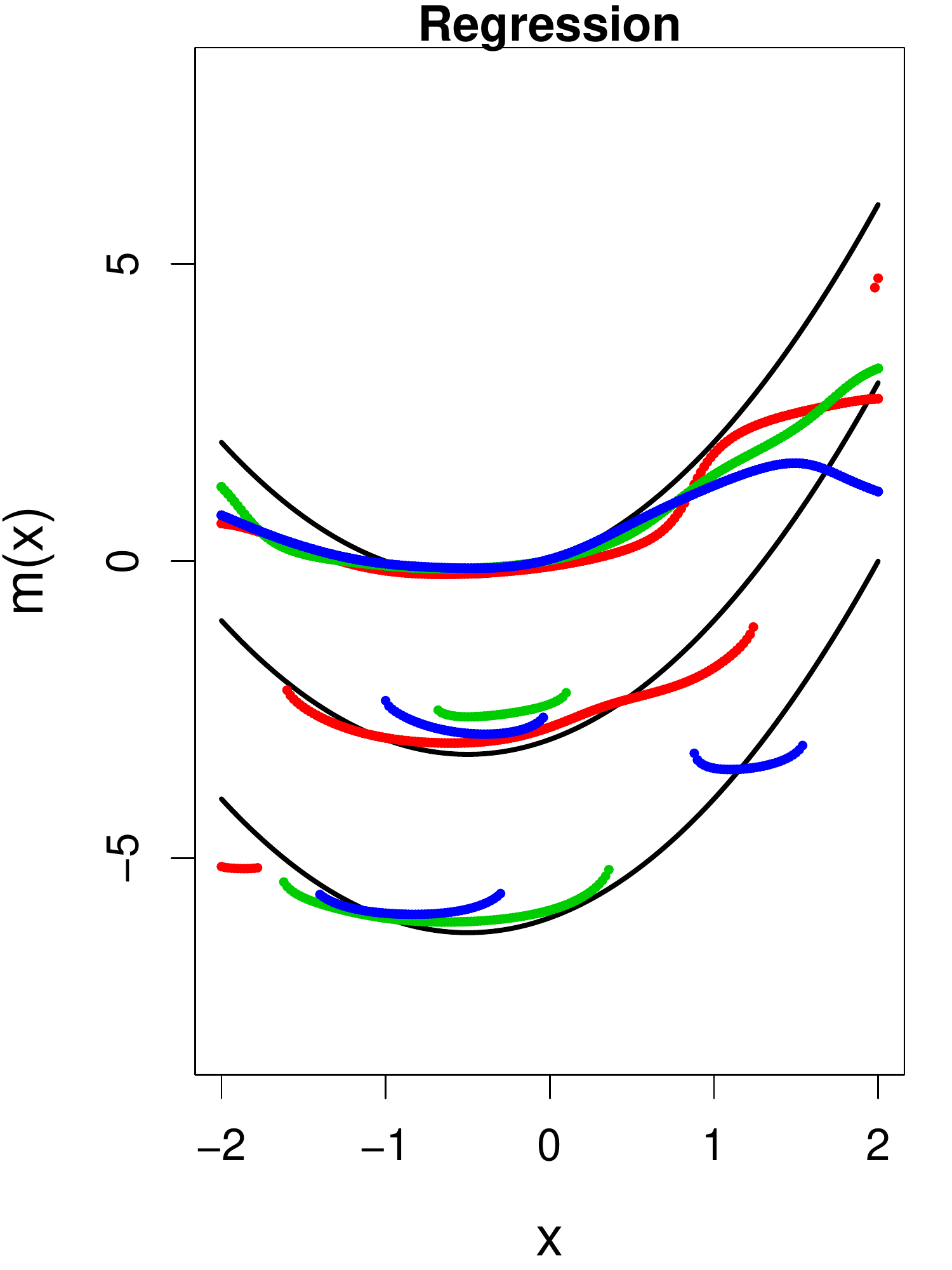} }
	\caption{Estimated mode curves resulting from eight choices of bandwidths $\bh$ under (C4). The correspondence of the eight panels to eight methods and the correspondence of different colors to different lines are the same as those in Figure~\ref{Sim1:curves}.}
	\label{Sim4:curves}
\end{figure}

\begin{figure}
	\centering
	\setlength{\linewidth}{0.2\textwidth}
	\subfigure[]{ \includegraphics[width=\linewidth]{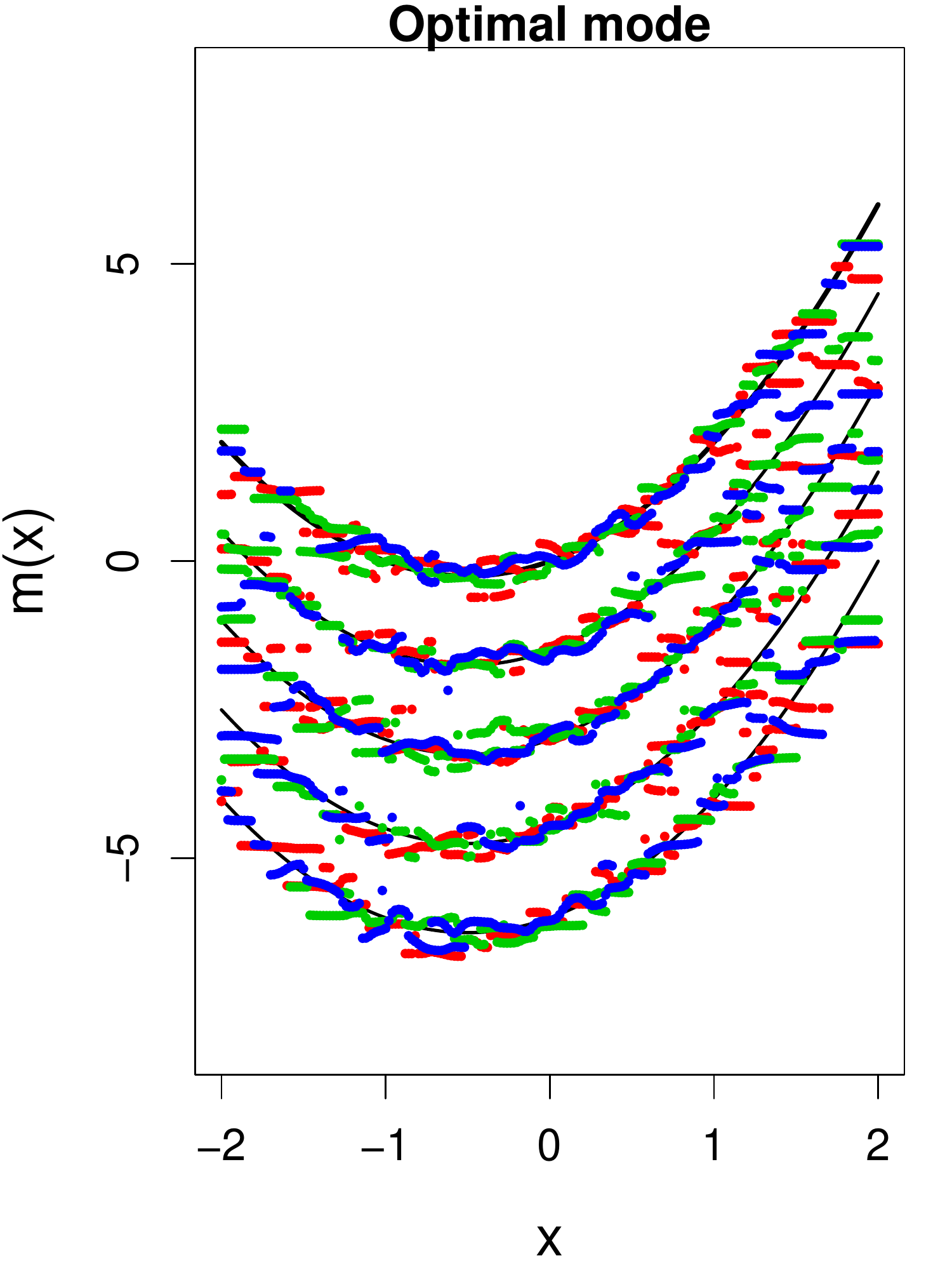} }
	\subfigure[]{ \includegraphics[width=\linewidth]{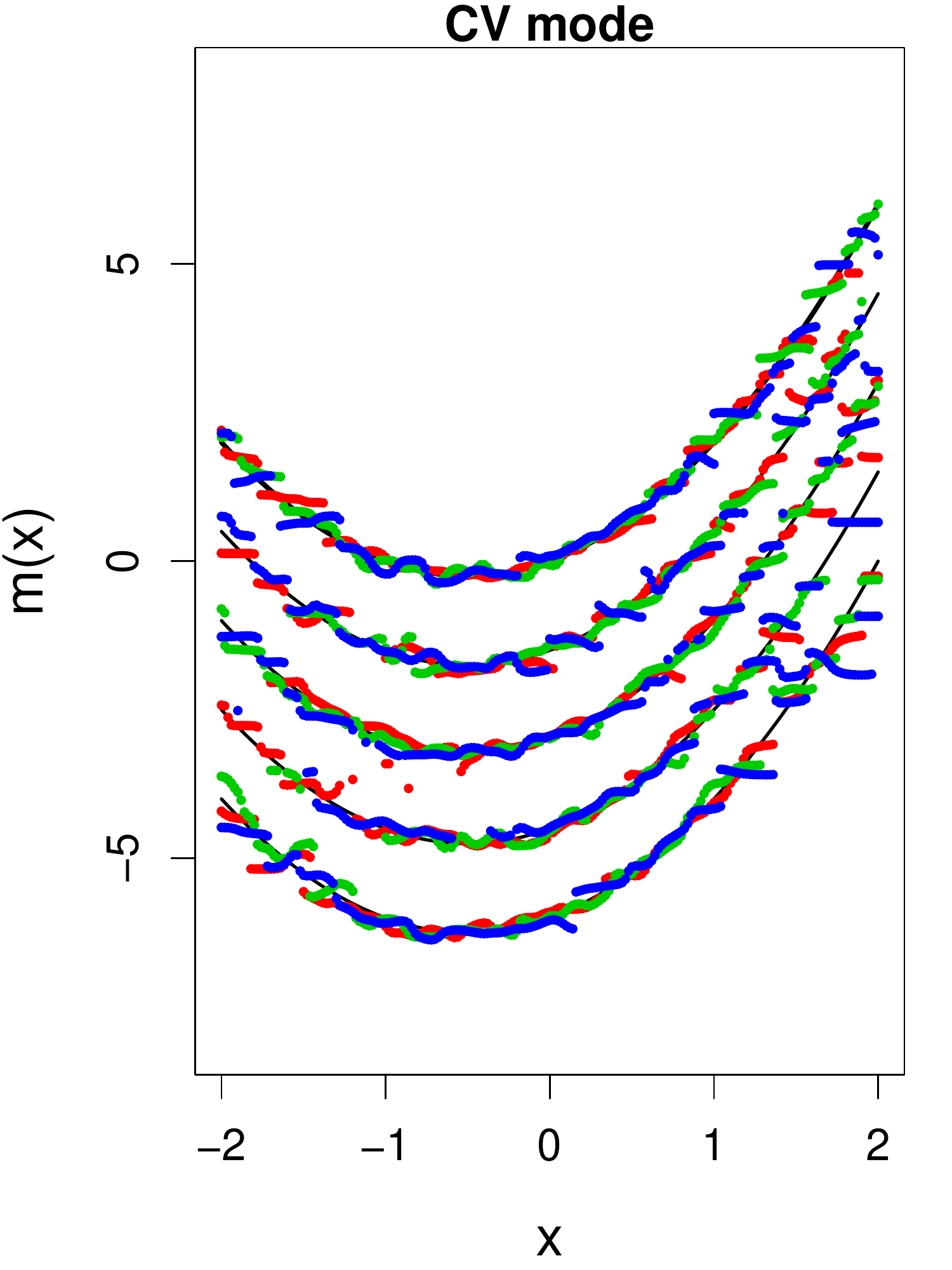} }
	\subfigure[]{ \includegraphics[width=\linewidth]{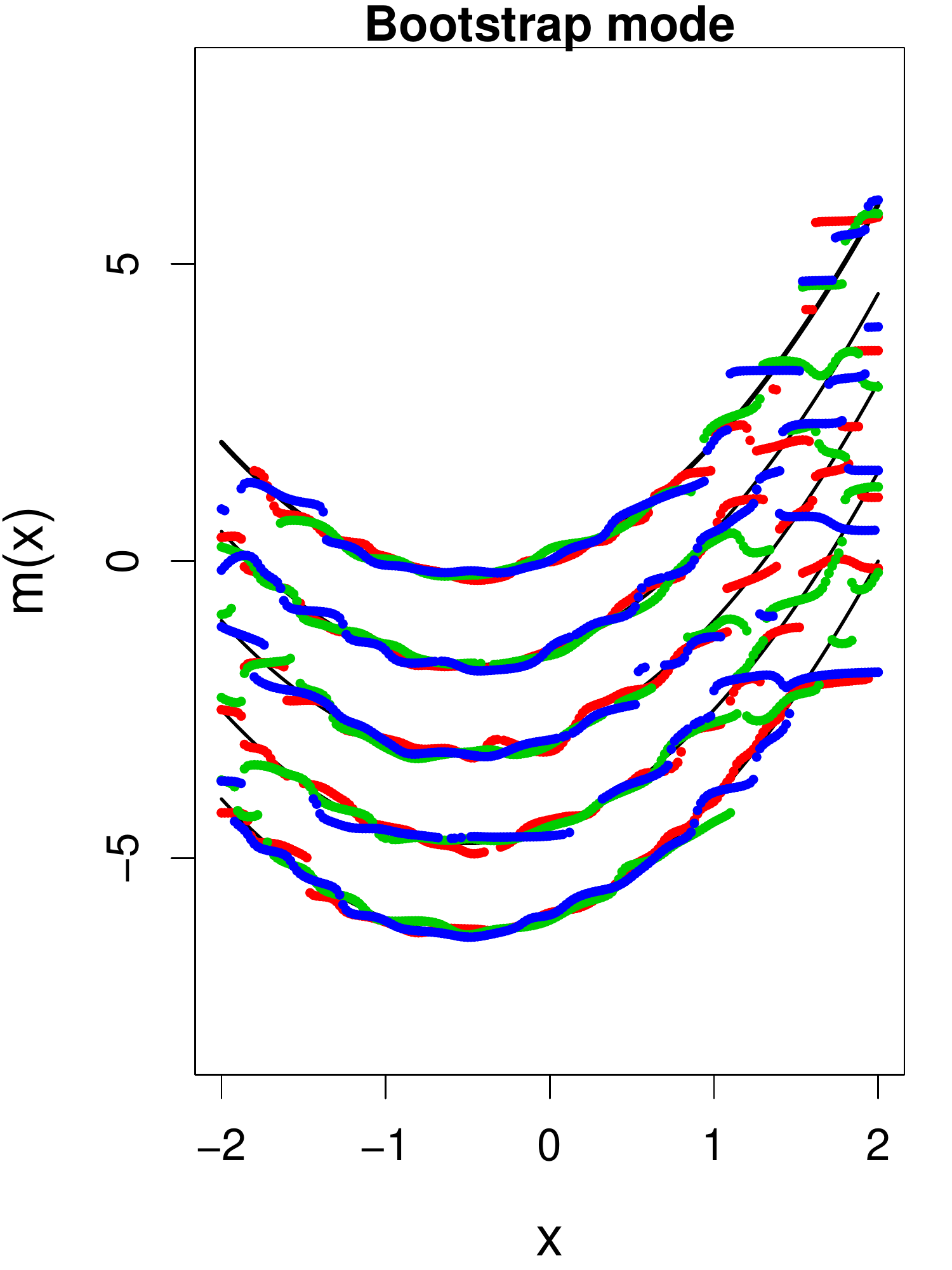} }
	\subfigure[]{ \includegraphics[width=\linewidth]{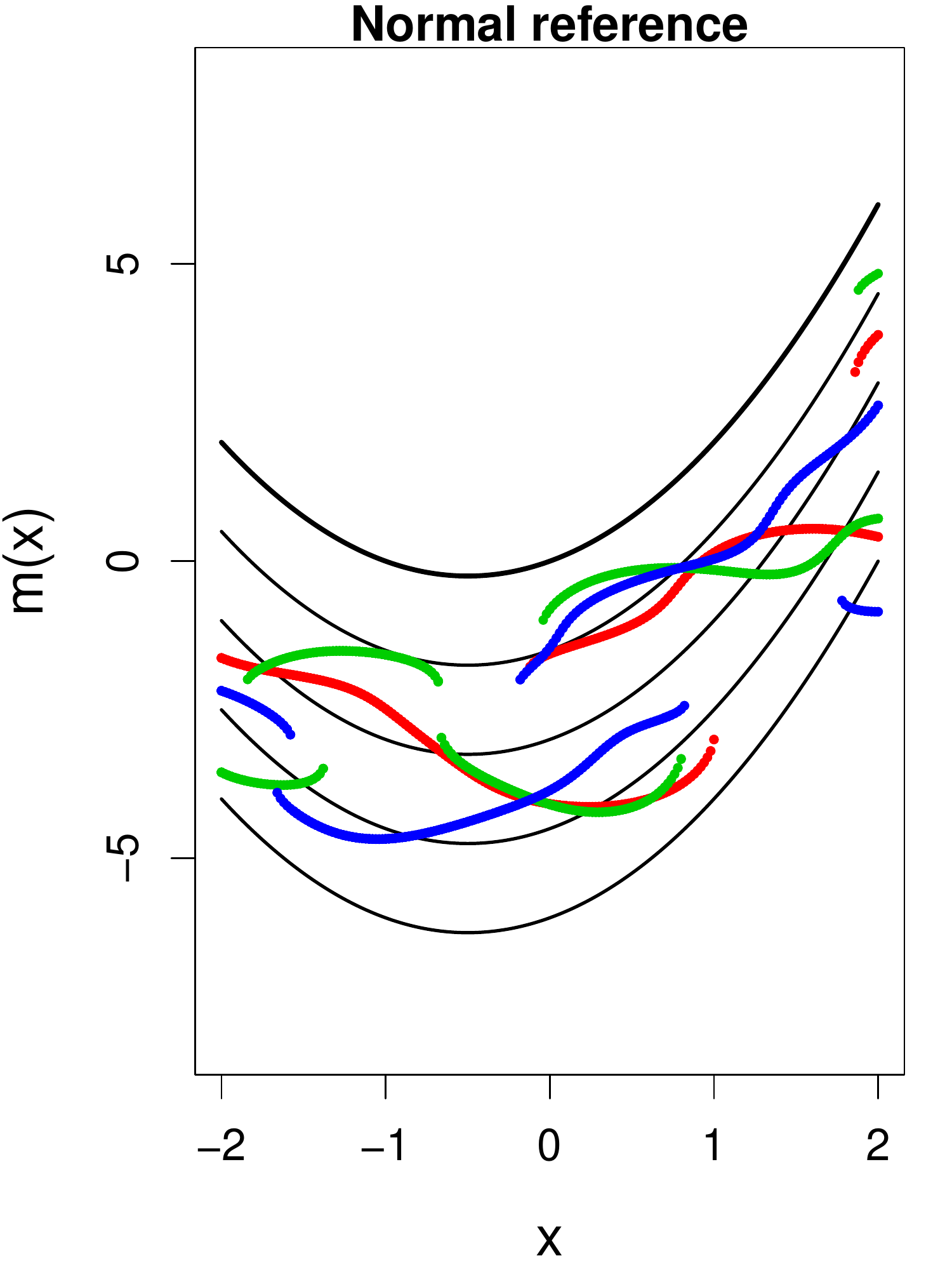} }\\
	\subfigure[]{ \includegraphics[width=\linewidth]{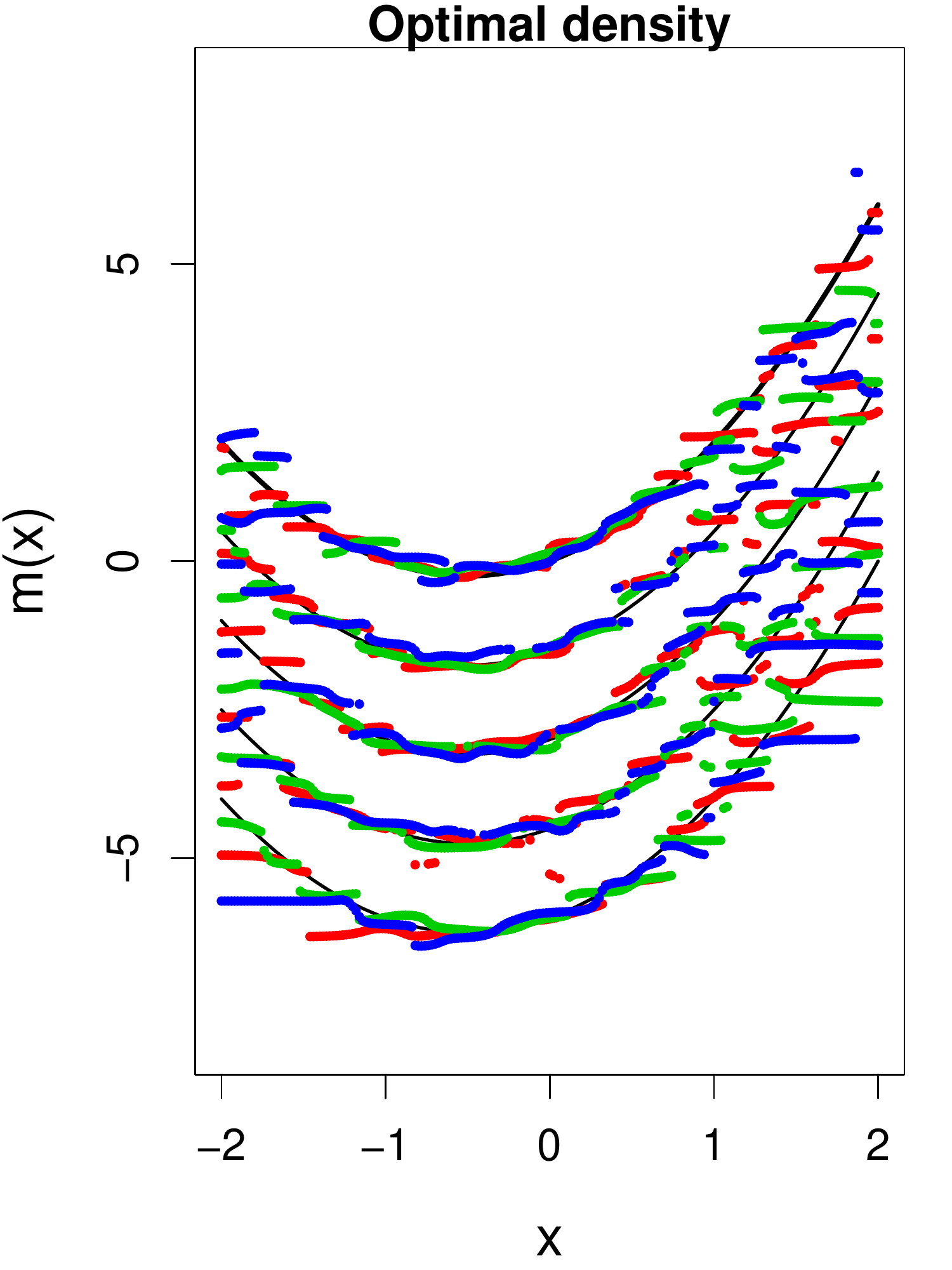} }
	\subfigure[]{ \includegraphics[width=\linewidth]{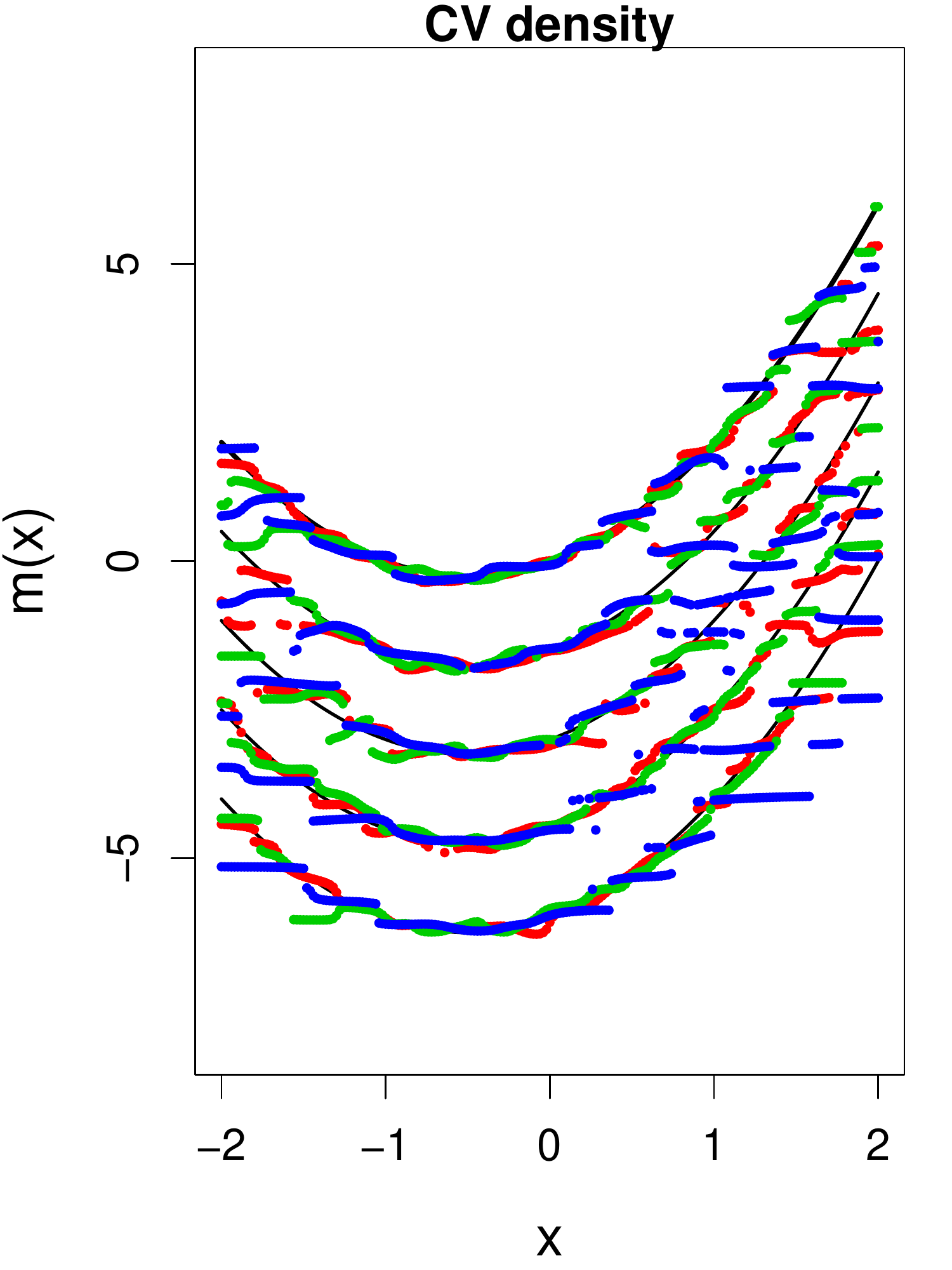} }
	\subfigure[]{ \includegraphics[width=\linewidth]{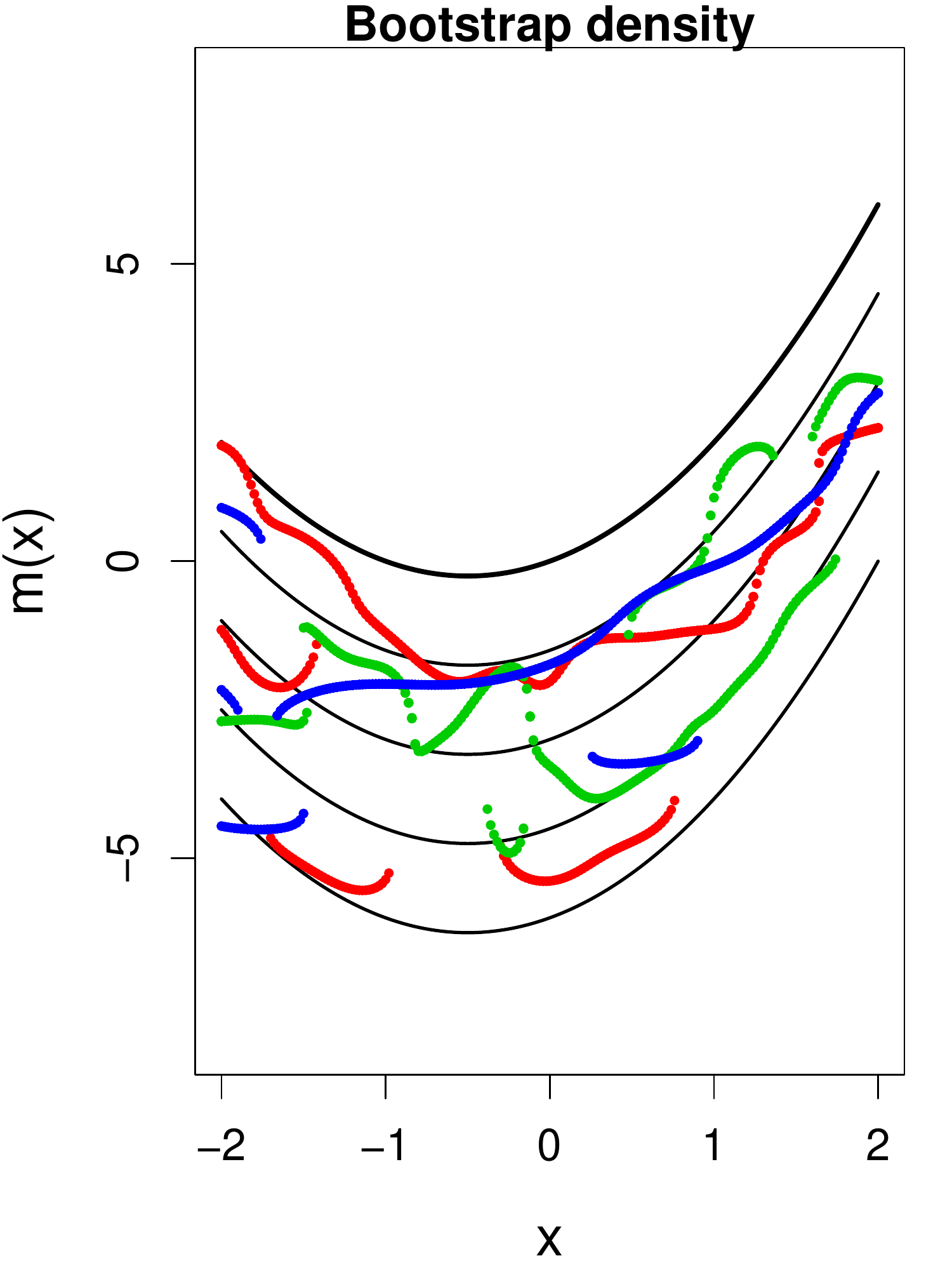} }
	\subfigure[]{ \includegraphics[width=\linewidth]{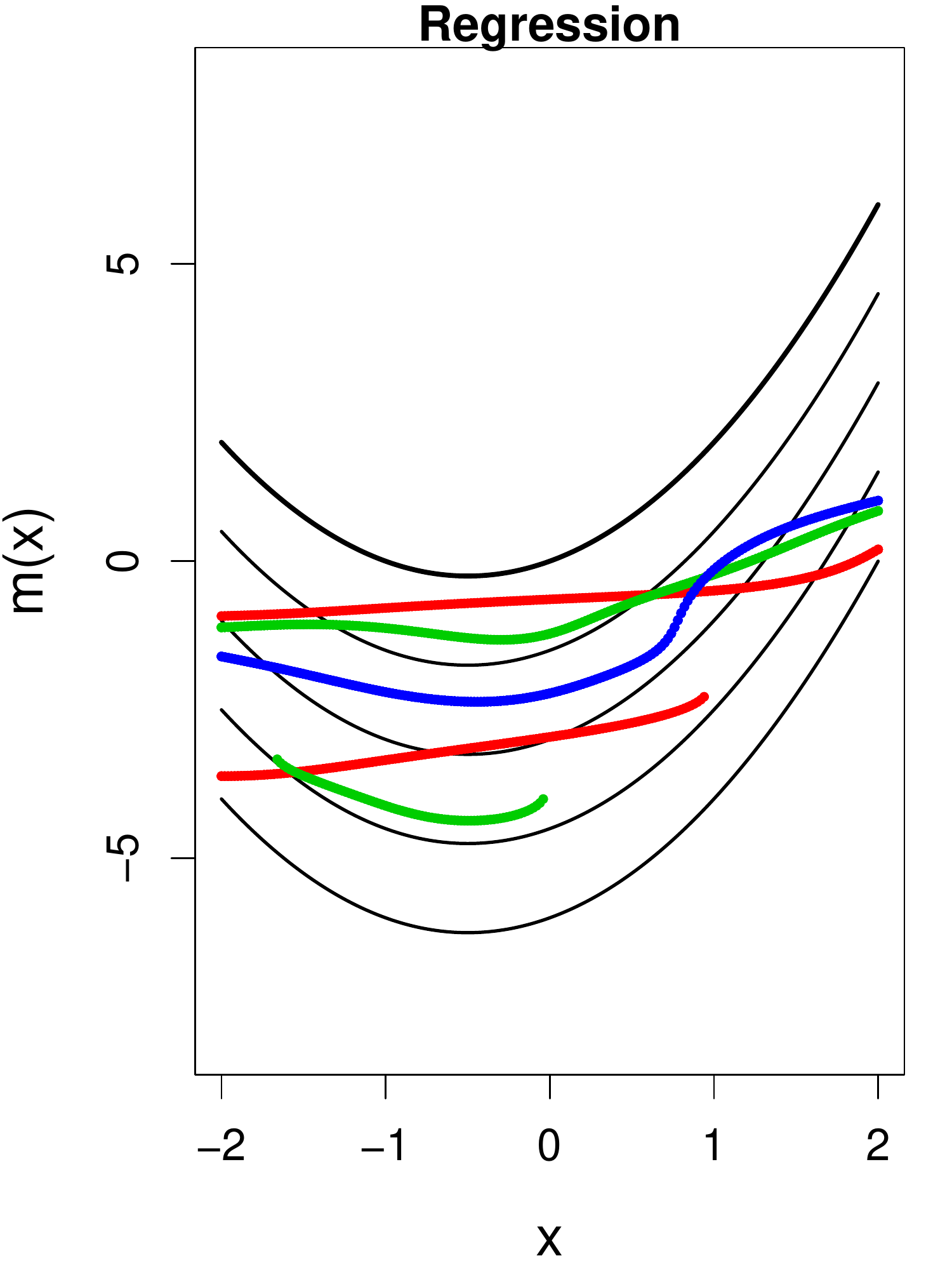} }
	\caption{Estimated mode curves resulting from eight choices of bandwidths $\bh$ under (C5). The correspondence of the eight panels to eight methods and the correspondence of different colors to different lines are the same as those in Figure~\ref{Sim1:curves}.}
	\label{Sim5:curves}
\end{figure}

Figures~\ref{Sim1:curves}--\ref{Sim5:curves} show the estimated mode curves under the five true model configurations with $[x_{\hbox {\tiny $L$}}, x_{\hbox {\tiny $U$}}]=[-2, 2]$. Under each true model configuration, the comparison between the mode estimates resulting from the two optimal (in different senses) bandwidths, $\tilde\bh_{\hbox {\tiny $D$}}$ and $\tilde\bh_{\hbox {\tiny $M$}}$ (see panels (a) and (e)), support our earlier conjecture that bandwidths suitable for density estimation are poor choices for mode estimation. In particular, using  $\tilde\bh_{\hbox {\tiny $D$}}$ results in overfitted mode curves, which are much more noisy than the estimated mode curves when $\tilde\bh_{\hbox {\tiny $M$}}$ is used. The phenomenon of overfitting is also evident in the estimated mode curves resulting from setting $\bh=\bh_{\hbox {\tiny $D$}}$, and also often seen when letting $\bh=\bh_{\hbox {\tiny $B$}}$, both choices of $\bh$ lead to mode estimates clearly outperformed by mode estimates when our proposed bandwidths, $\bh_{\hbox {\tiny $M$}}$ or $\bh^*_{\hbox {\tiny $B$}}$, are used. The overfitting trend observed when using the two density-based bandwidths, $\tilde \bh_{\hbox {\tiny $D$}}$ and $\bh_{\hbox {\tiny $D$}}$, does less harm when $p(y|x)$ has many modes, as seen under (C5), where mode estimates resulting from using $\tilde \bh_{\hbox {\tiny $D$}}$ and $\bh_{\hbox {\tiny $D$}}$ are relatively comparable with estimates resulting from using $\tilde \bh_{\hbox {\tiny $M$}}$ and $\bh_{\hbox {\tiny $M$}}$, respectively (see panels (a), (b), (e) and (f) in Figure~\ref{Sim5:curves}). When the normal reference $\bh_{\hbox {\tiny $N$}}$ or the regression-based bandwidths $\bh_{\hbox {\tiny $R$}}$ are used, although less noisy, the resultant estimated mode curves exhibit underfitting and fail to capture key features of the true conditional mode curves around the boundary or at the valley. In fact, they fail miserably at other regions of $\mathscr{X}$ too when there are many modes as in (C5) (see panels (d) and (h) in Figure~\ref{Sim5:curves}). 

\begin{figure}[p]
	\centering
	\setlength{\linewidth}{0.2\textwidth}
	\subfigure[]{ \includegraphics[width=\linewidth]{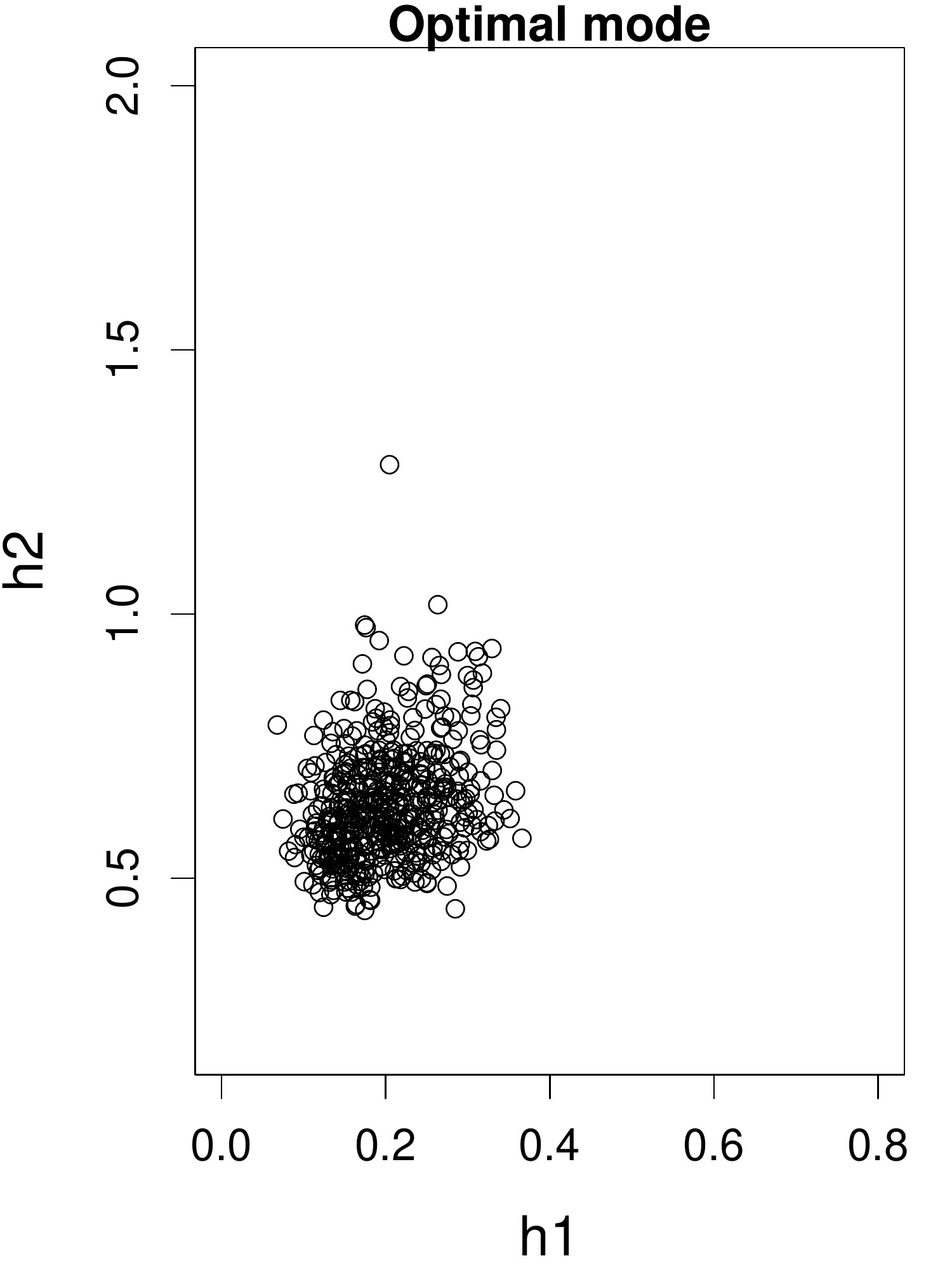} }
	\subfigure[]{ \includegraphics[width=\linewidth]{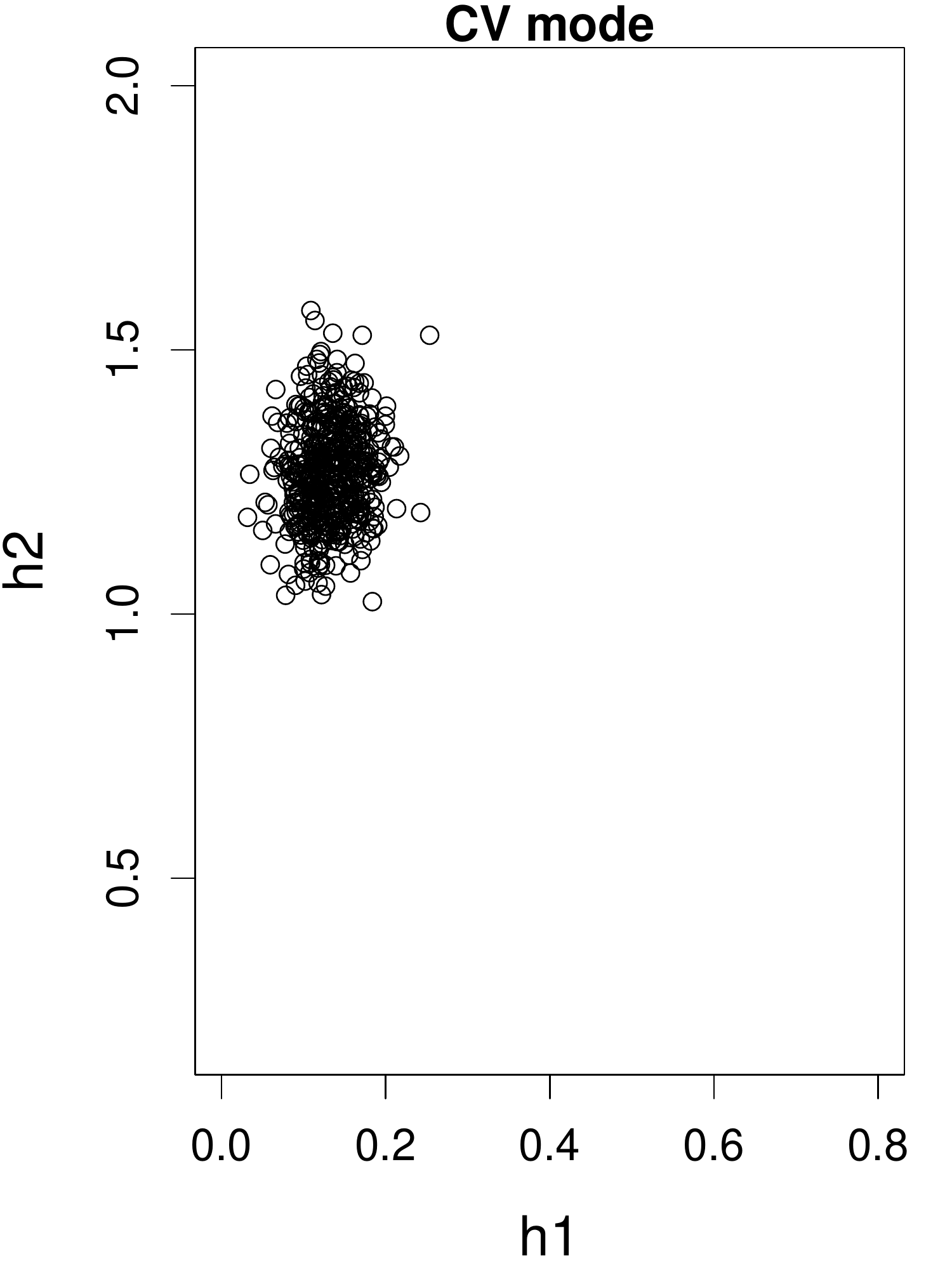} }
	\subfigure[]{ \includegraphics[width=\linewidth]{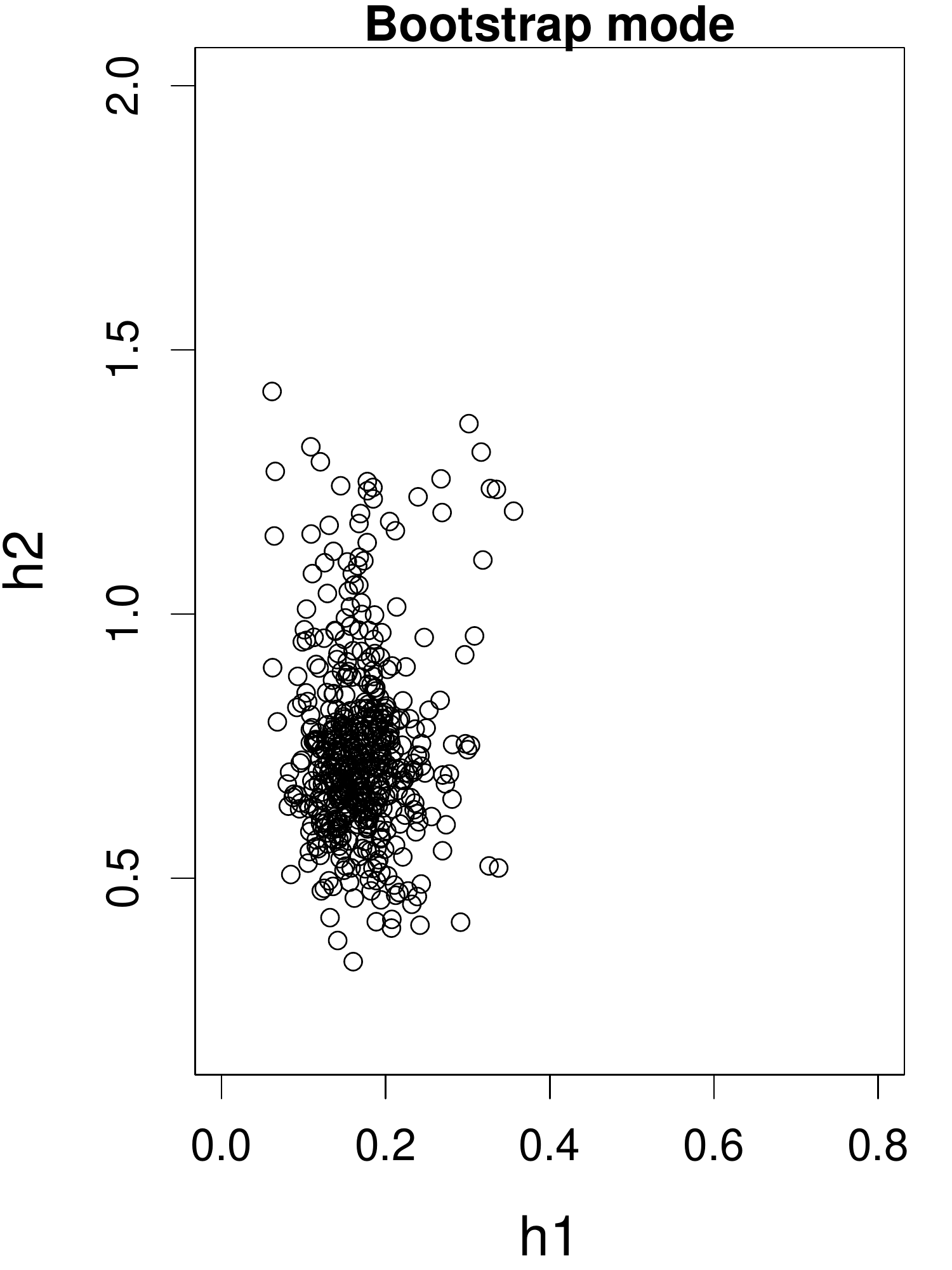} }
	\subfigure[]{ \includegraphics[width=\linewidth]{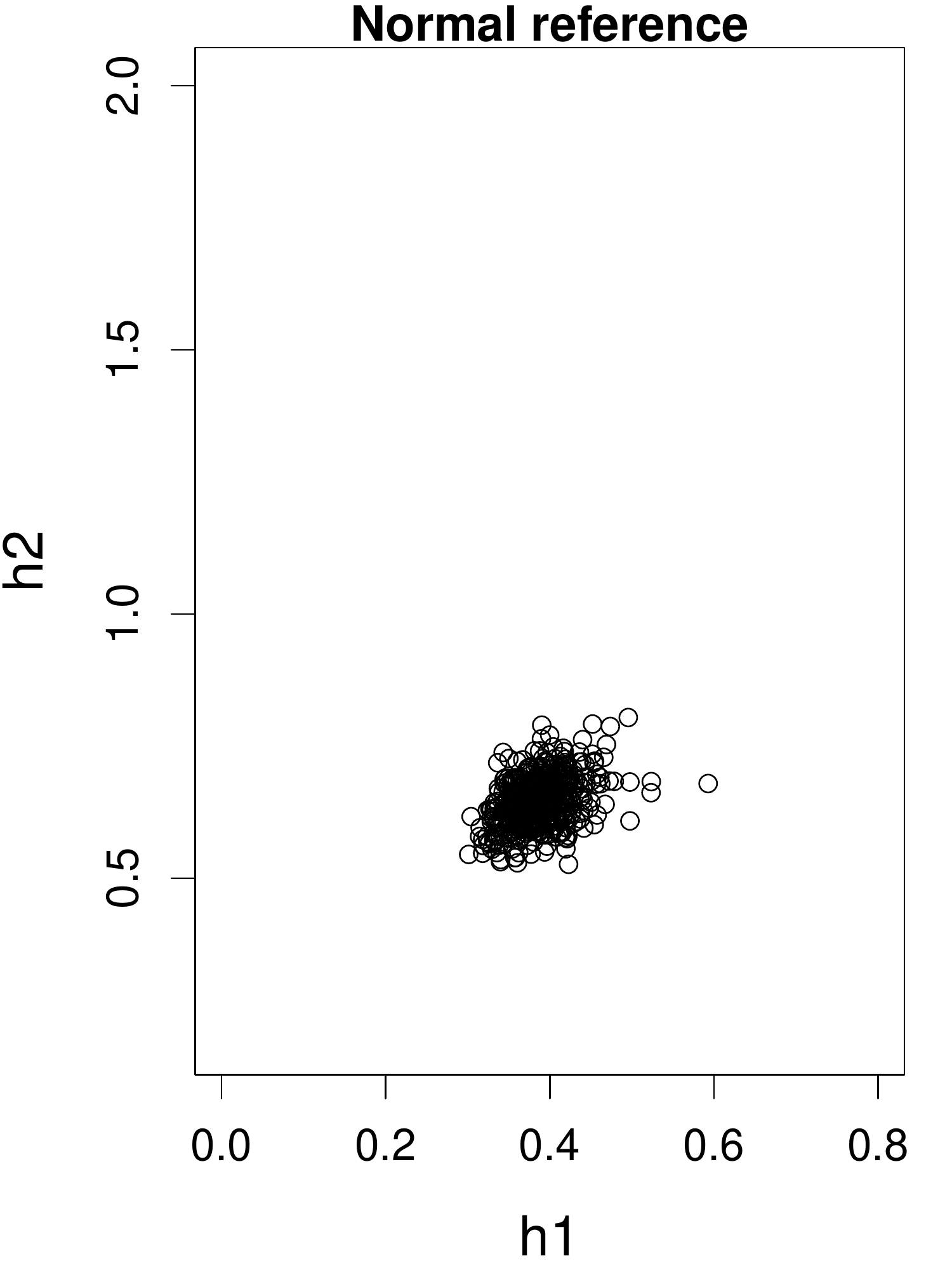} }\\
	\subfigure[]{ \includegraphics[width=\linewidth]{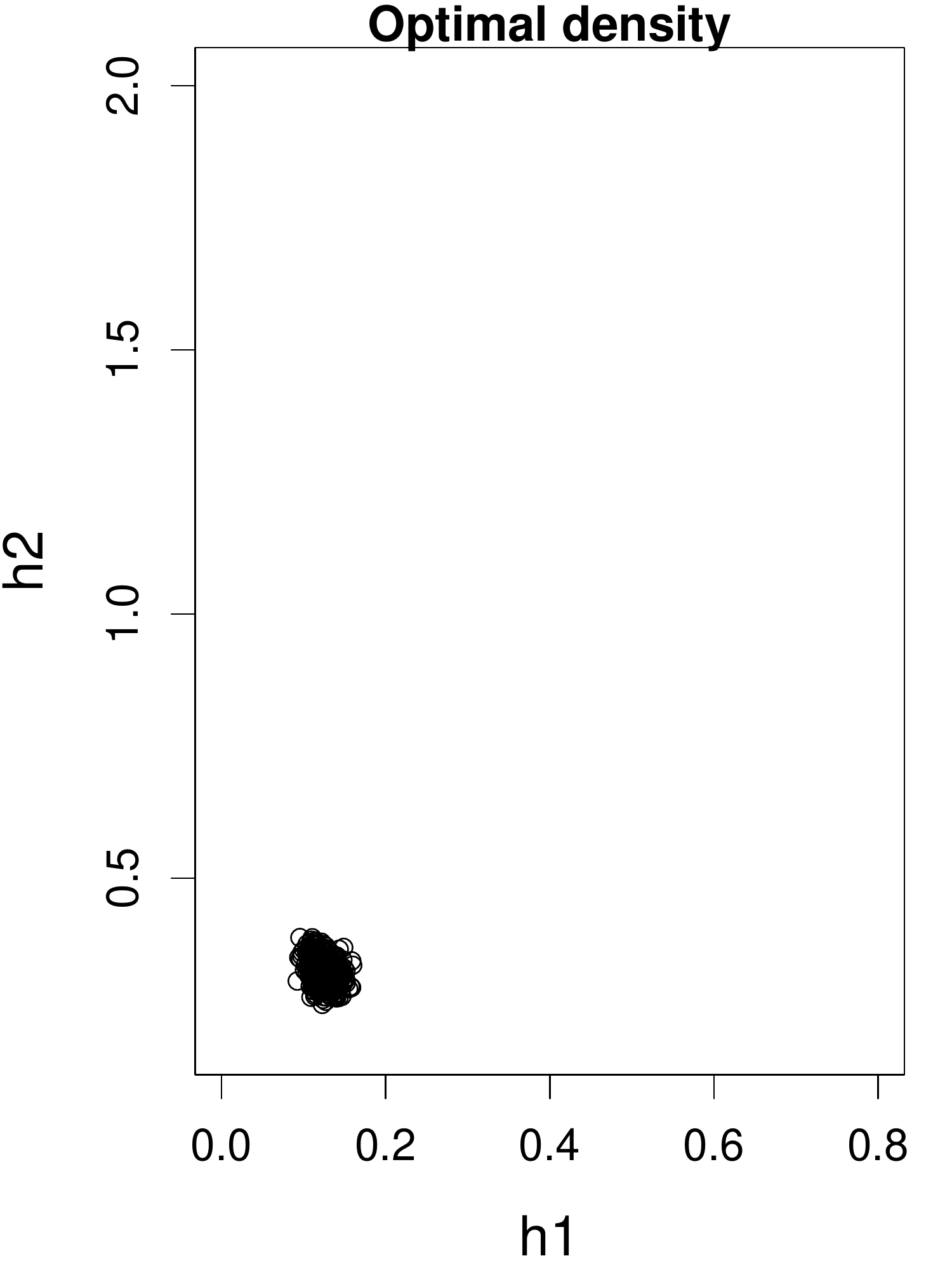} }
	\subfigure[]{ \includegraphics[width=\linewidth]{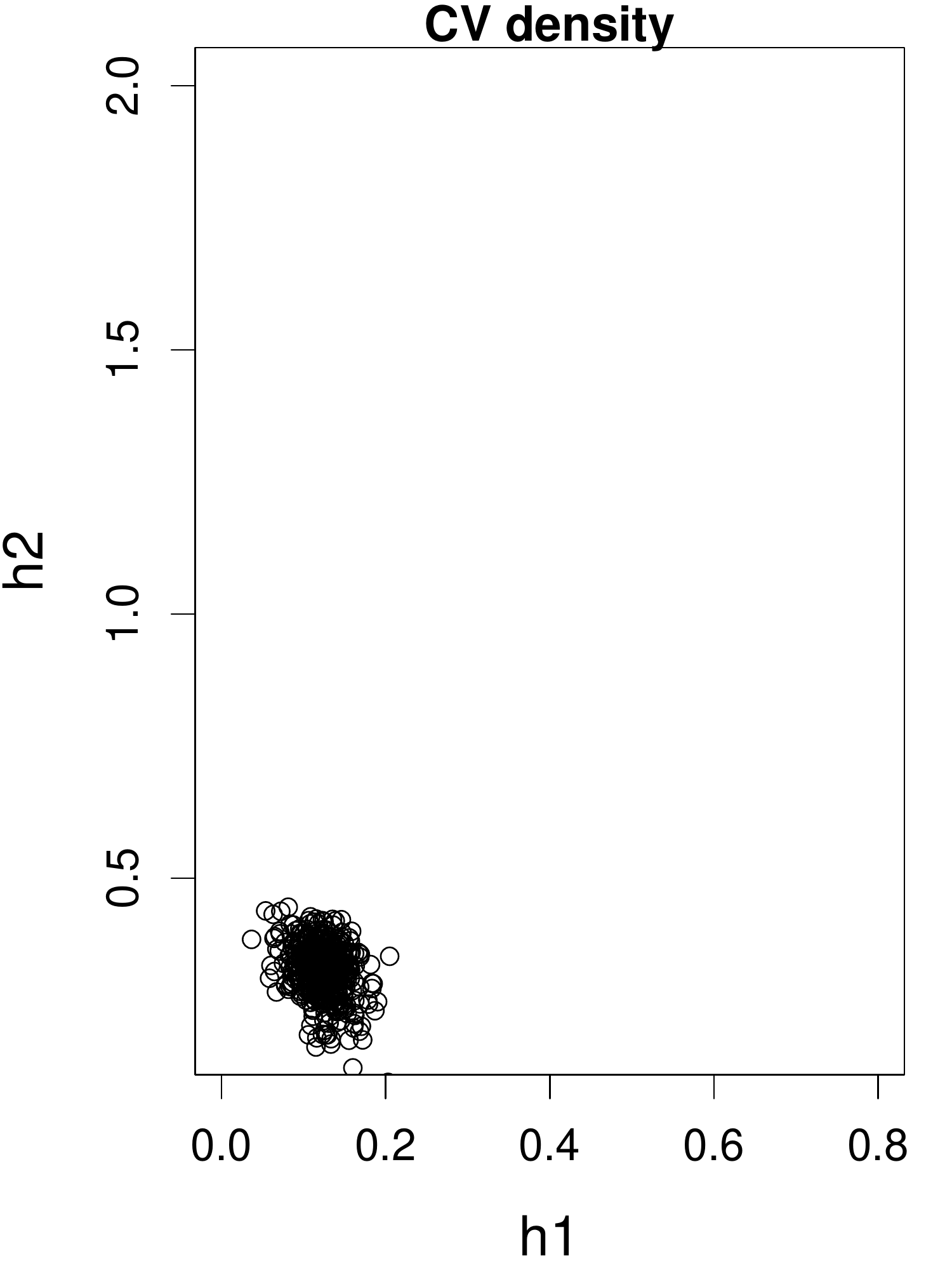} }
	\subfigure[]{ \includegraphics[width=\linewidth]{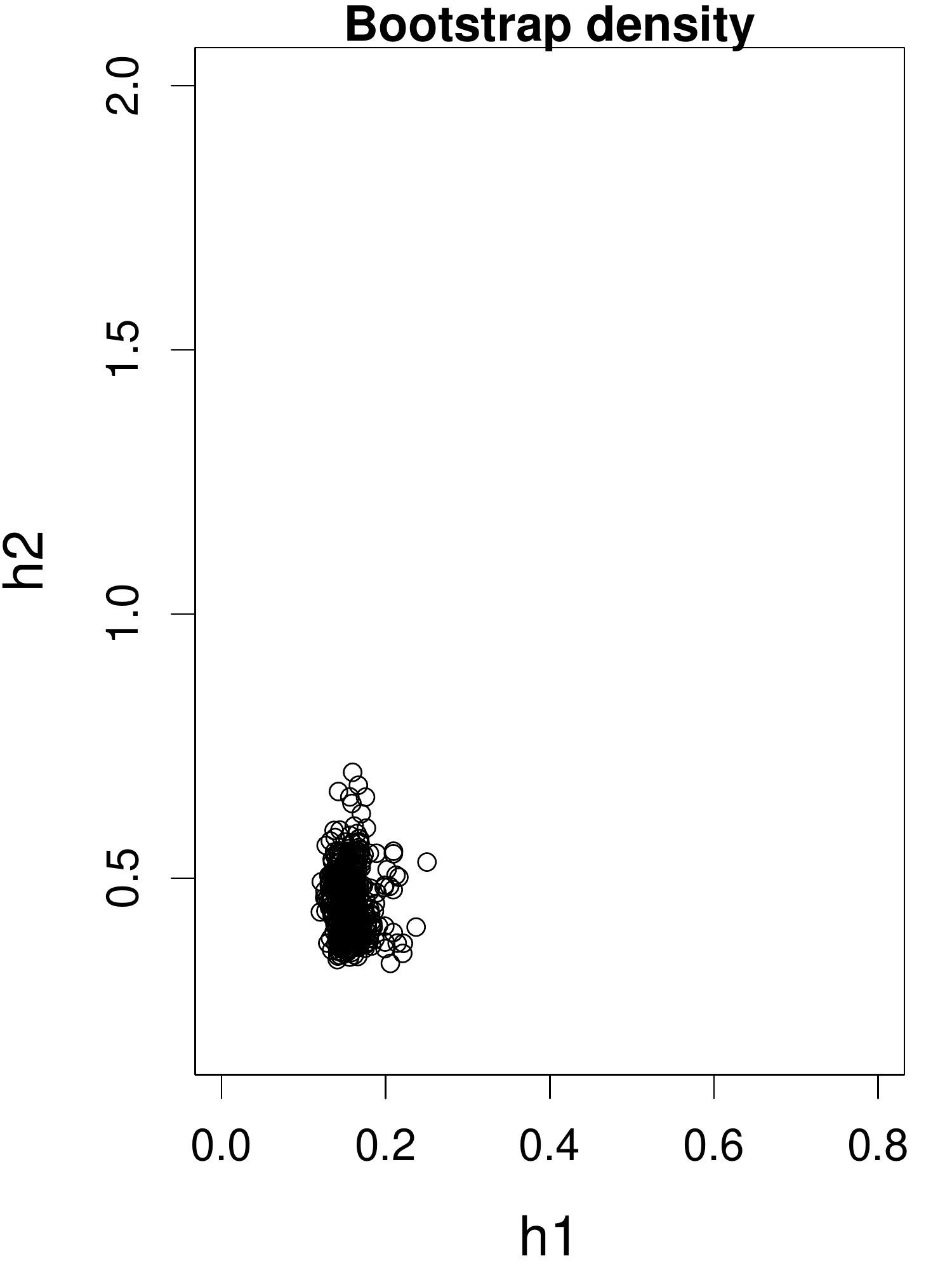} }
	\subfigure[]{ \includegraphics[width=\linewidth]{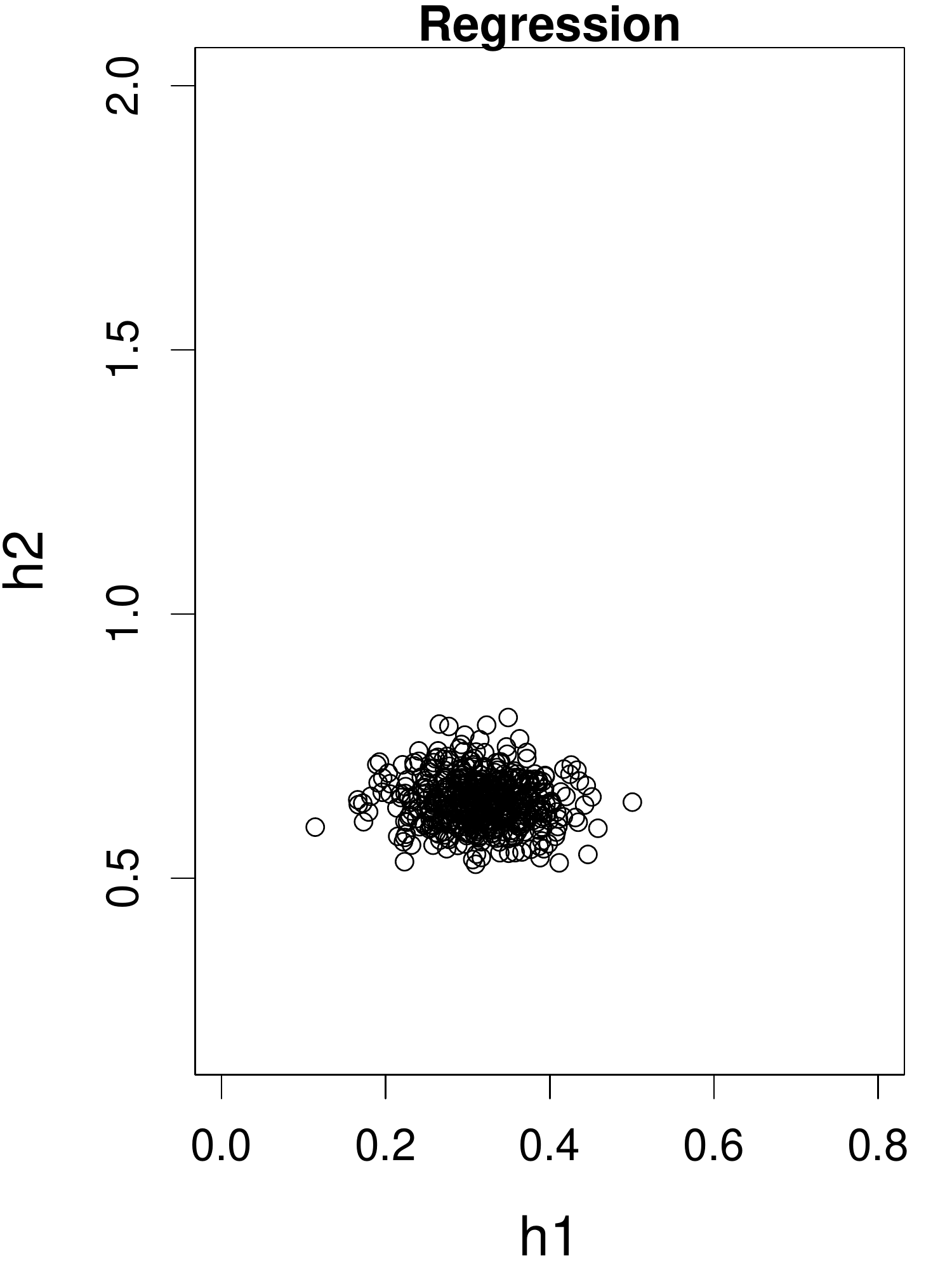} }
	\caption{Scatter plots of selected bandwidths across 500 MC replicates under (C1) corresponding to eight ways of bandwidth selection. The correspondence of the eight panels to eight methods is the same as that in Figure~\ref{Sim1:curves}.} 
	\label{Sim1:bandwidths}
\end{figure}

\begin{figure}[p]
	\centering
	\setlength{\linewidth}{0.2\textwidth}
	\subfigure[]{ \includegraphics[width=\linewidth]{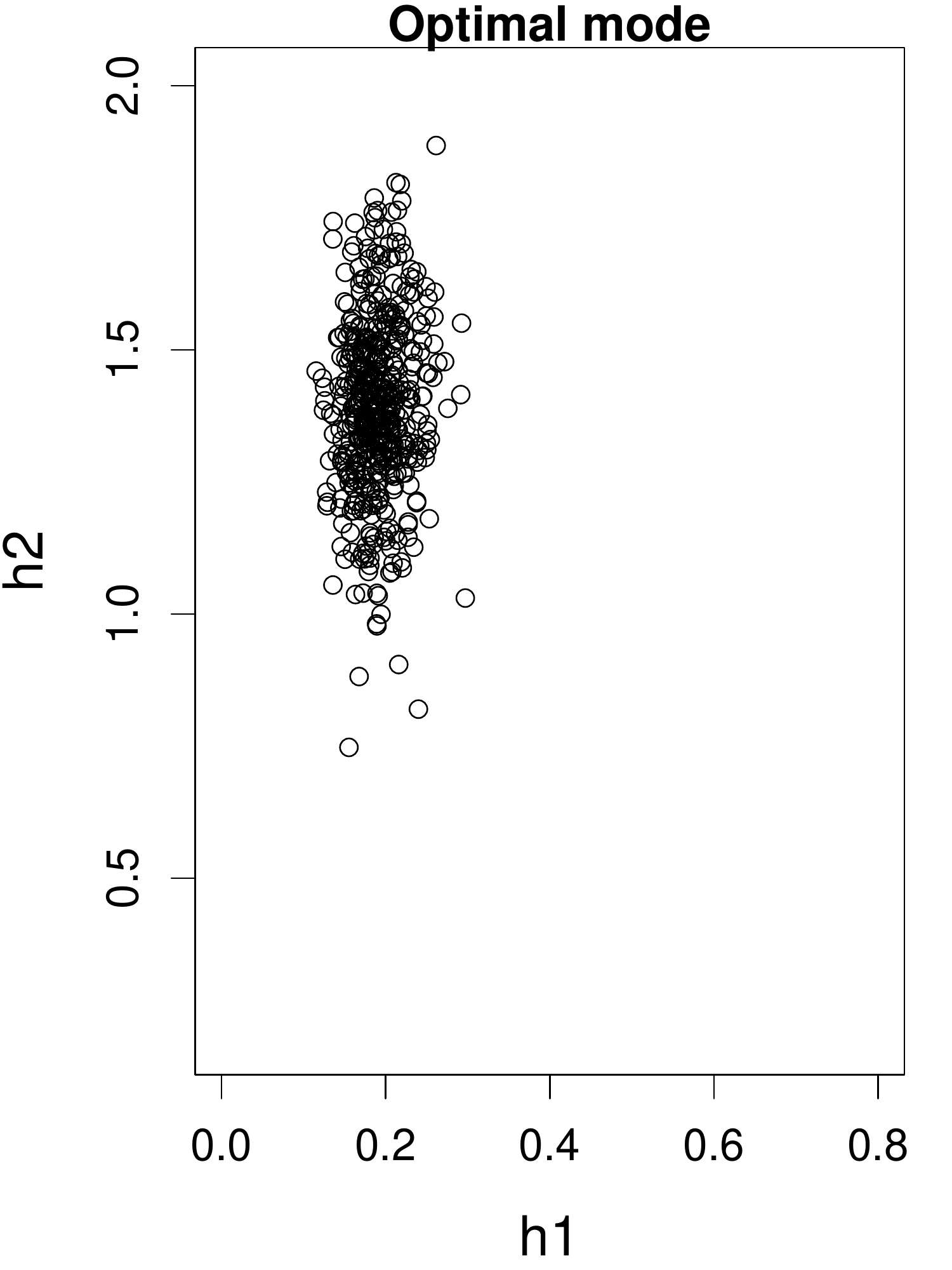} }
	\subfigure[]{ \includegraphics[width=\linewidth]{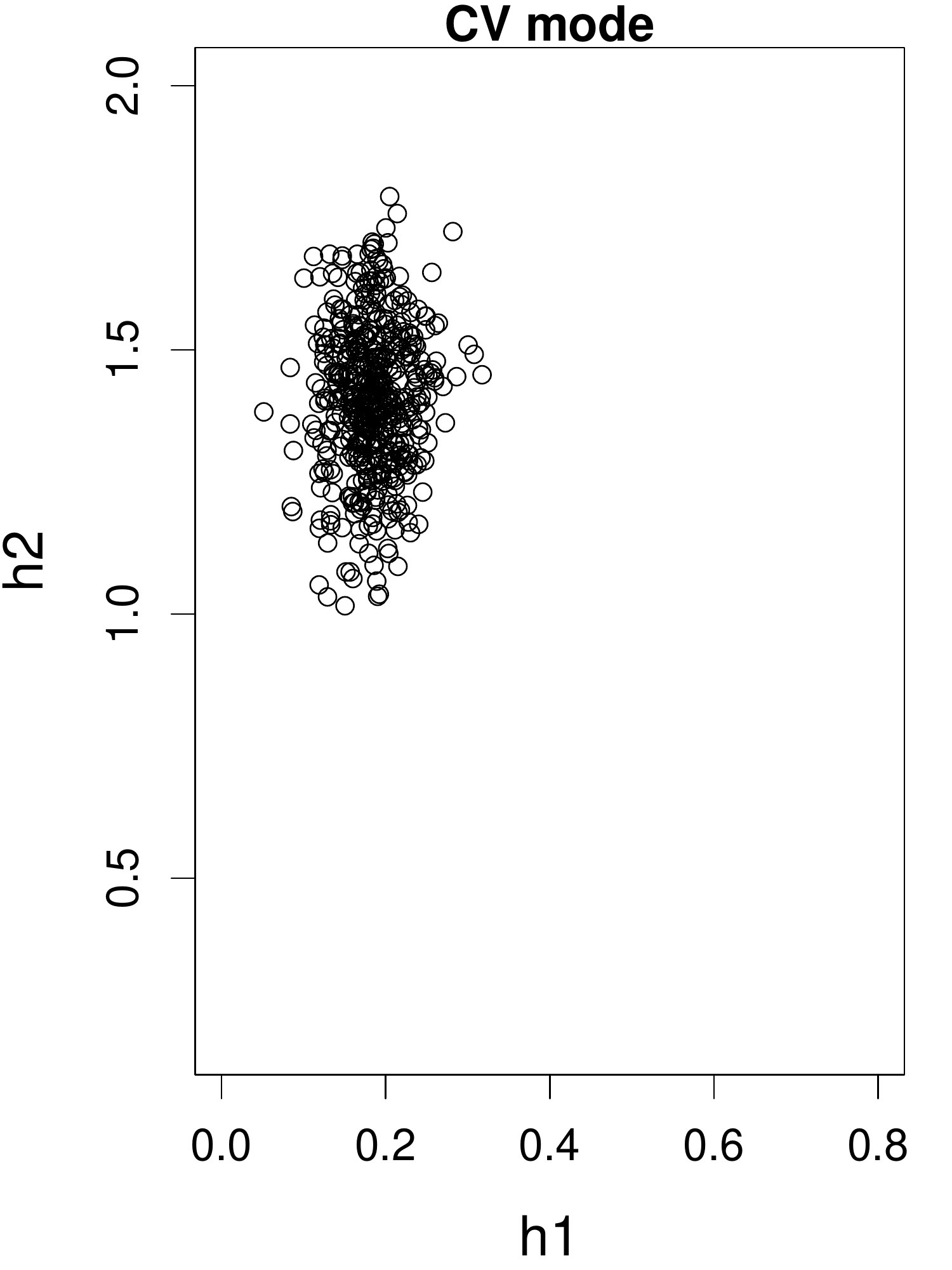} }
	\subfigure[]{ \includegraphics[width=\linewidth]{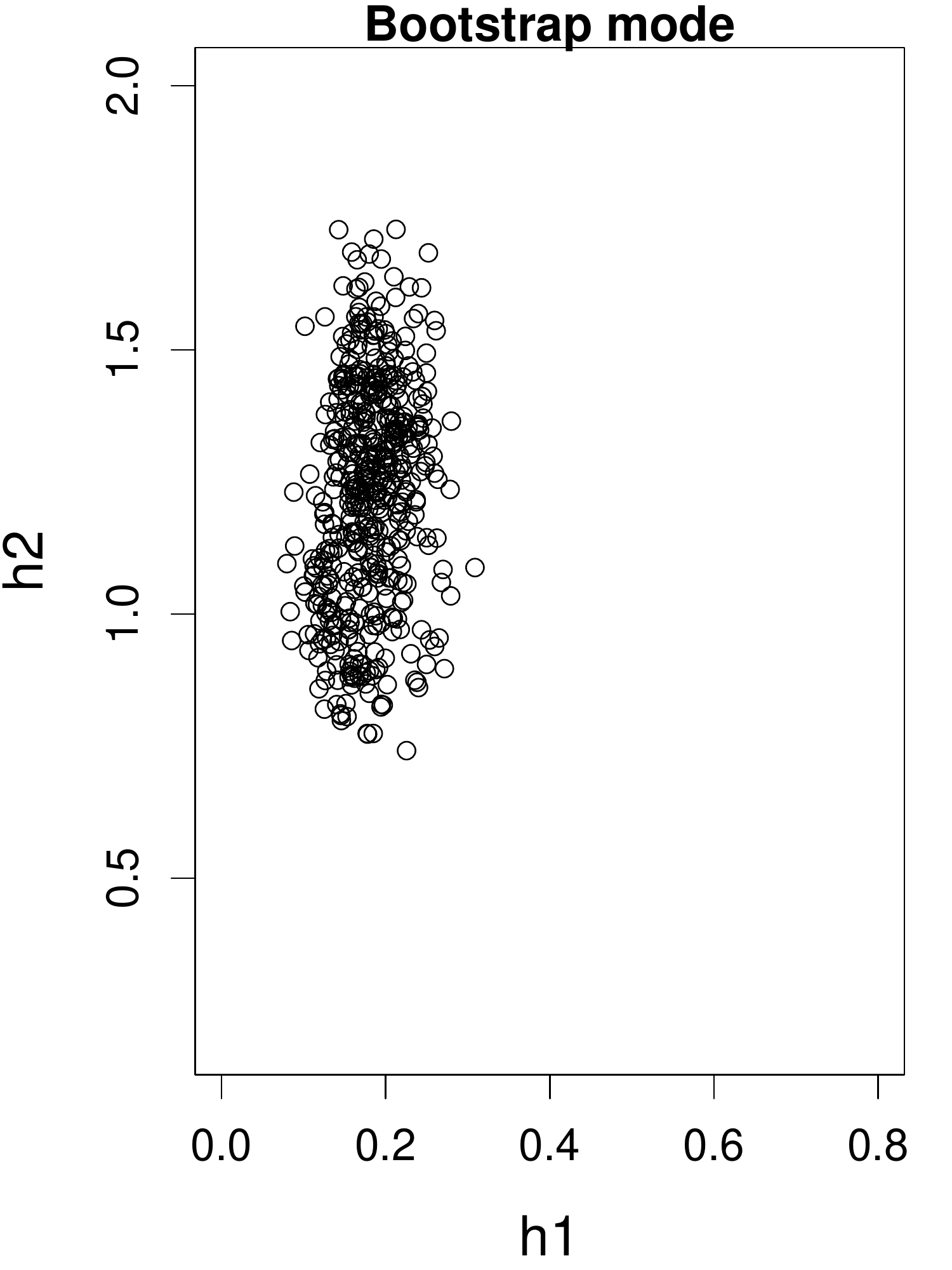} }
	\subfigure[]{ \includegraphics[width=\linewidth]{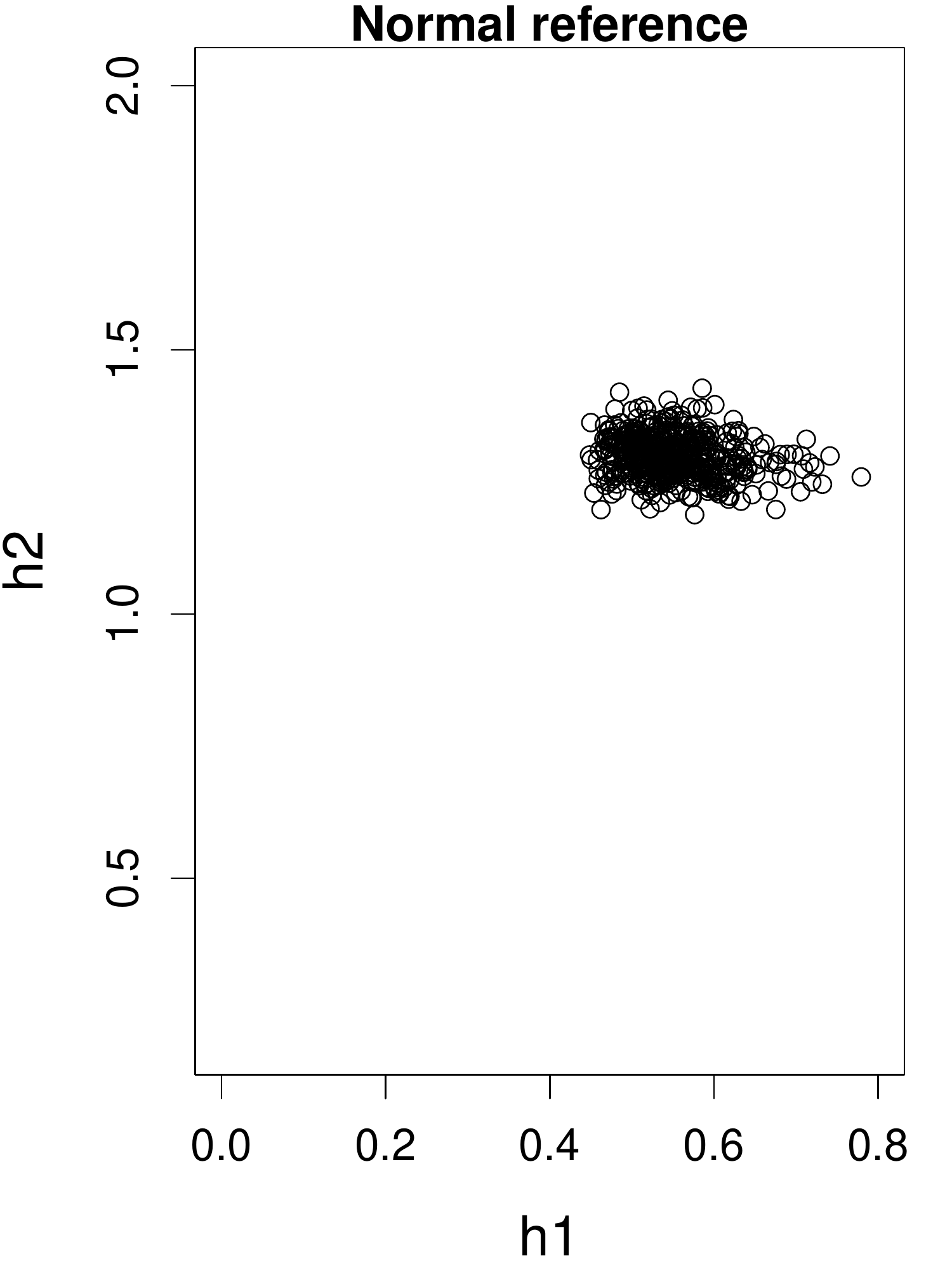} }\\
	\subfigure[]{ \includegraphics[width=\linewidth]{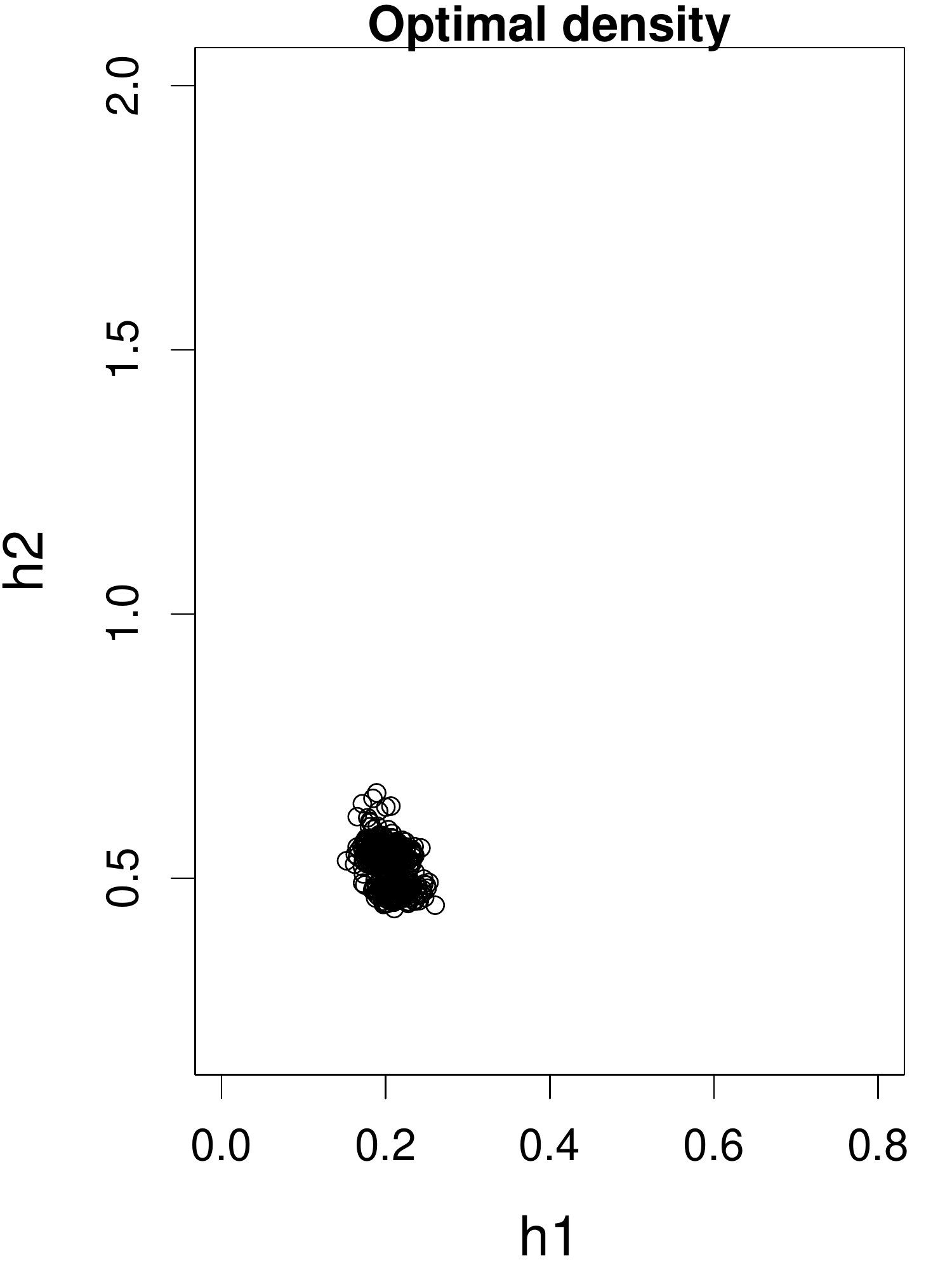} }
	\subfigure[]{ \includegraphics[width=\linewidth]{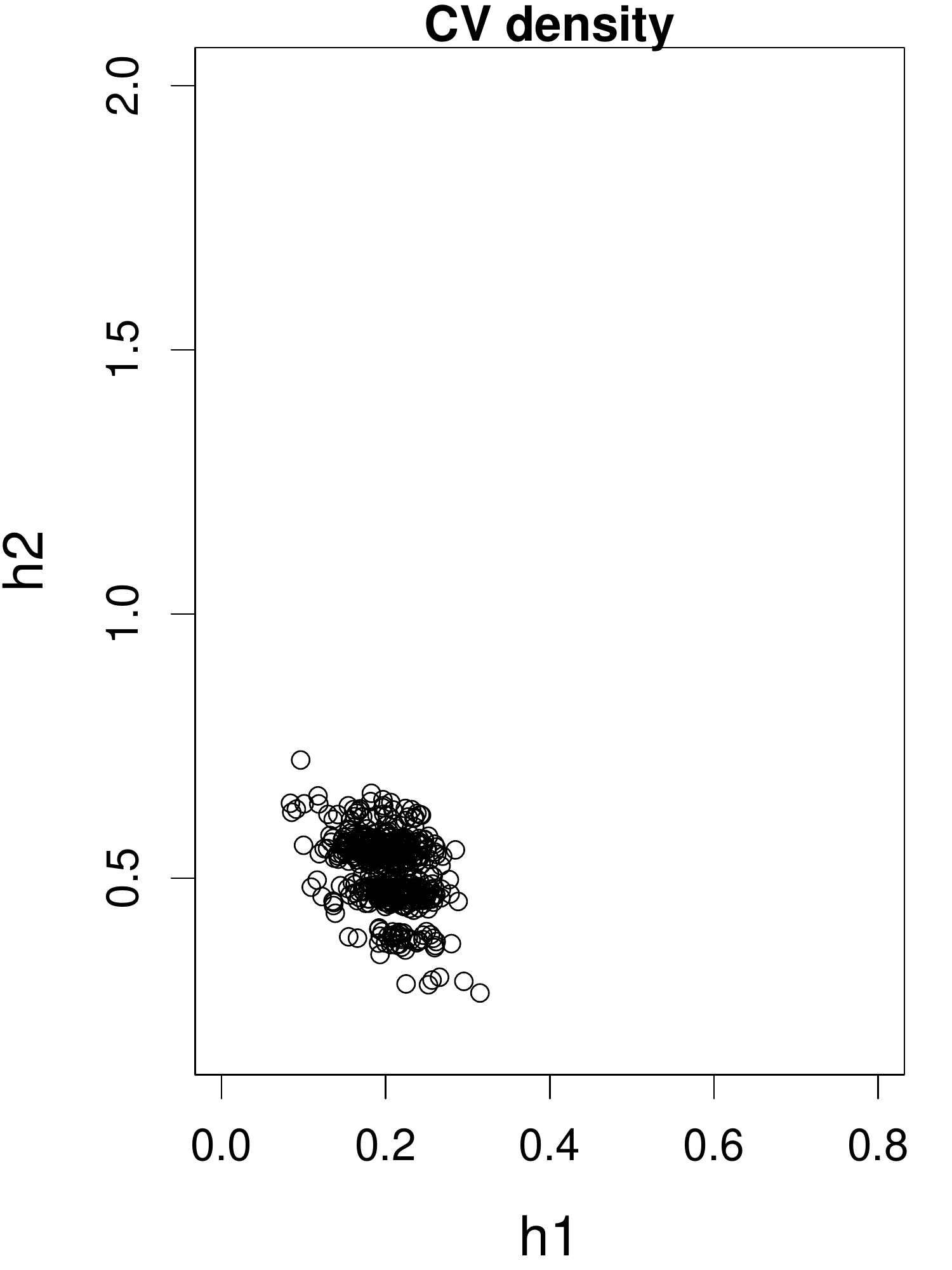} }
	\subfigure[]{ \includegraphics[width=\linewidth]{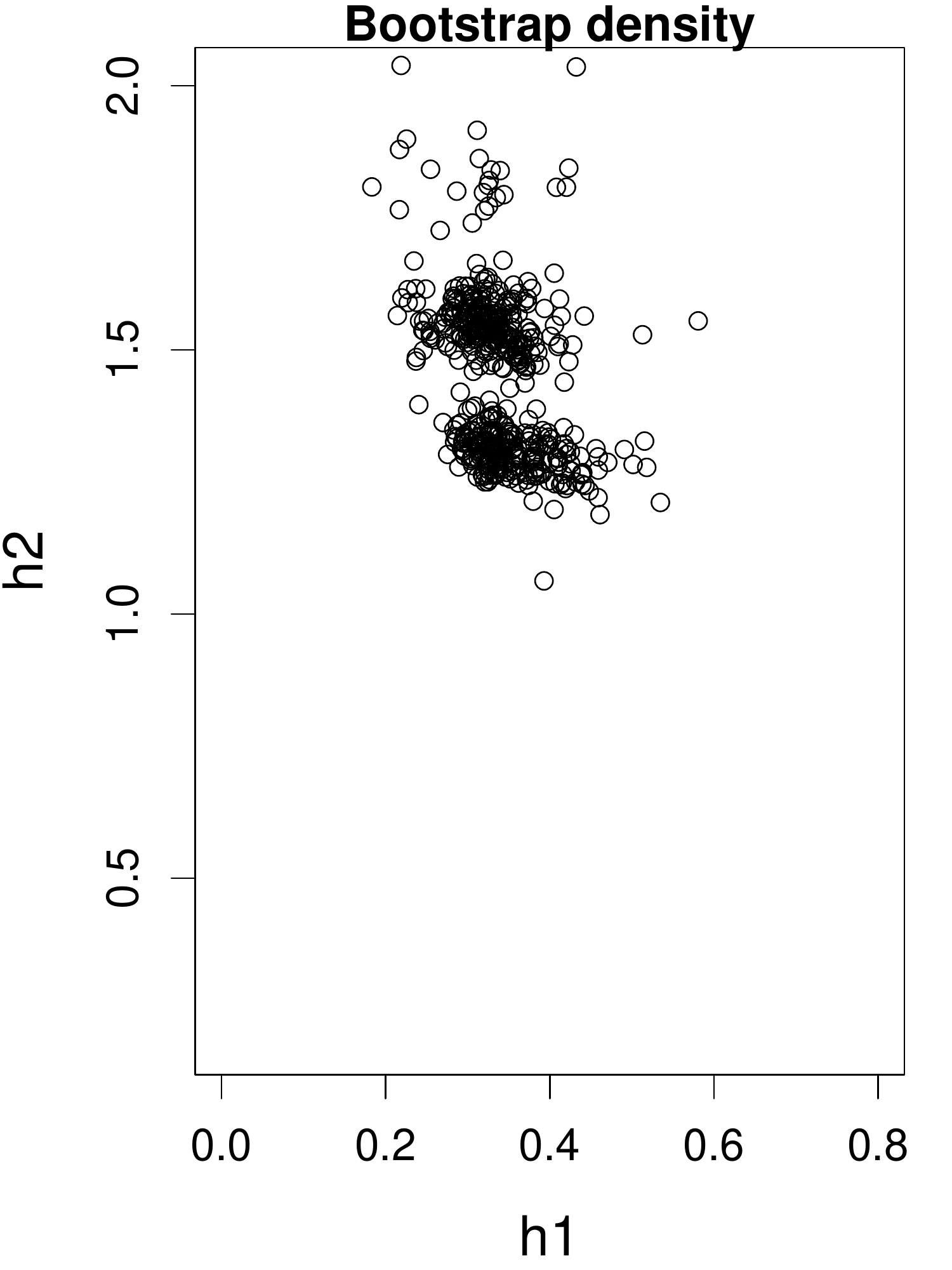} }
	\subfigure[]{ \includegraphics[width=\linewidth]{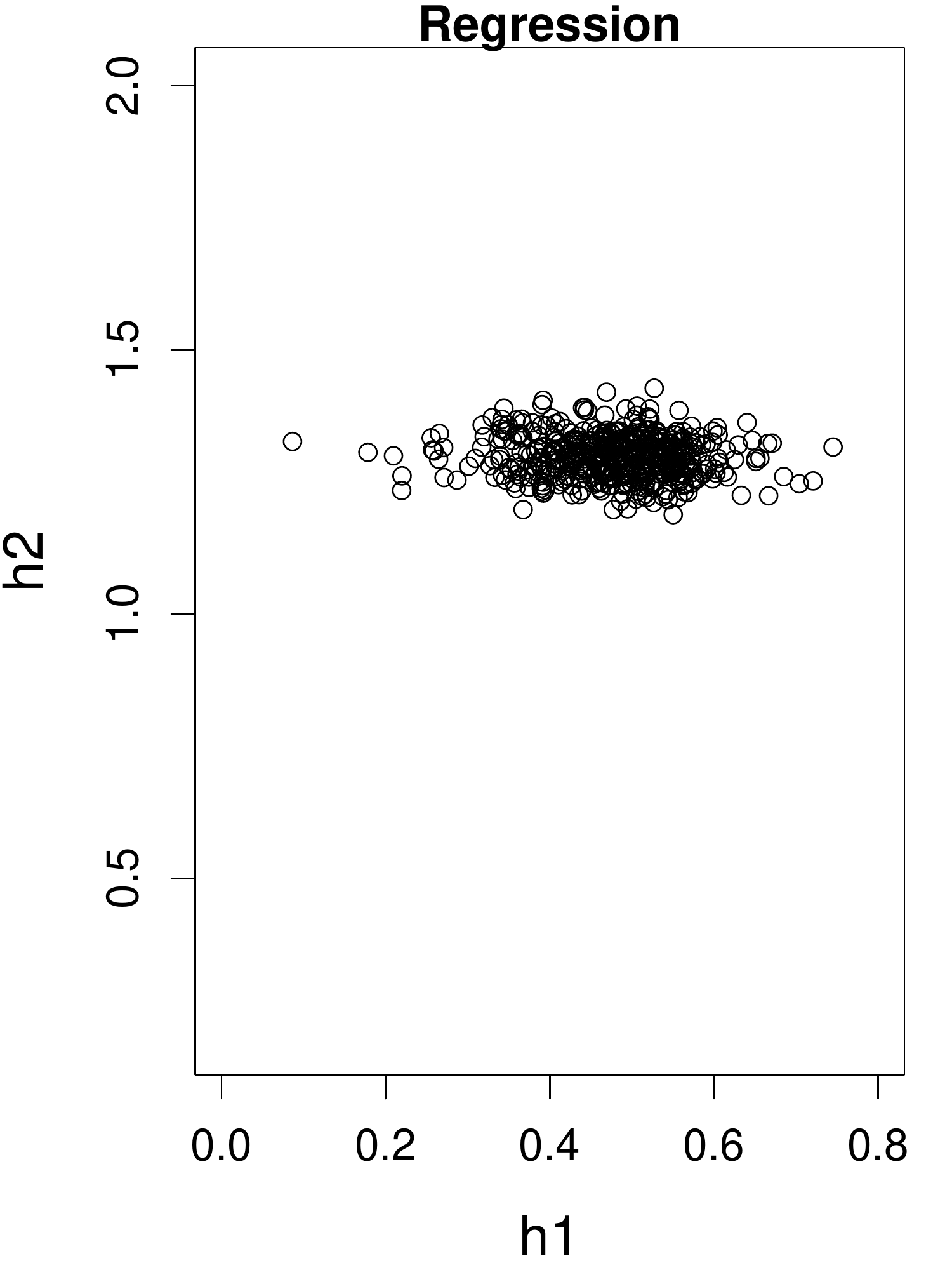} }
	\caption{Scatter plots of selected bandwidths across 500 MC replicates under (C2) corresponding to eight ways of bandwidth selection. The correspondence of the eight panels to eight methods is the same as that in Figure~\ref{Sim1:curves}.}
	\label{Sim2:bandwidths}
\end{figure}

\begin{figure}[p]
	\centering
	\setlength{\linewidth}{0.2\textwidth}
	\subfigure[]{ \includegraphics[width=\linewidth]{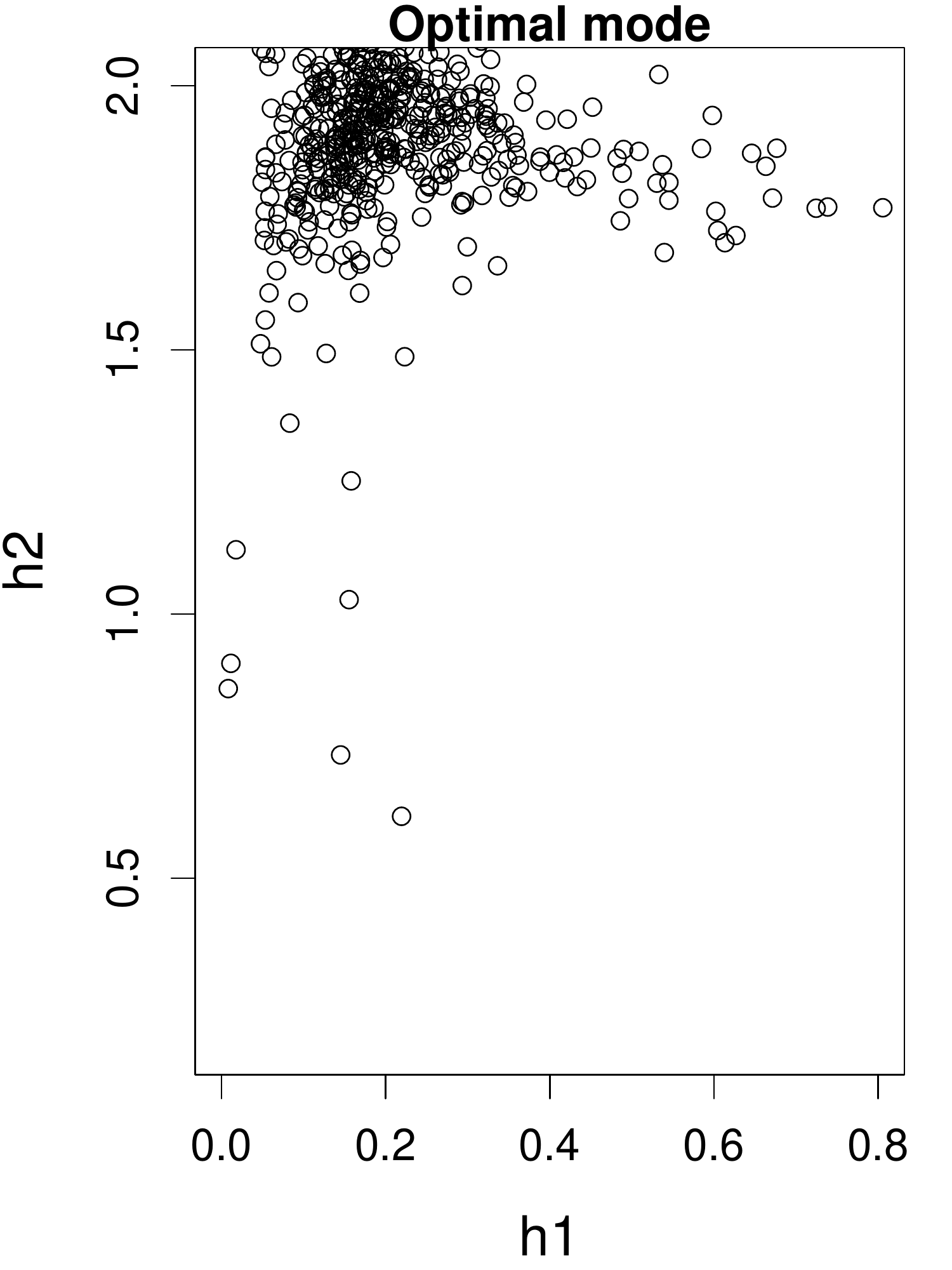} }
	\subfigure[]{ \includegraphics[width=\linewidth]{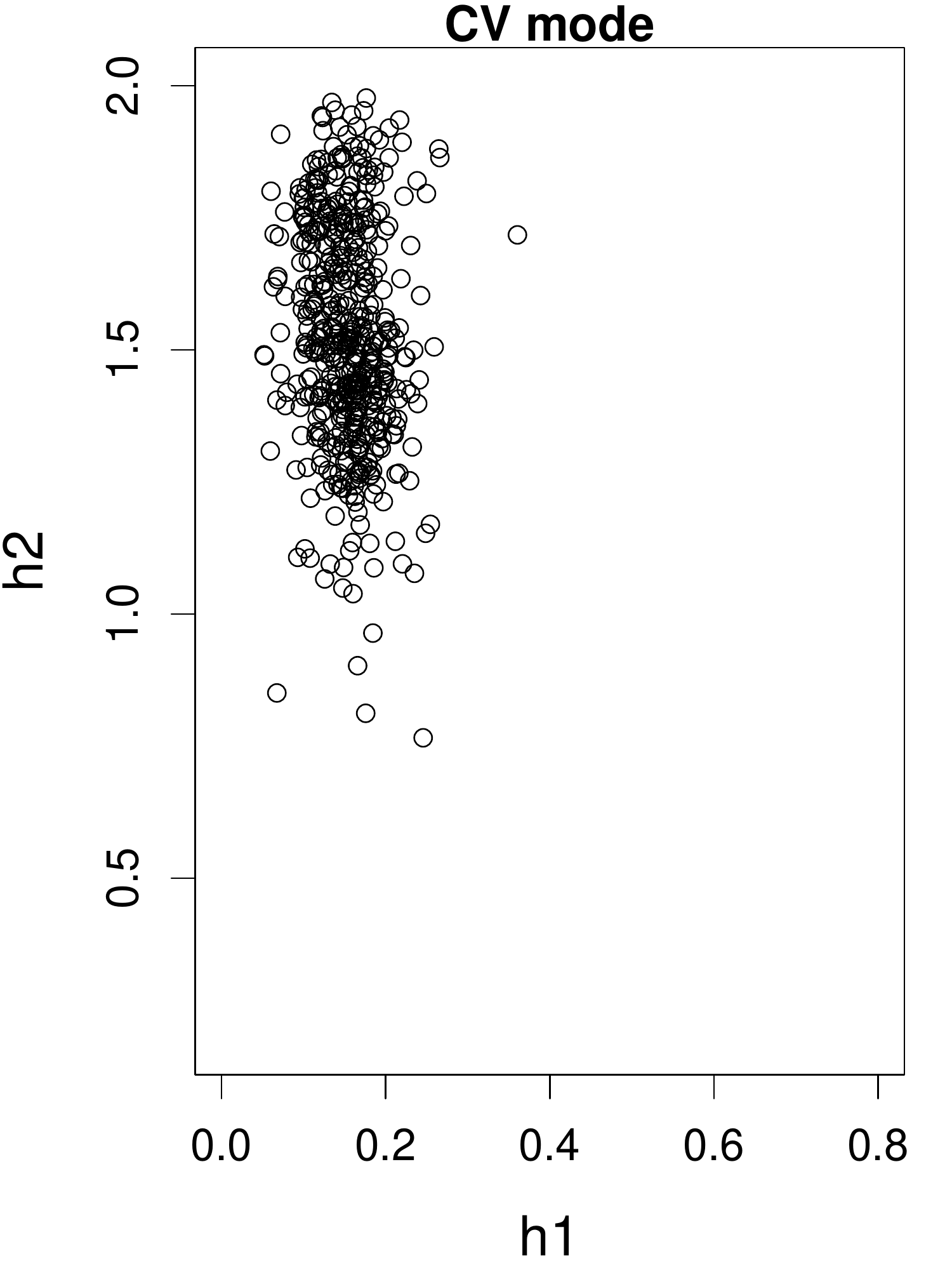} }
	\subfigure[]{ \includegraphics[width=\linewidth]{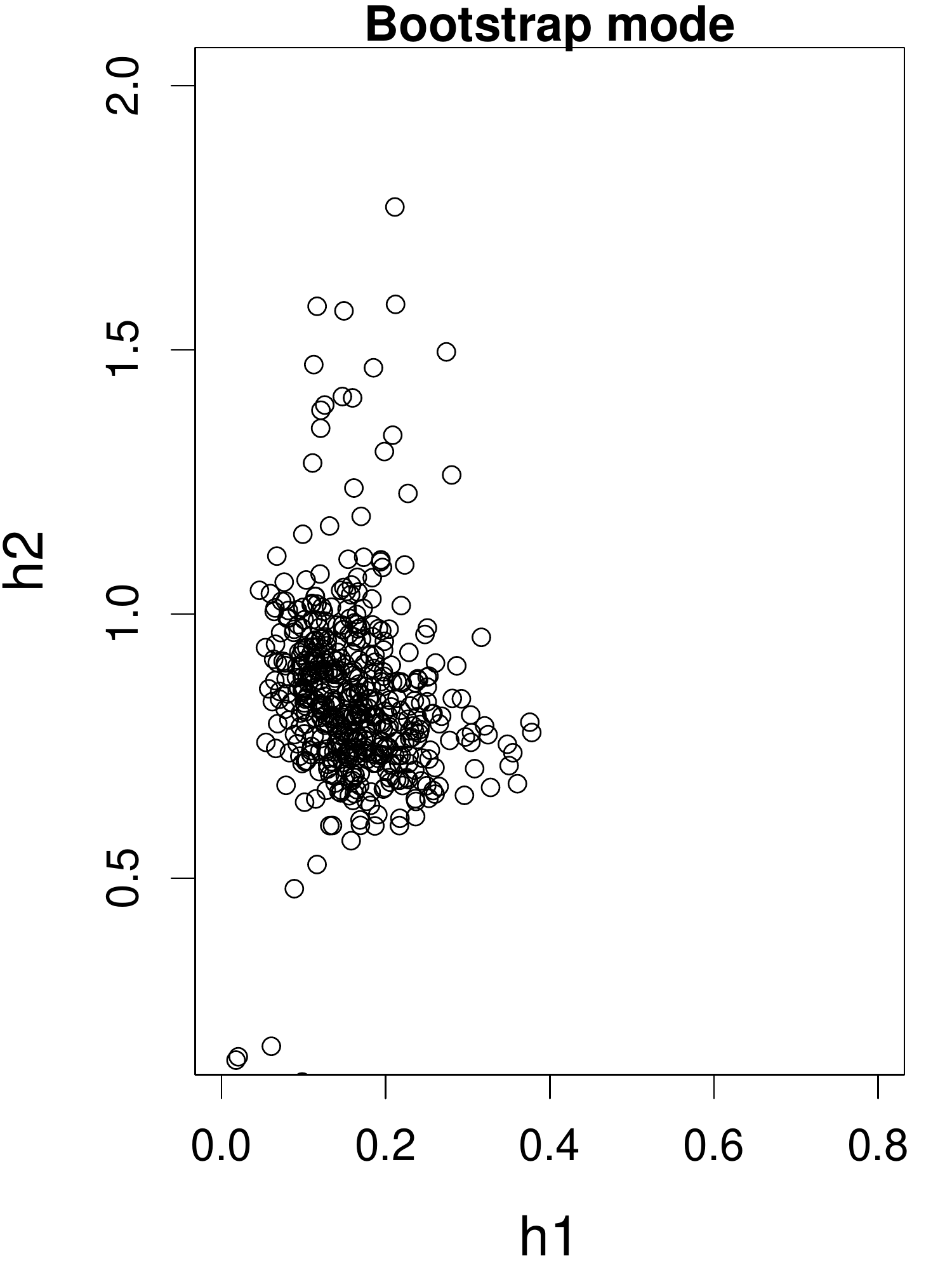} }
	\subfigure[]{ \includegraphics[width=\linewidth]{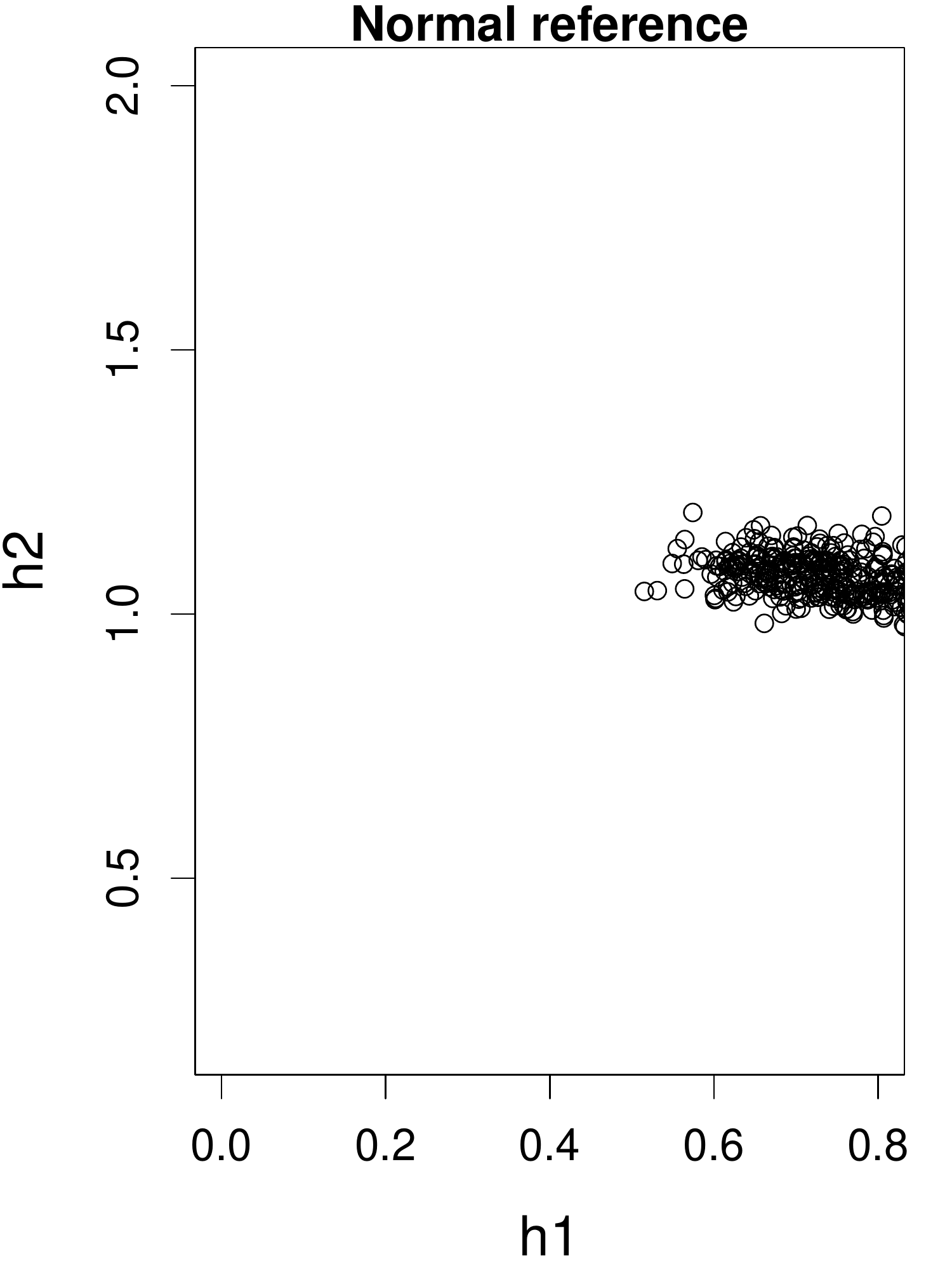} }\\
	\subfigure[]{ \includegraphics[width=\linewidth]{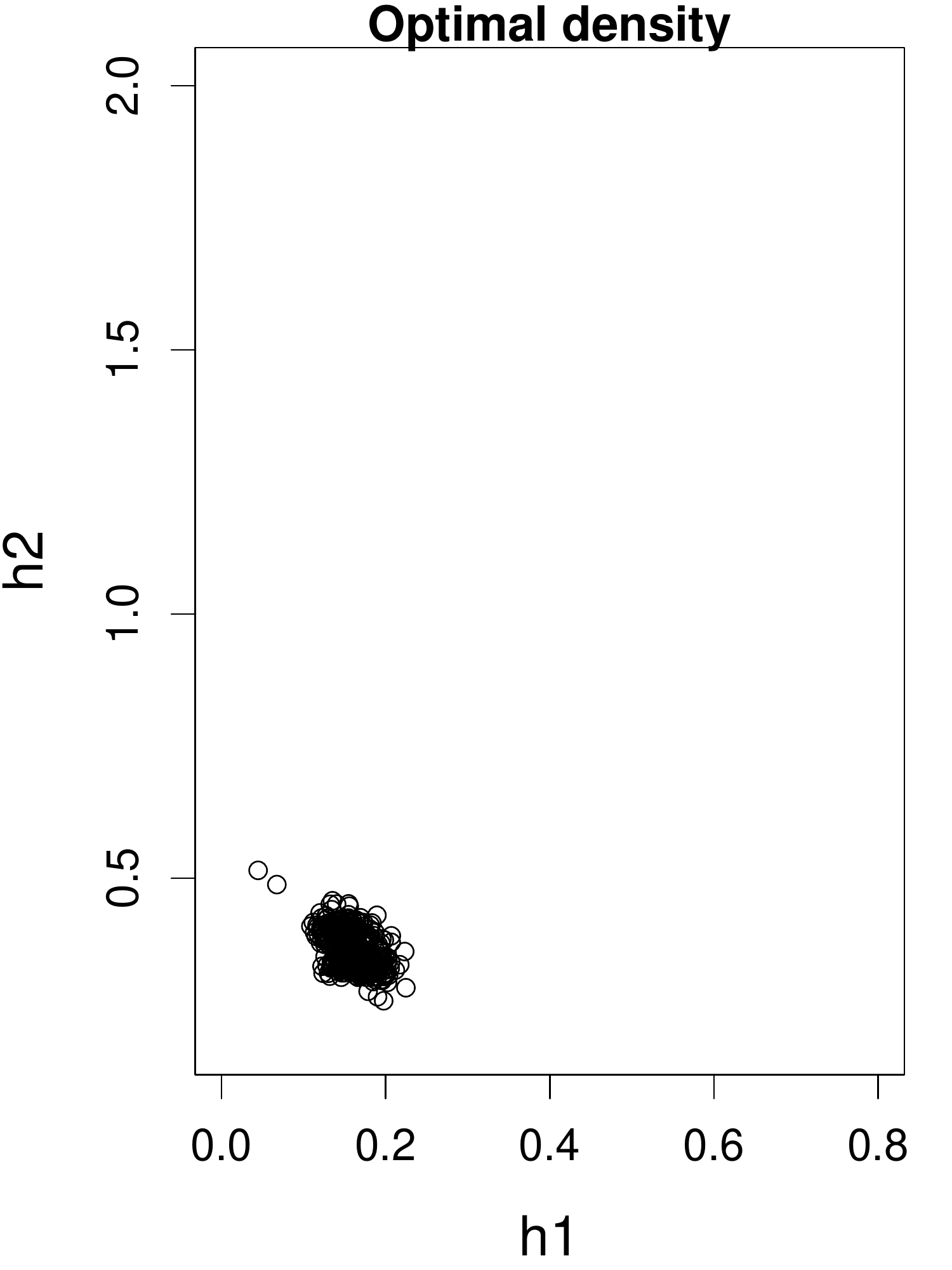} }
	\subfigure[]{ \includegraphics[width=\linewidth]{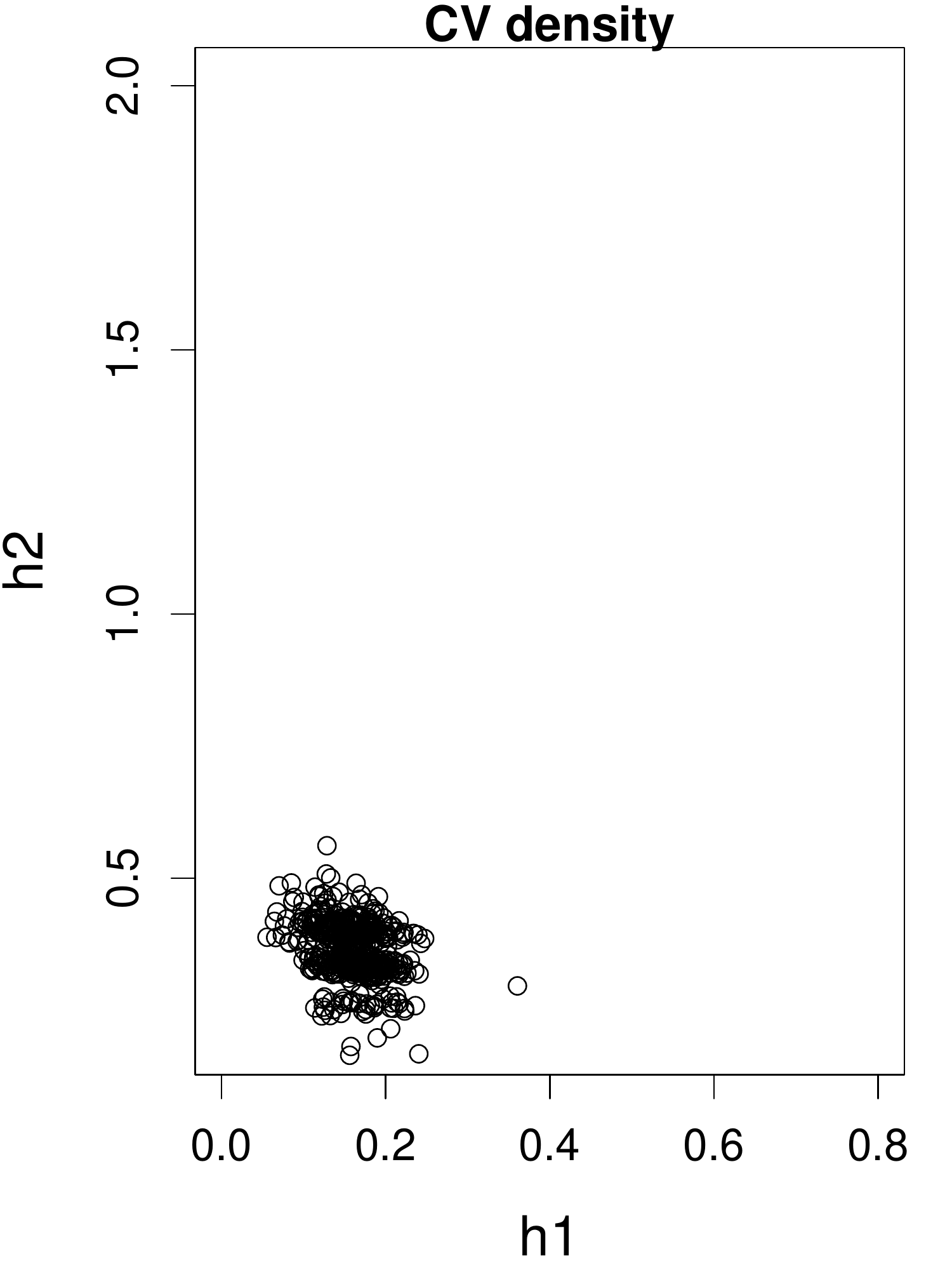} }
	\subfigure[]{ \includegraphics[width=\linewidth]{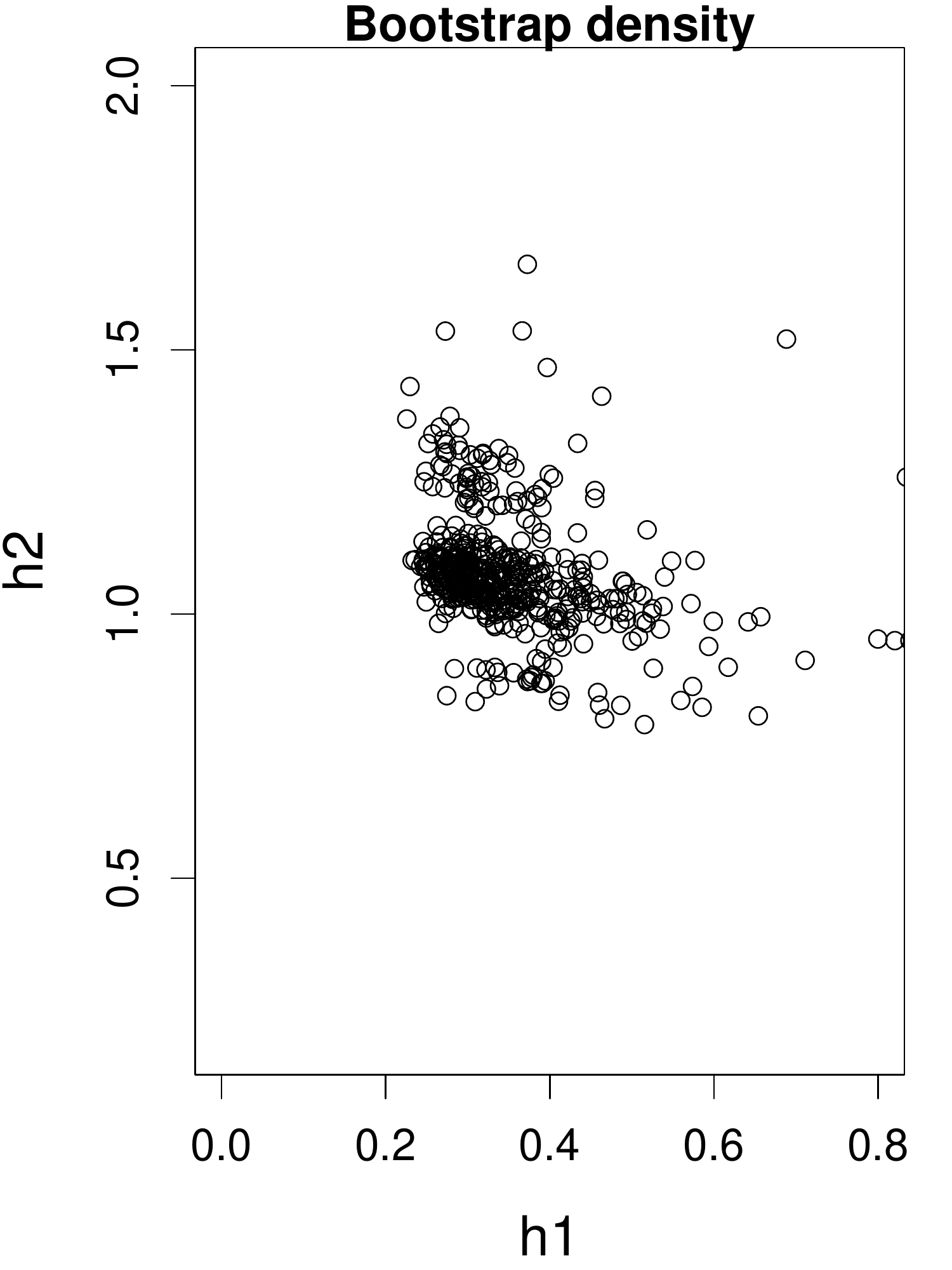} }
	\subfigure[]{ \includegraphics[width=\linewidth]{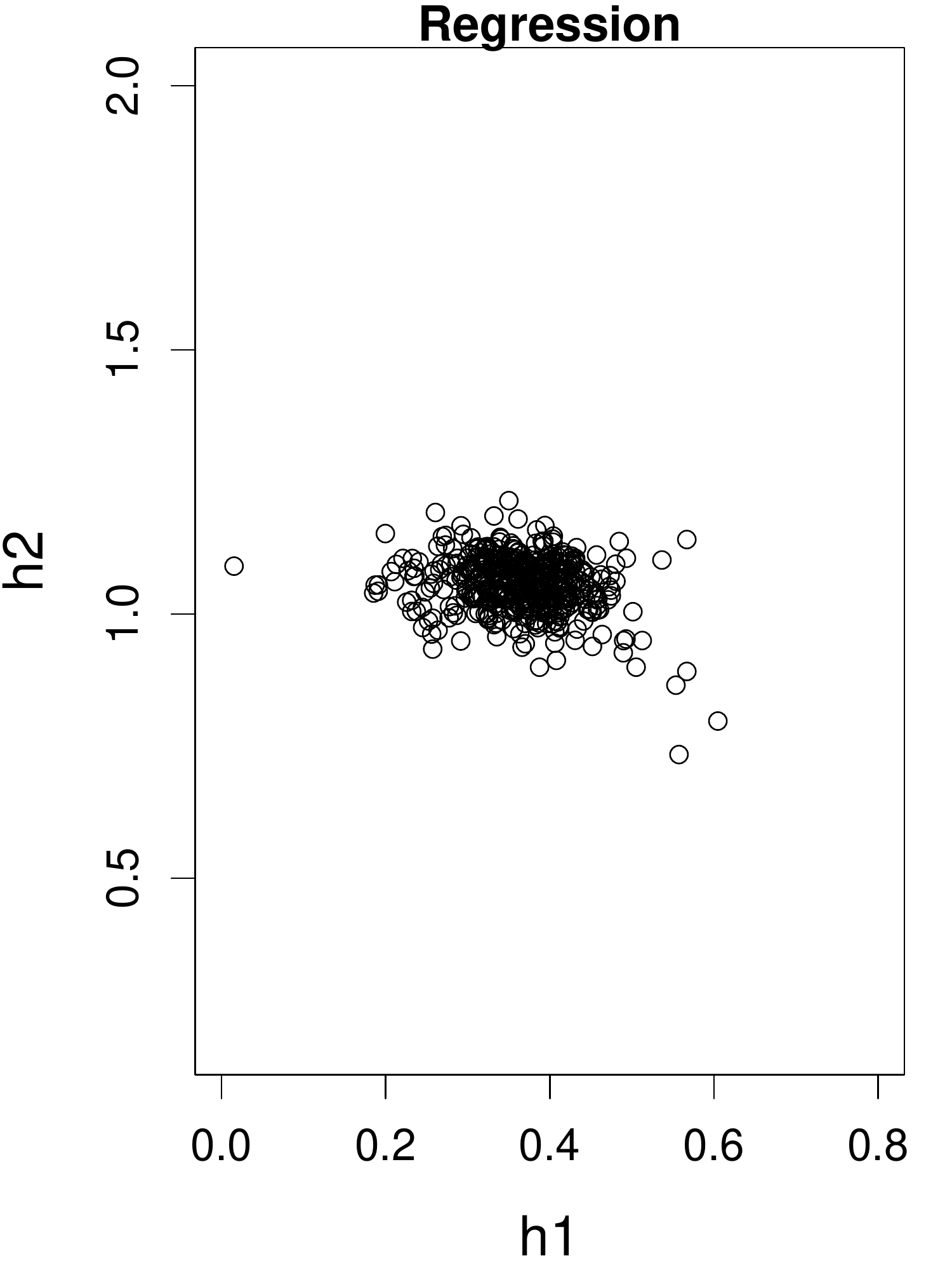} }
	\caption{Scatter plots of selected bandwidths across 500 MC replicates under (C3) corresponding to eight ways of bandwidth selection. The correspondence of the eight panels to eight methods is the same as that in Figure~\ref{Sim1:curves}.}
	\label{Sim3:bandwidths}
\end{figure}

\begin{figure}
	\centering
	\setlength{\linewidth}{0.2\textwidth}
	\subfigure[]{ \includegraphics[width=\linewidth]{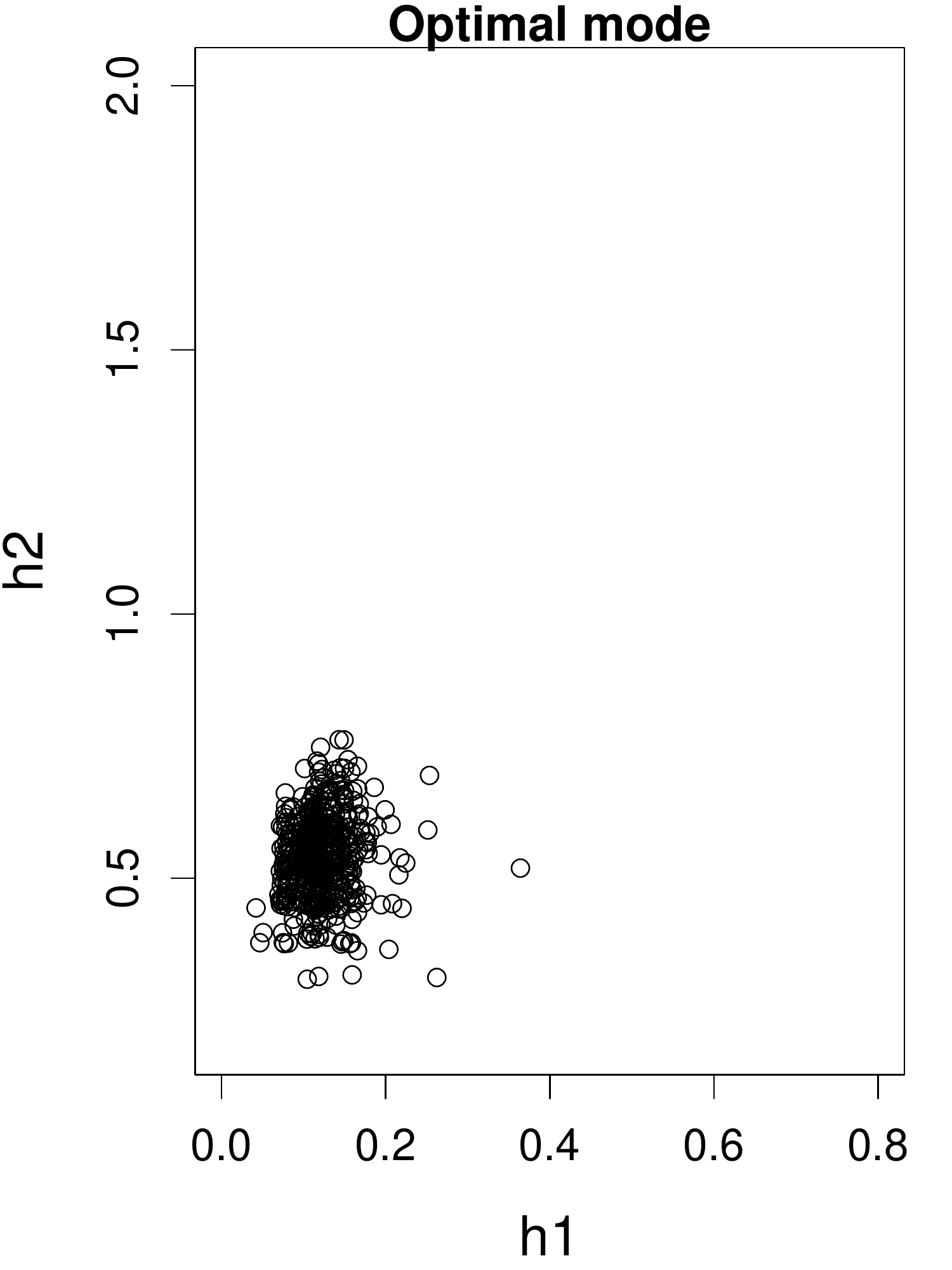} }
	\subfigure[]{ \includegraphics[width=\linewidth]{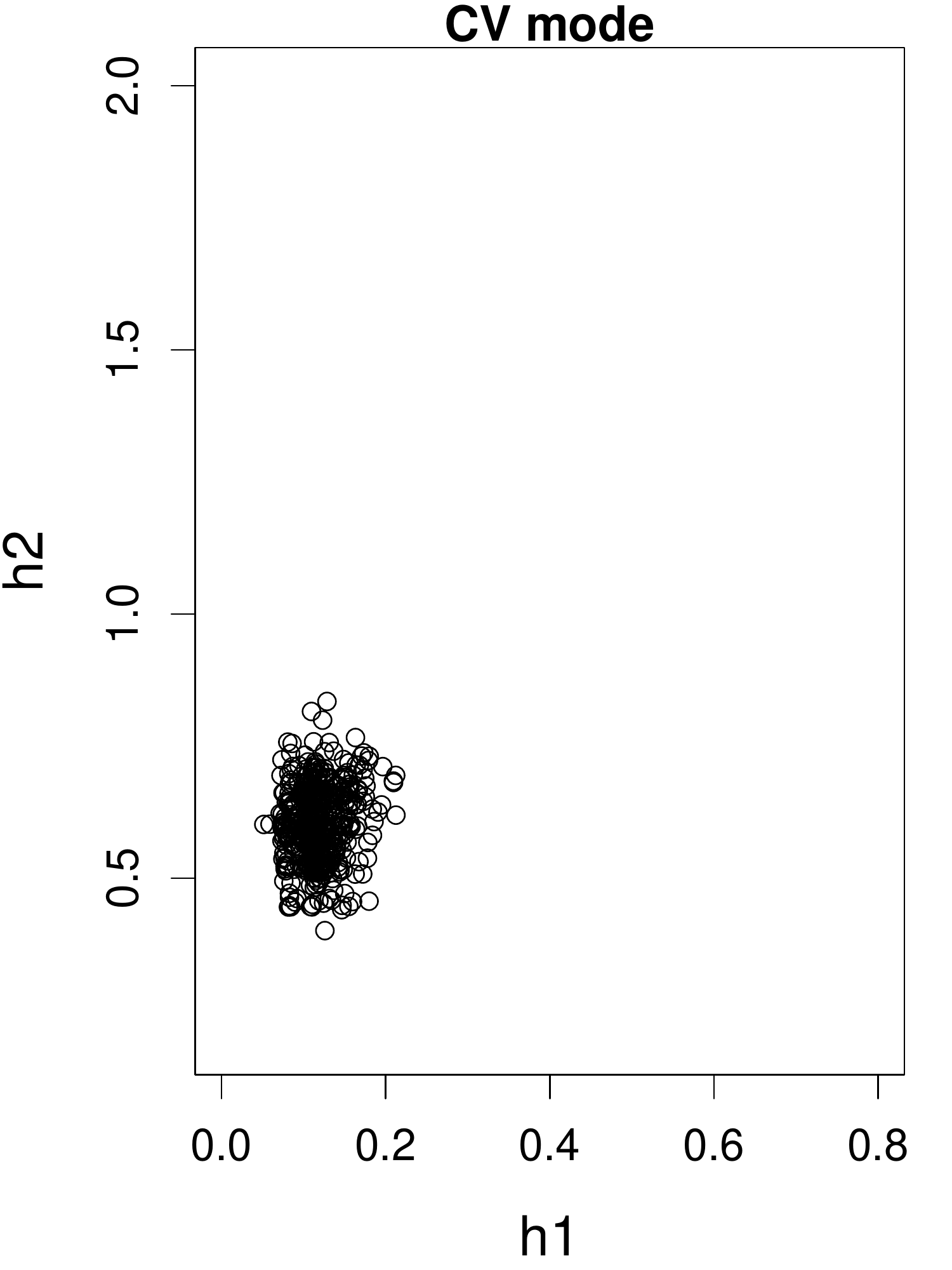} }
	\subfigure[]{ \includegraphics[width=\linewidth]{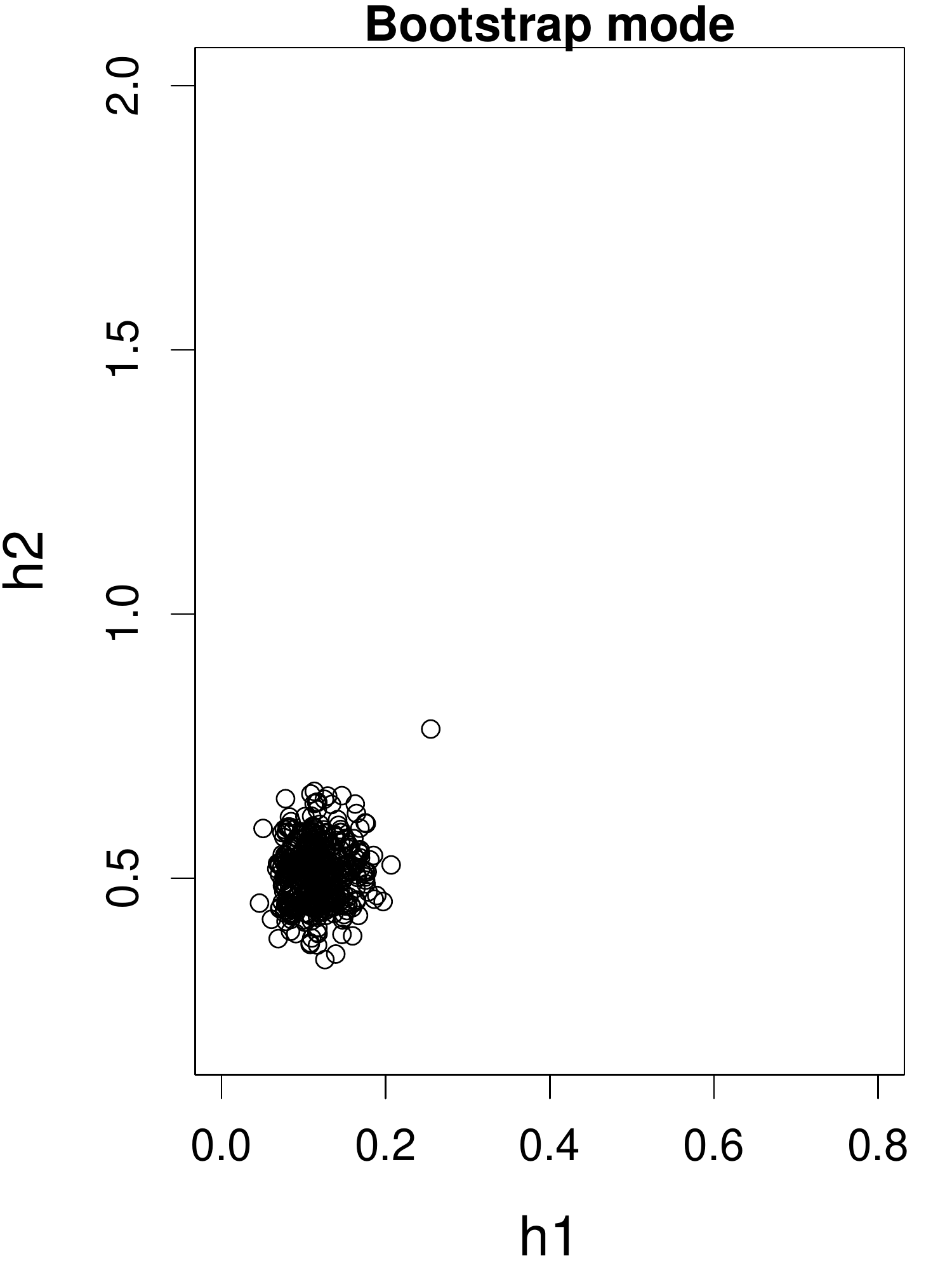} }
	\subfigure[]{ \includegraphics[width=\linewidth]{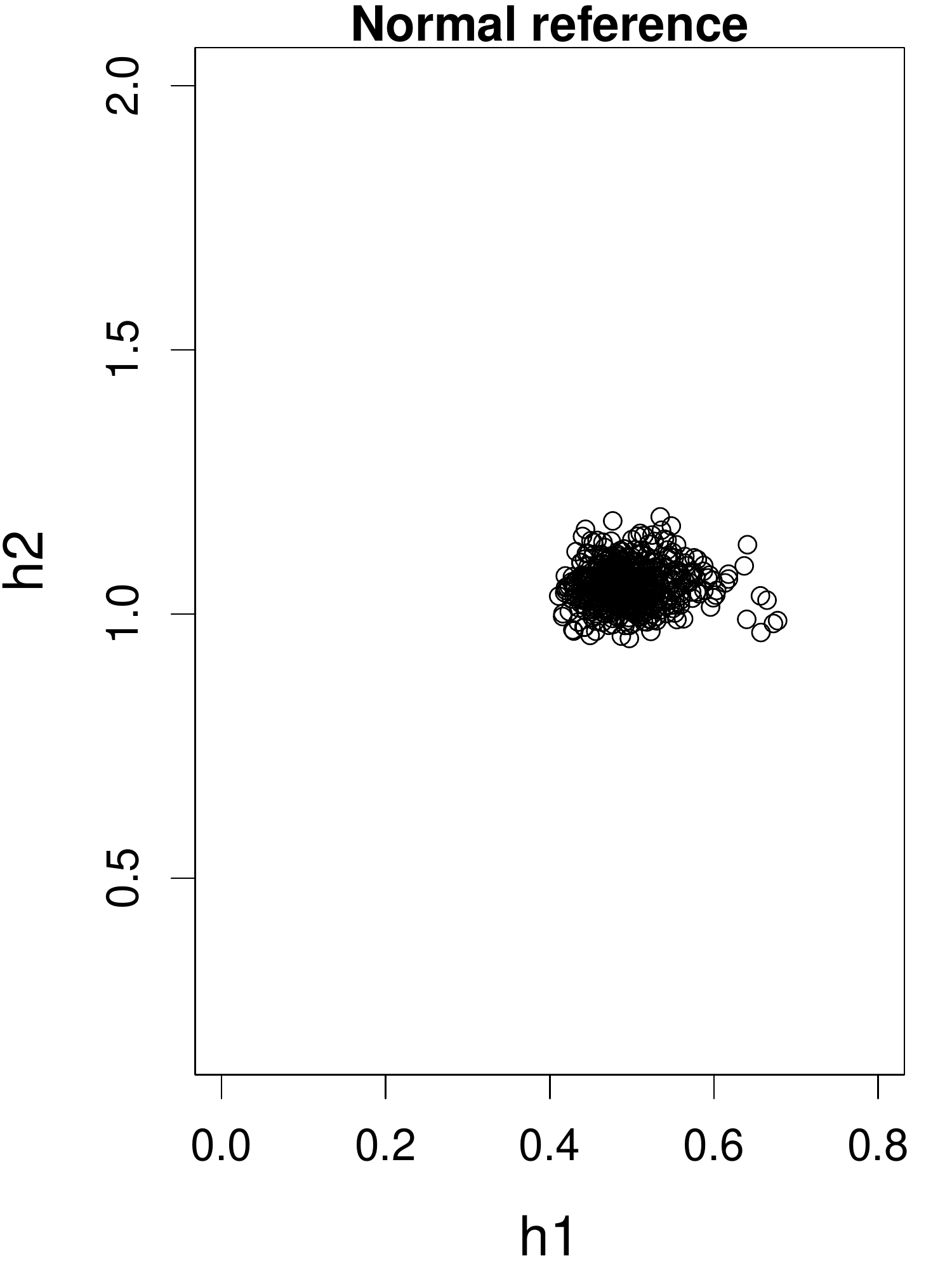} }\\
	\subfigure[]{ \includegraphics[width=\linewidth]{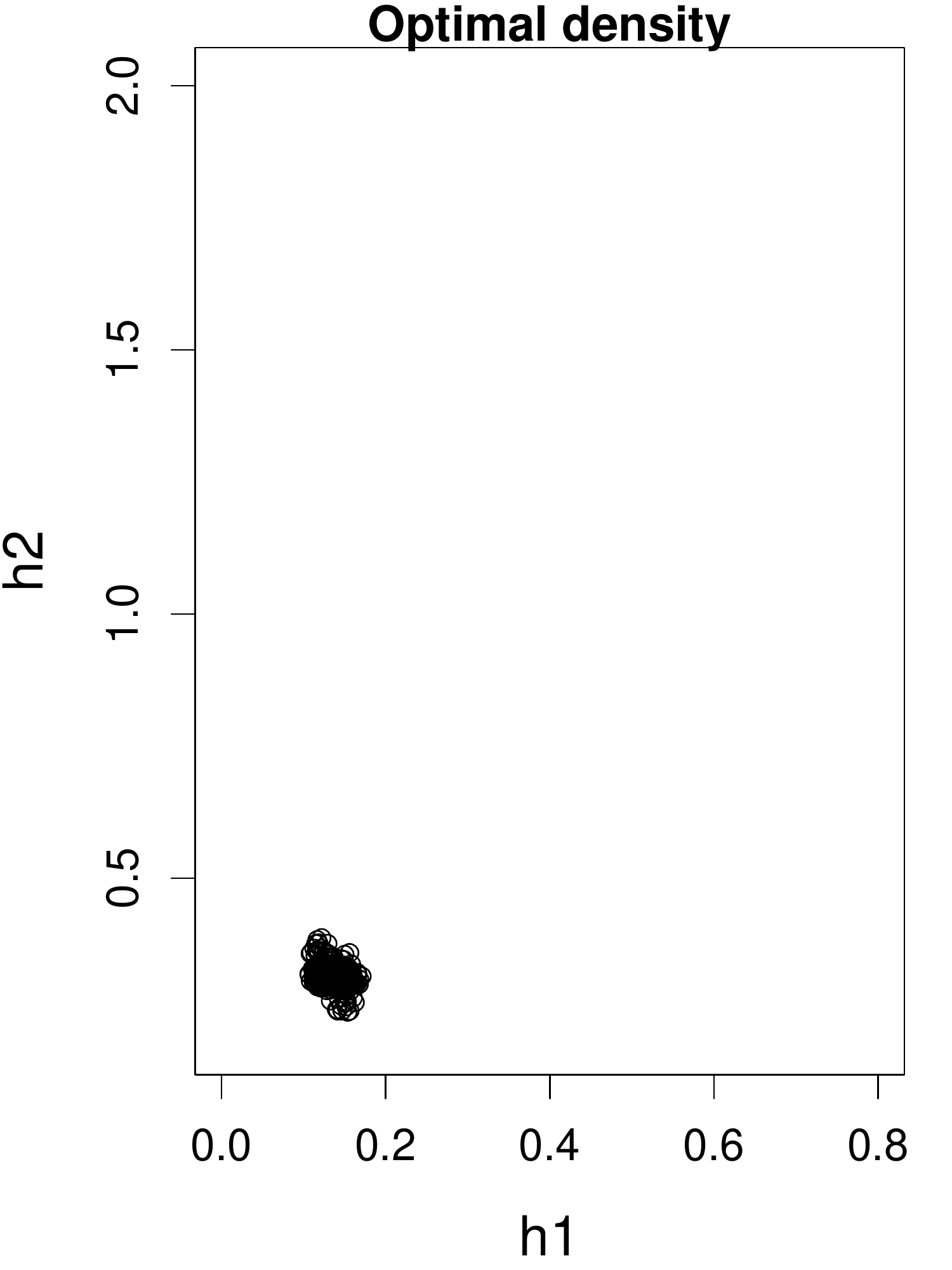} }
	\subfigure[]{ \includegraphics[width=\linewidth]{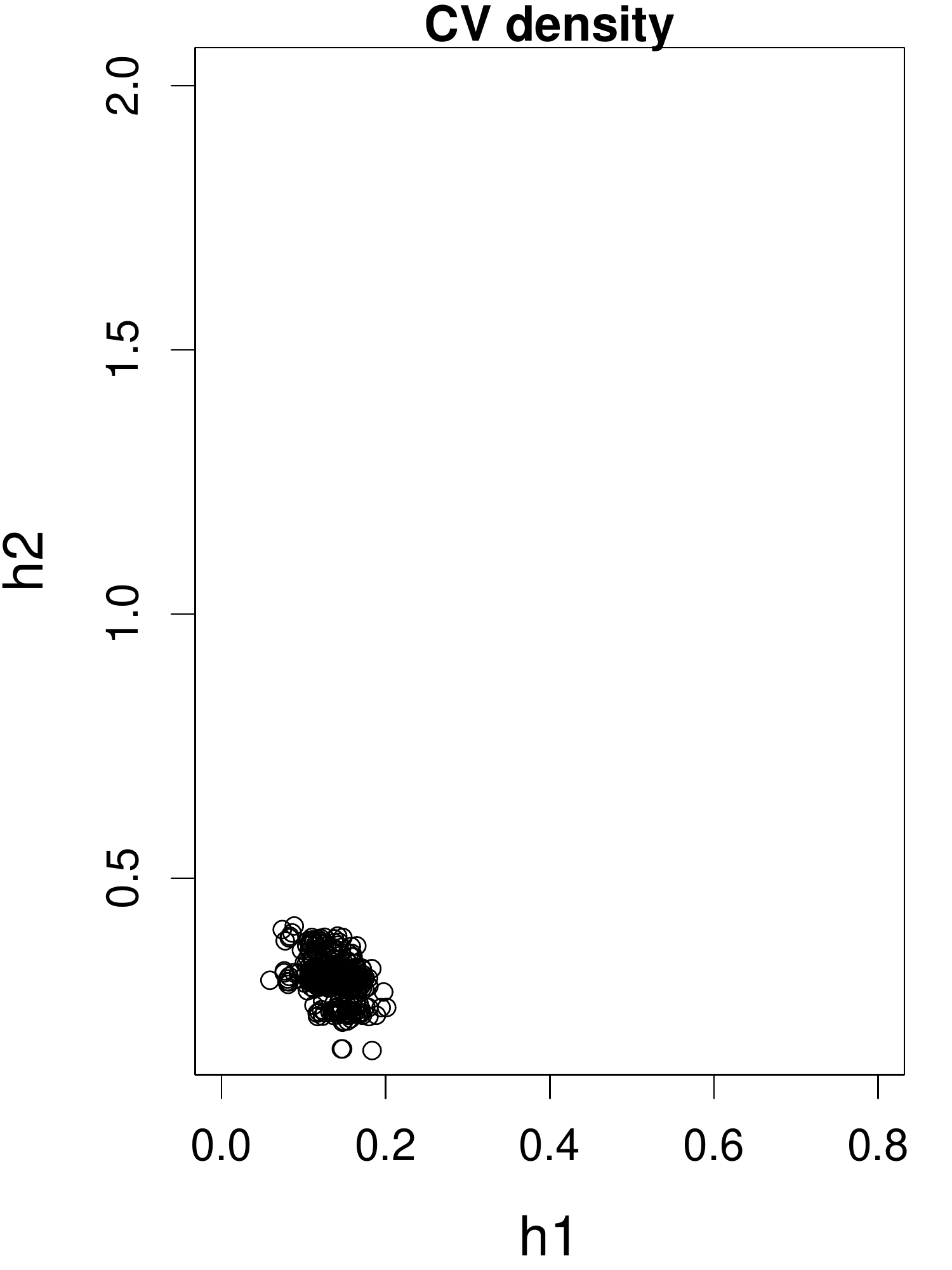} }
	\subfigure[]{ \includegraphics[width=\linewidth]{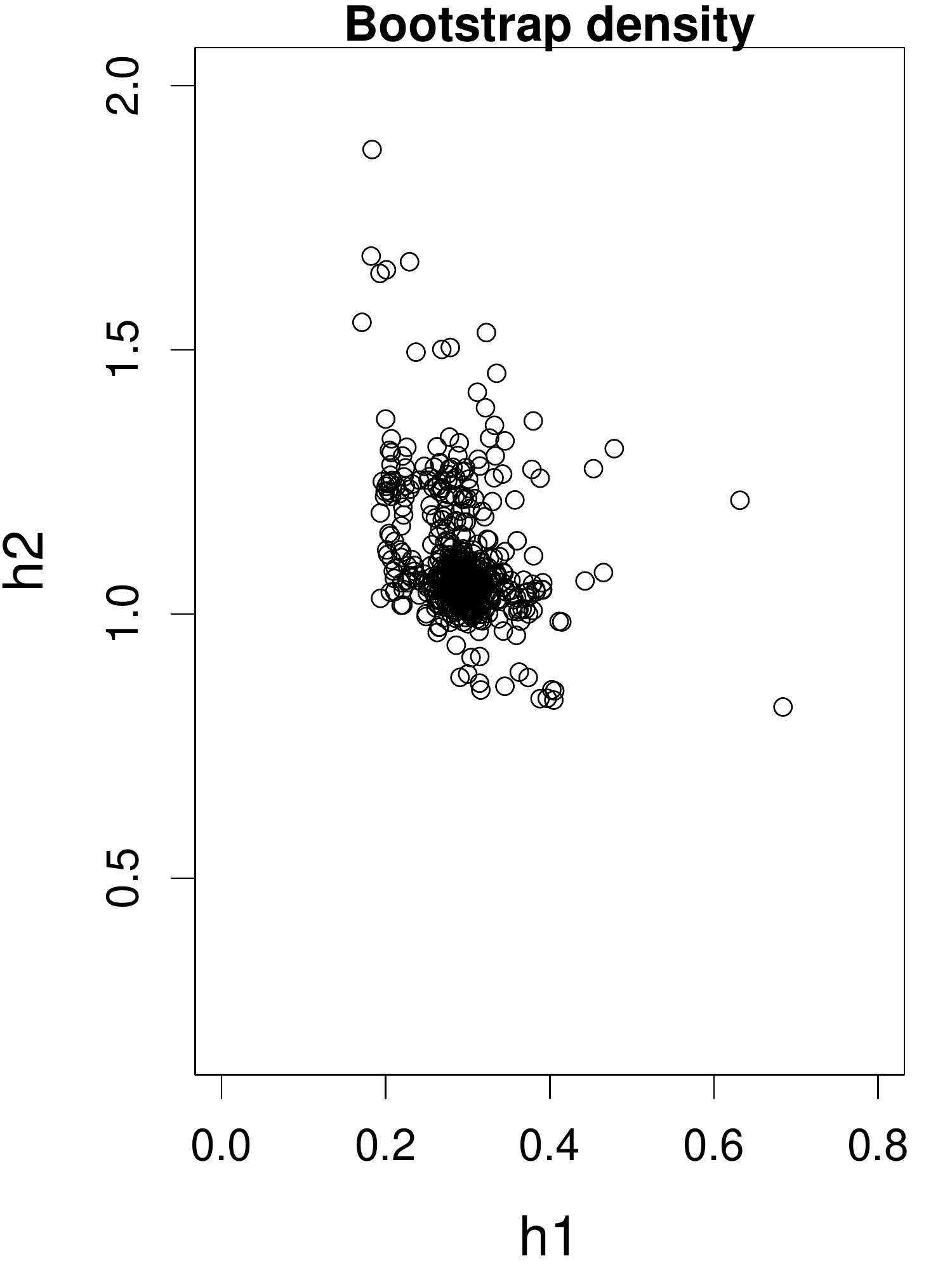} }
	\subfigure[]{ \includegraphics[width=\linewidth]{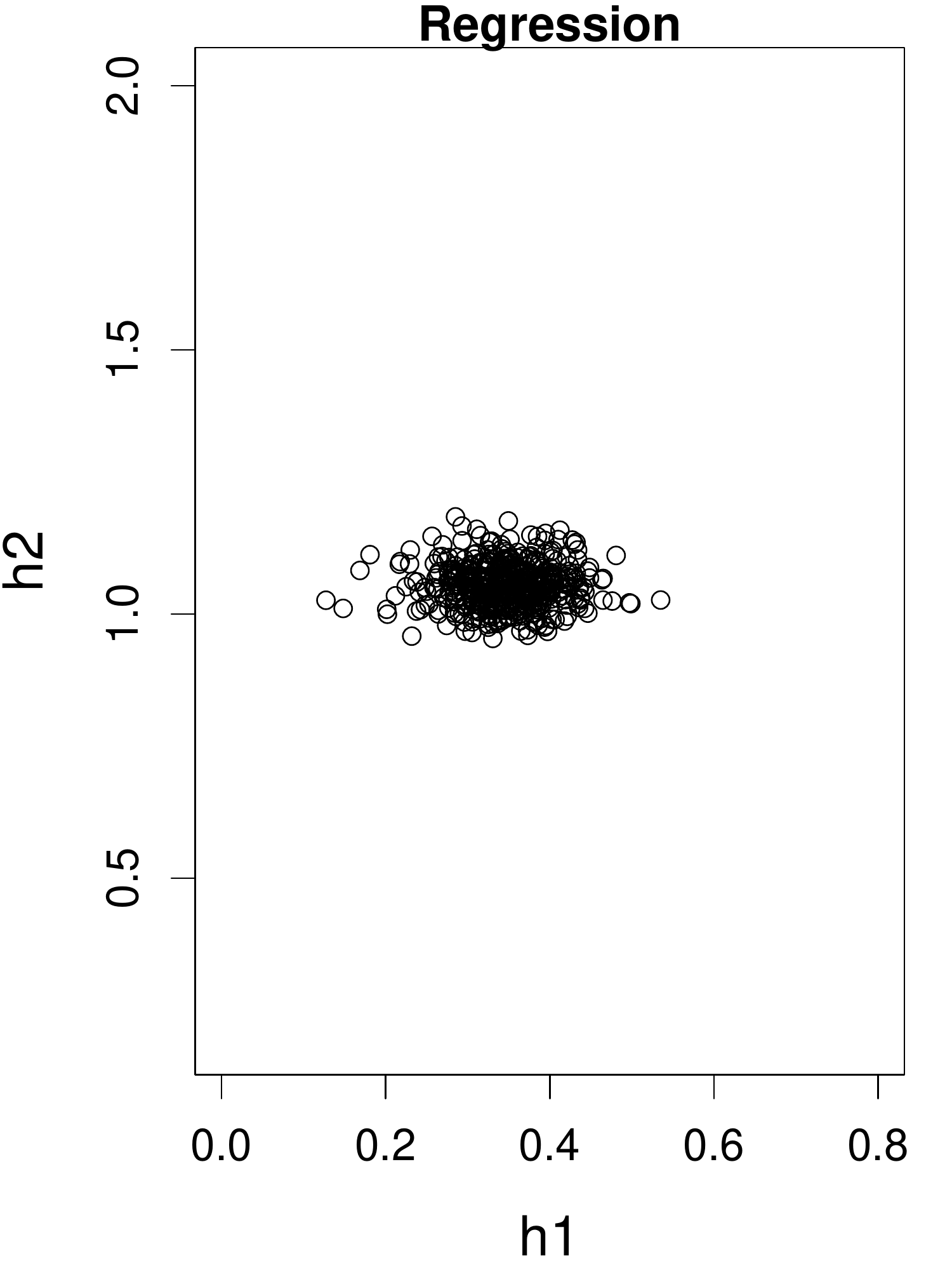} }
	\caption{Scatter plots of selected bandwidths across 500 MC replicates under (C4) corresponding to eight ways of bandwidth selection. The correspondence of the eight panels to eight methods is the same as that in Figure~\ref{Sim1:curves}.}
	\label{Sim4:bandwidths}
\end{figure}

\begin{figure}
	\centering
	\setlength{\linewidth}{0.2\textwidth}
	\subfigure[]{ \includegraphics[width=\linewidth]{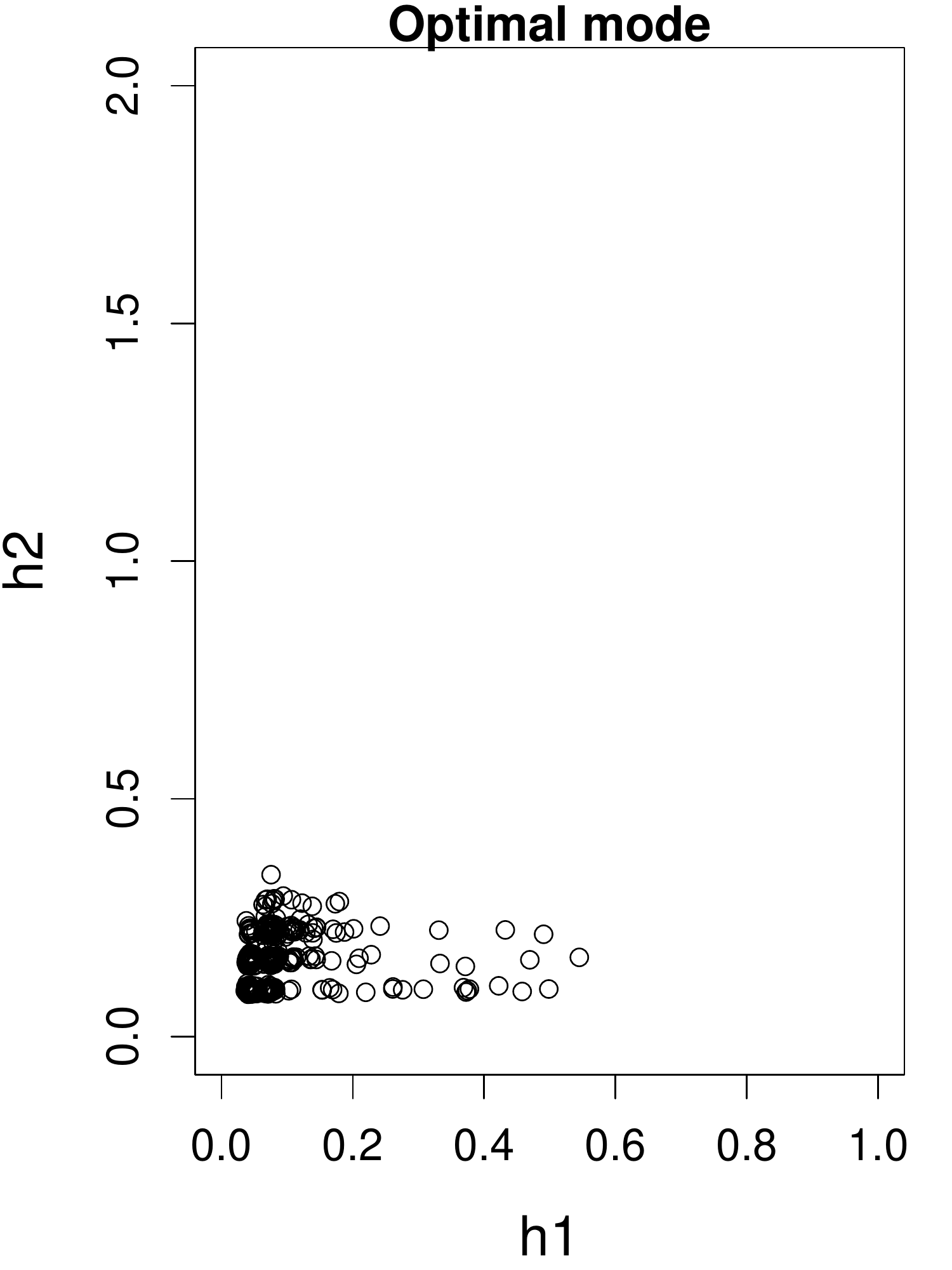} }
	\subfigure[]{ \includegraphics[width=\linewidth]{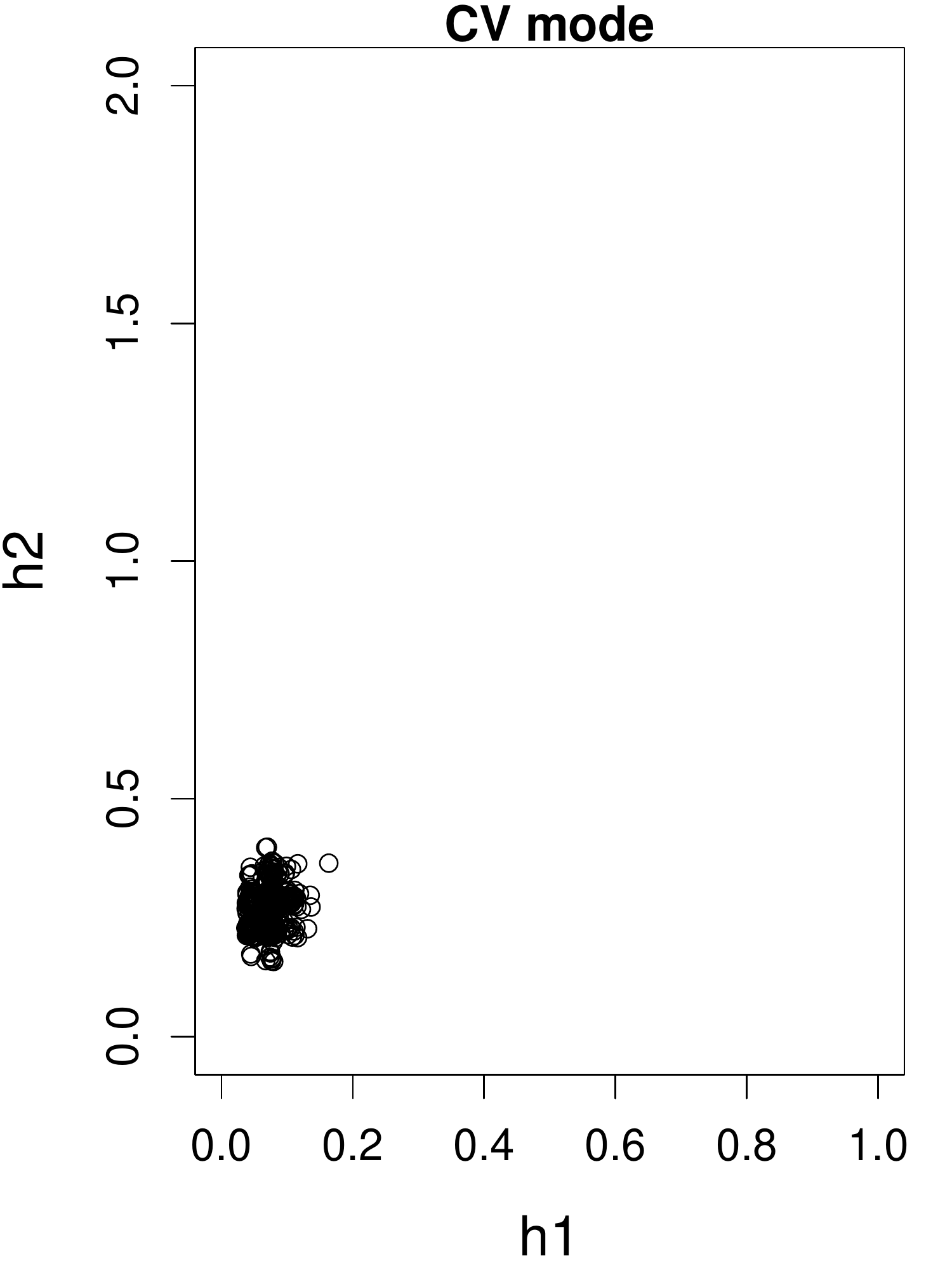} }
	\subfigure[]{ \includegraphics[width=\linewidth]{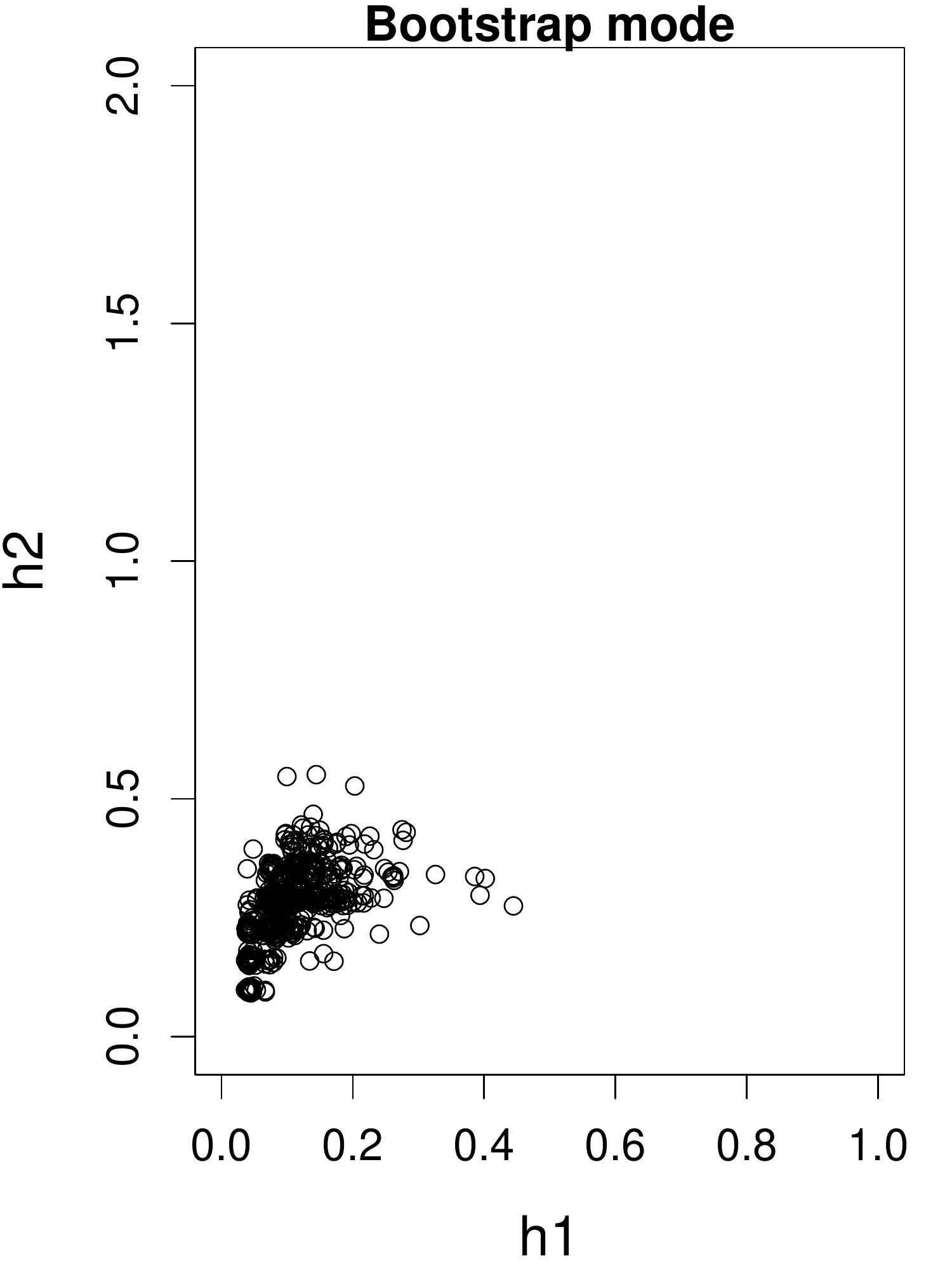} }
	\subfigure[]{ \includegraphics[width=\linewidth]{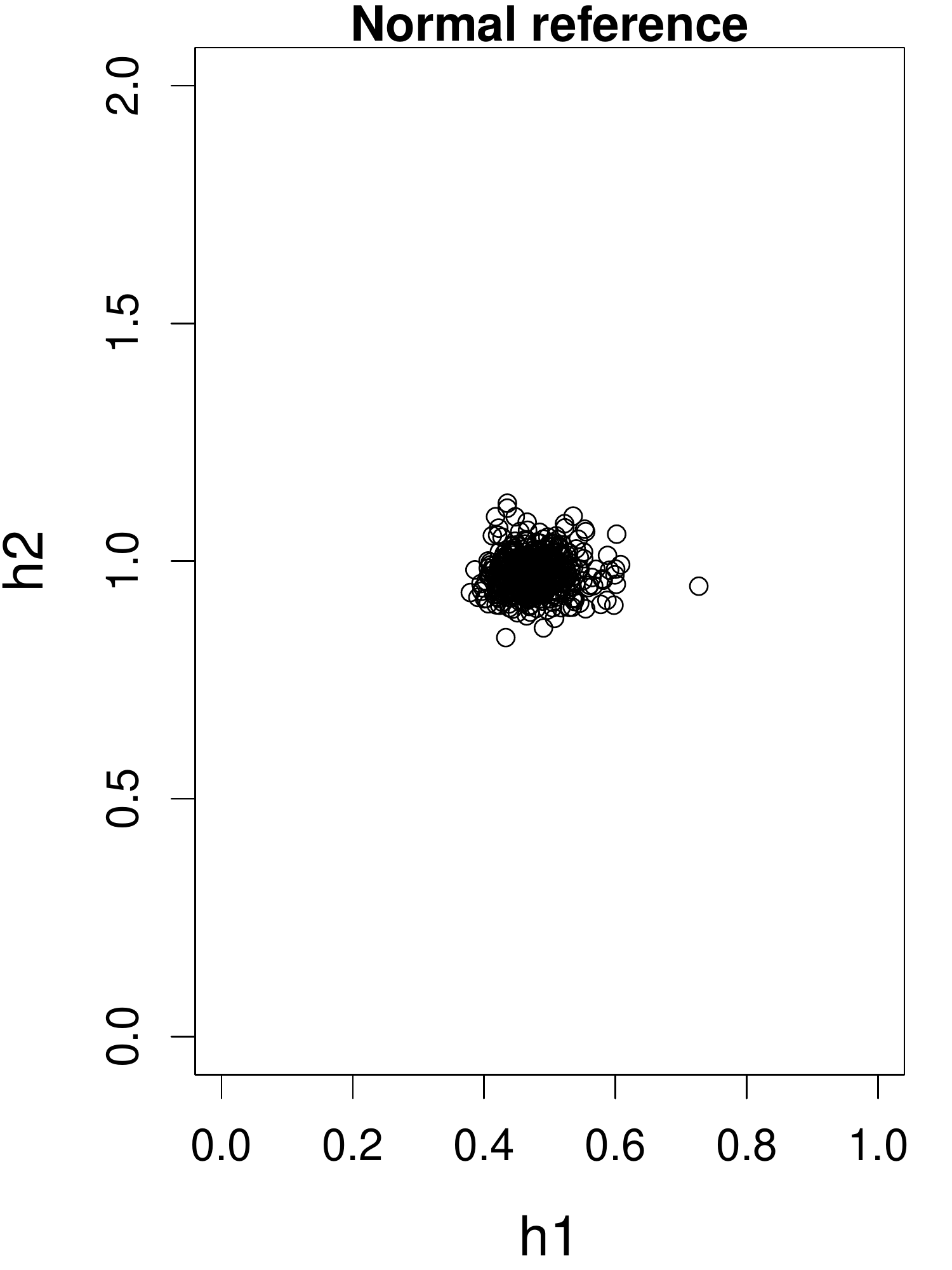} }\\
	\subfigure[]{ \includegraphics[width=\linewidth]{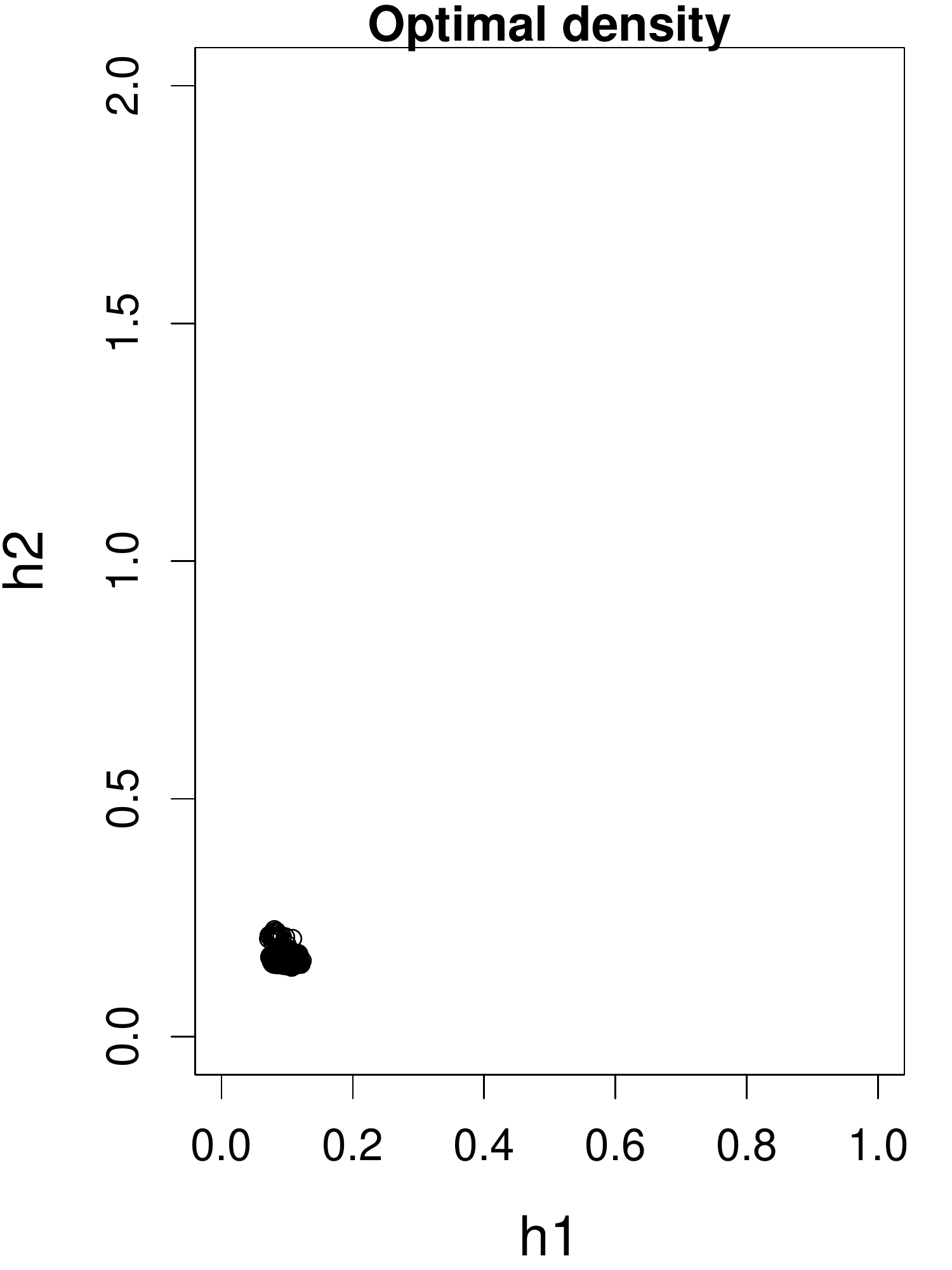} }
	\subfigure[]{ \includegraphics[width=\linewidth]{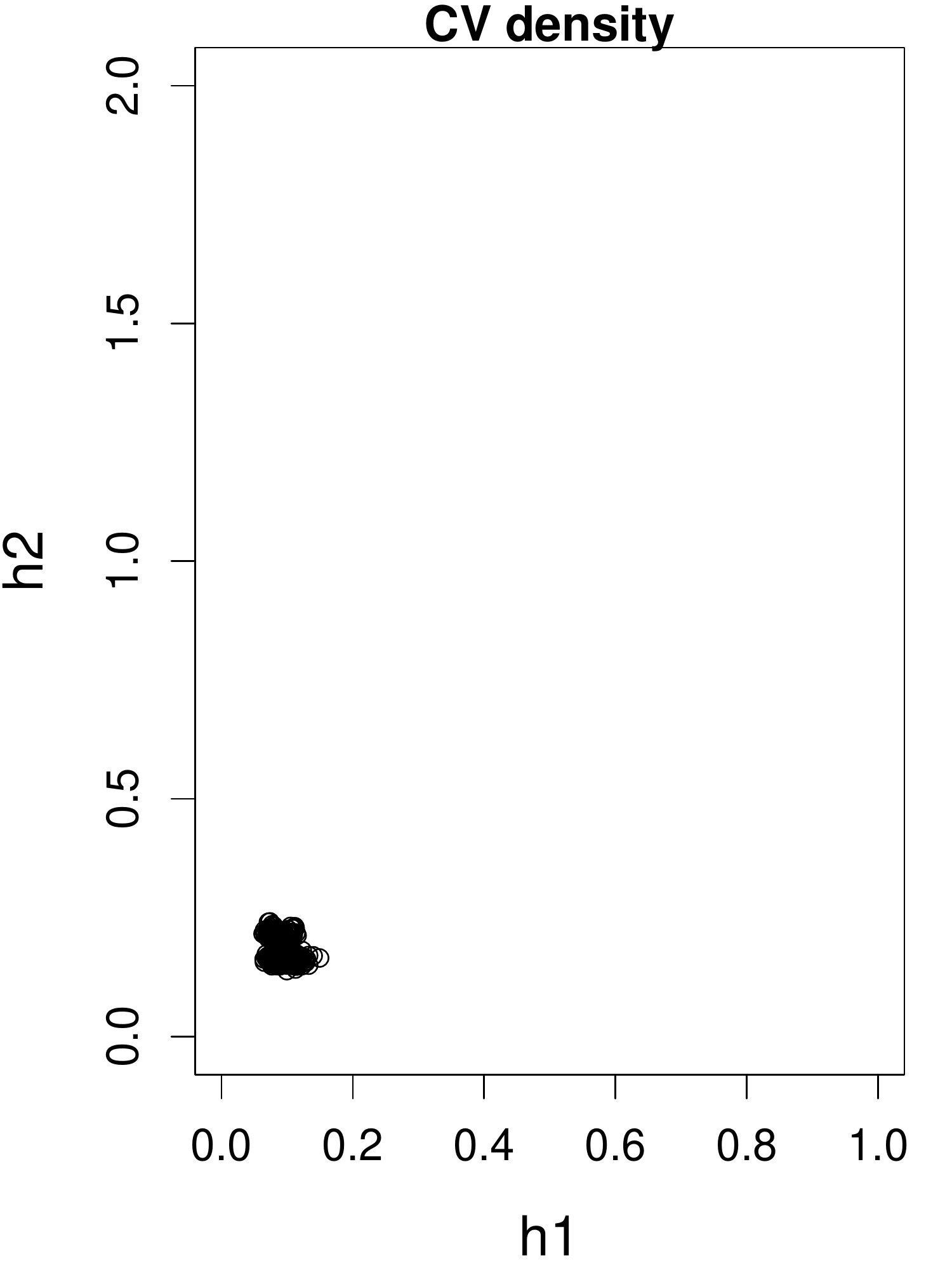} }
	\subfigure[]{ \includegraphics[width=\linewidth]{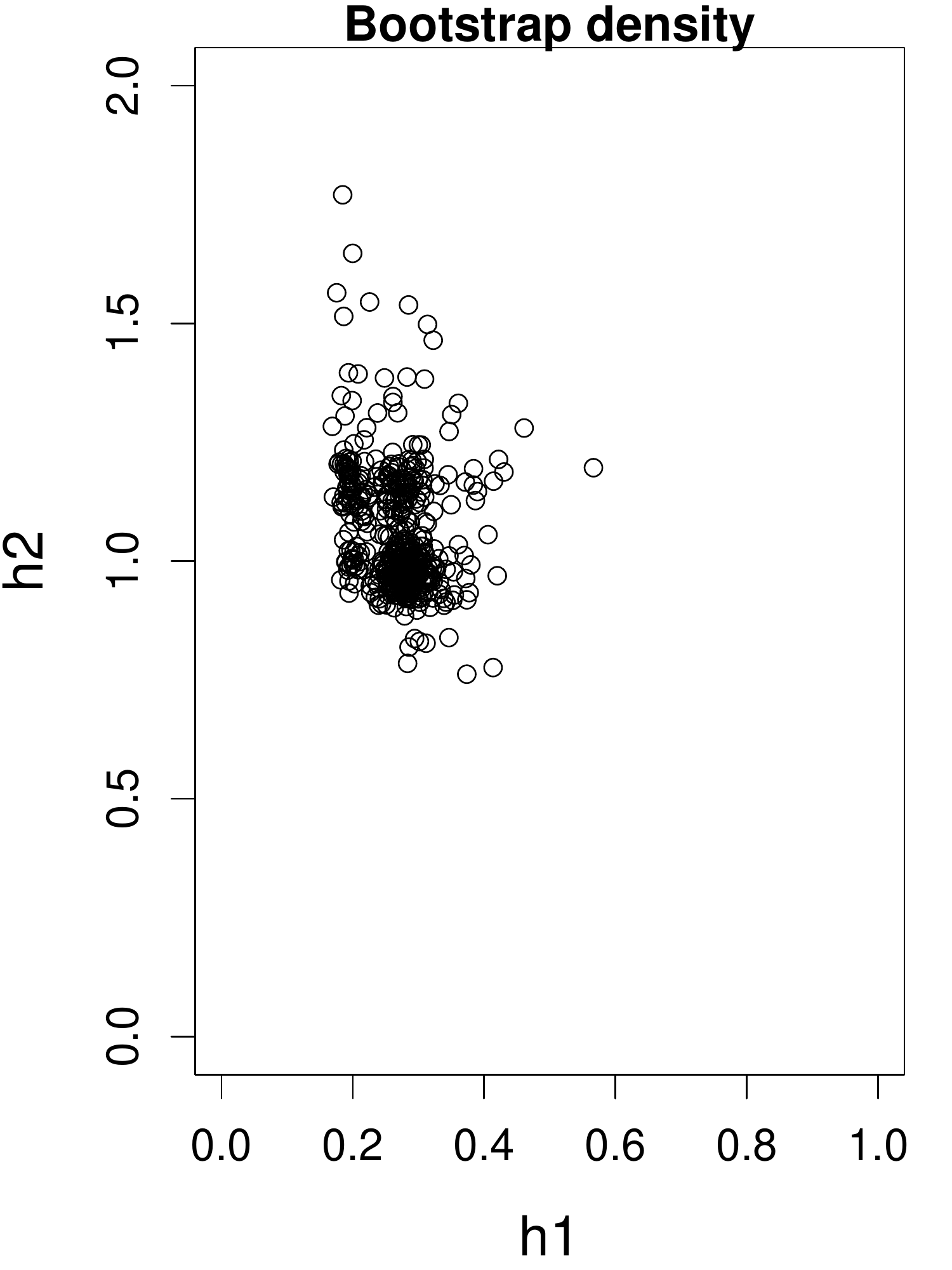} }
	\subfigure[]{ \includegraphics[width=\linewidth]{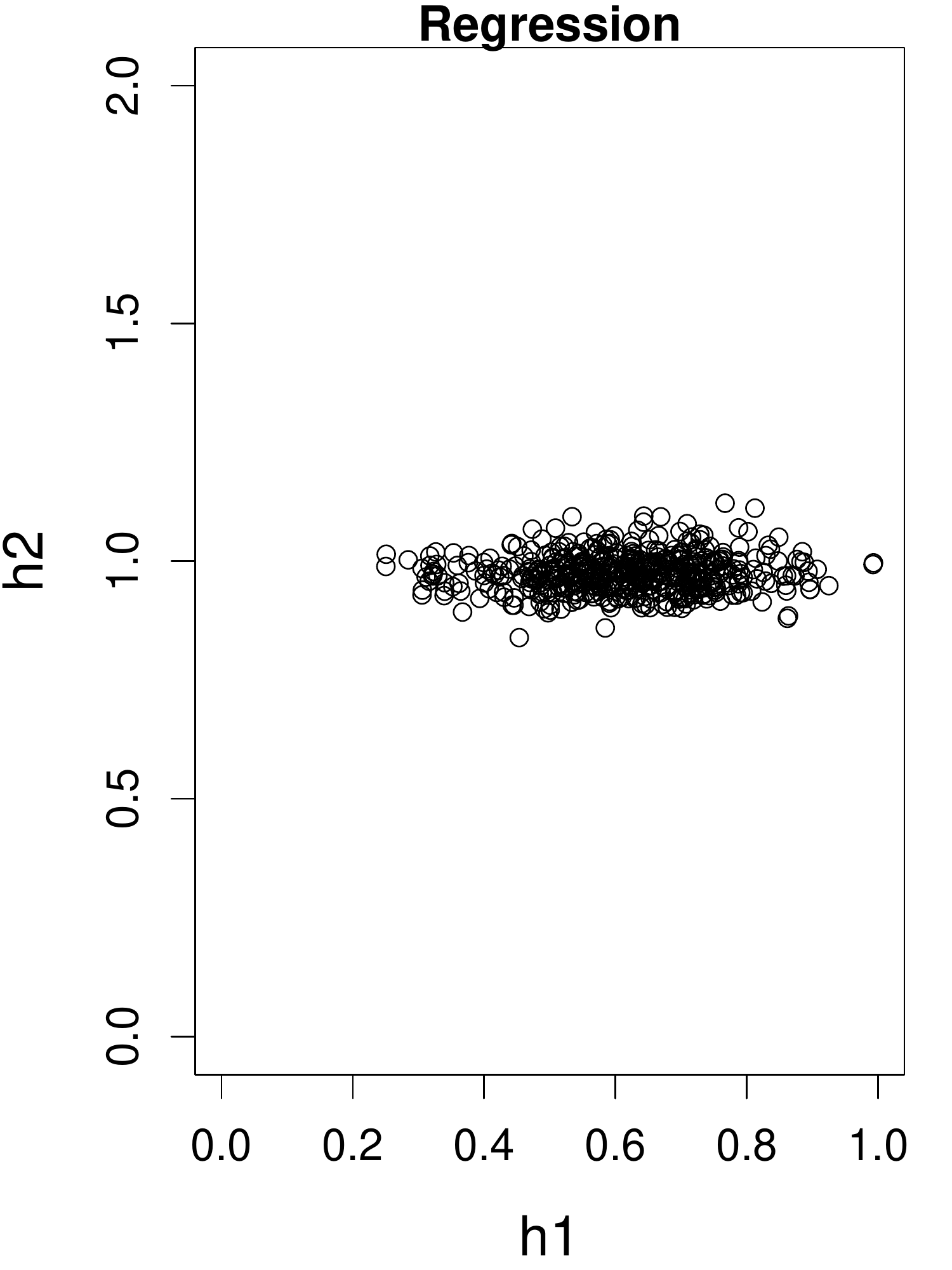} }
	\caption{Scatter plots of selected bandwidths across 500 MC replicates under (C5) corresponding to eight ways of bandwidth selection. The correspondence of the eight panels to eight methods is the same as that in Figure~\ref{Sim1:curves}.}
	\label{Sim5:bandwidths}
\end{figure}

To address the first aim of the simulation experiment, Figures~\ref{Sim1:bandwidths}--\ref{Sim5:bandwidths} present the scatter plots of chosen bandwidths across 500 MC replicates. The depicted optimal (in different senses) bandwidths (see panels (a) and (e)) reveal that $\tilde \bh_{\hbox {\tiny $D$}}$ and $\tilde \bh_{\hbox {\tiny $M$}}$ are indeed very different, especially in the bandwidth associated with $y$, except for (C5) that has the largest number of modes among the fives cases. In particular, $\hat h_{2, \hbox {\tiny $D$}}$ is substantially smaller than $\hat h_{2, \hbox {\tiny $M$}}$ under (C1)--(C4), which explains the severe overfitting when setting $\bh=\tilde \bh_{\hbox {\tiny $D$}}$ observed in Figures~\ref{Sim1:curves}--\ref{Sim4:curves}. If the interest lies in density estimation, among the four density-based methods, the one yielding $\bh_{\hbox {\tiny $D$}}$ (see panel (f)) is a clear winner since $\bh_{\hbox {\tiny $D$}}$ is closer to $\tilde\bh_{\hbox {\tiny $D$}}$ than $\bh_{\hbox {\tiny $N$}}$, $\bh_{\hbox {\tiny $R$}}$, and $\bh_{\hbox {\tiny $B$}}$. But if one is interested in inferring conditional modes, all four existing methods miss the mark since none of them yield bandwidths approximating $\tilde \bh_{\hbox {\tiny $M$}}$ well. In contrast, our proposed mode-based bandwidths, $\bh_{\hbox {\tiny $M$}}$ and $\bh^*_{\hbox {\tiny $B$}}$, are much more promising in approximating $\tilde \bh_{\hbox {\tiny $M$}}$, especially under (C2). Between these two, $\bh^*_{\hbox {\tiny $B$}}$ tends to be more variable than $\bh_{\hbox {\tiny $M$}}$, which can be due to the mixture model estimation. 

\begin{sidewaysfigure}[p]
	\centering
	\setlength{\linewidth}{4cm}
	\subfigure[]{ \includegraphics[width=\linewidth]{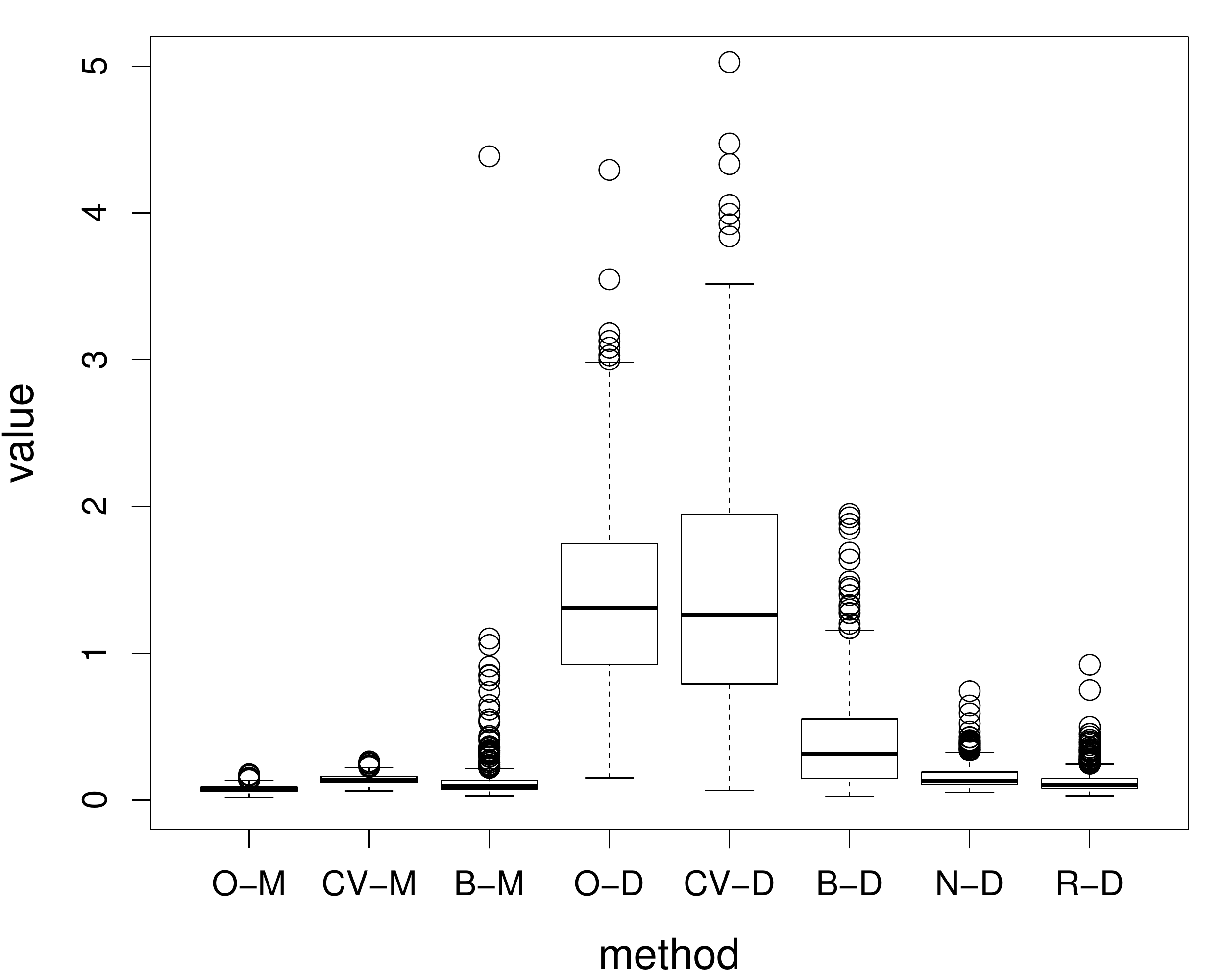} }
	\subfigure[]{ \includegraphics[width=\linewidth]{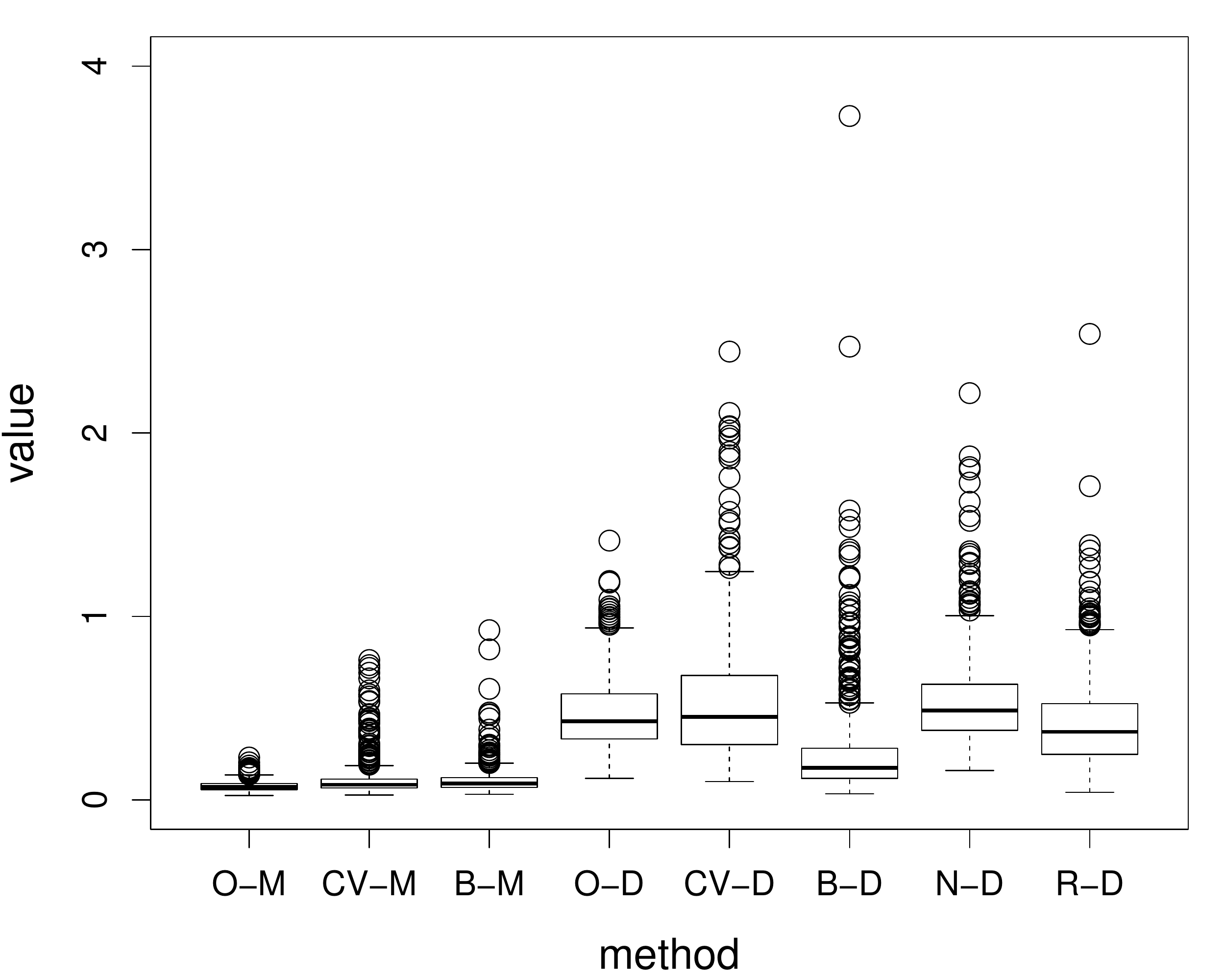} }
	\subfigure[]{ \includegraphics[width=\linewidth]{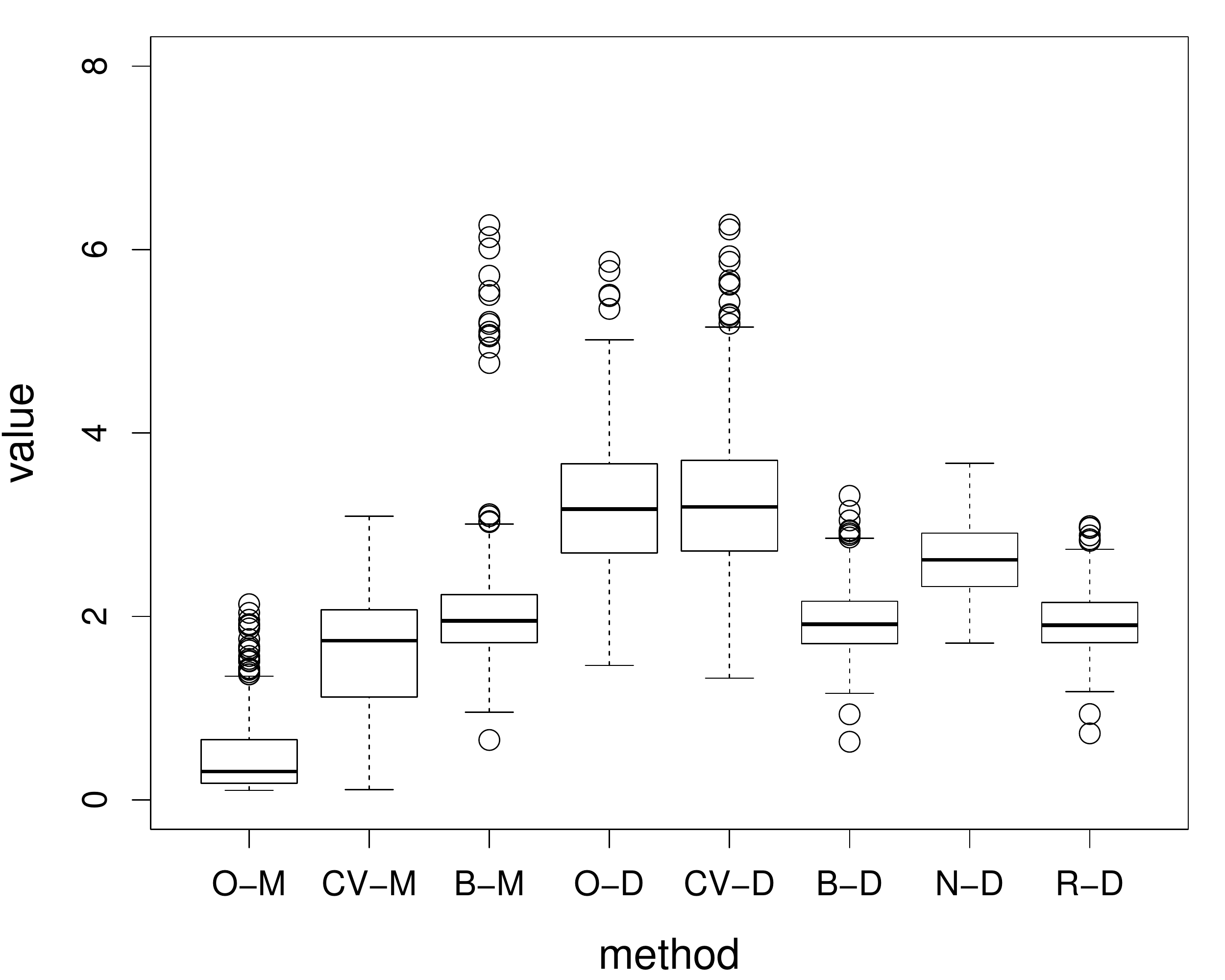} }
	\subfigure[]{ \includegraphics[width=\linewidth]{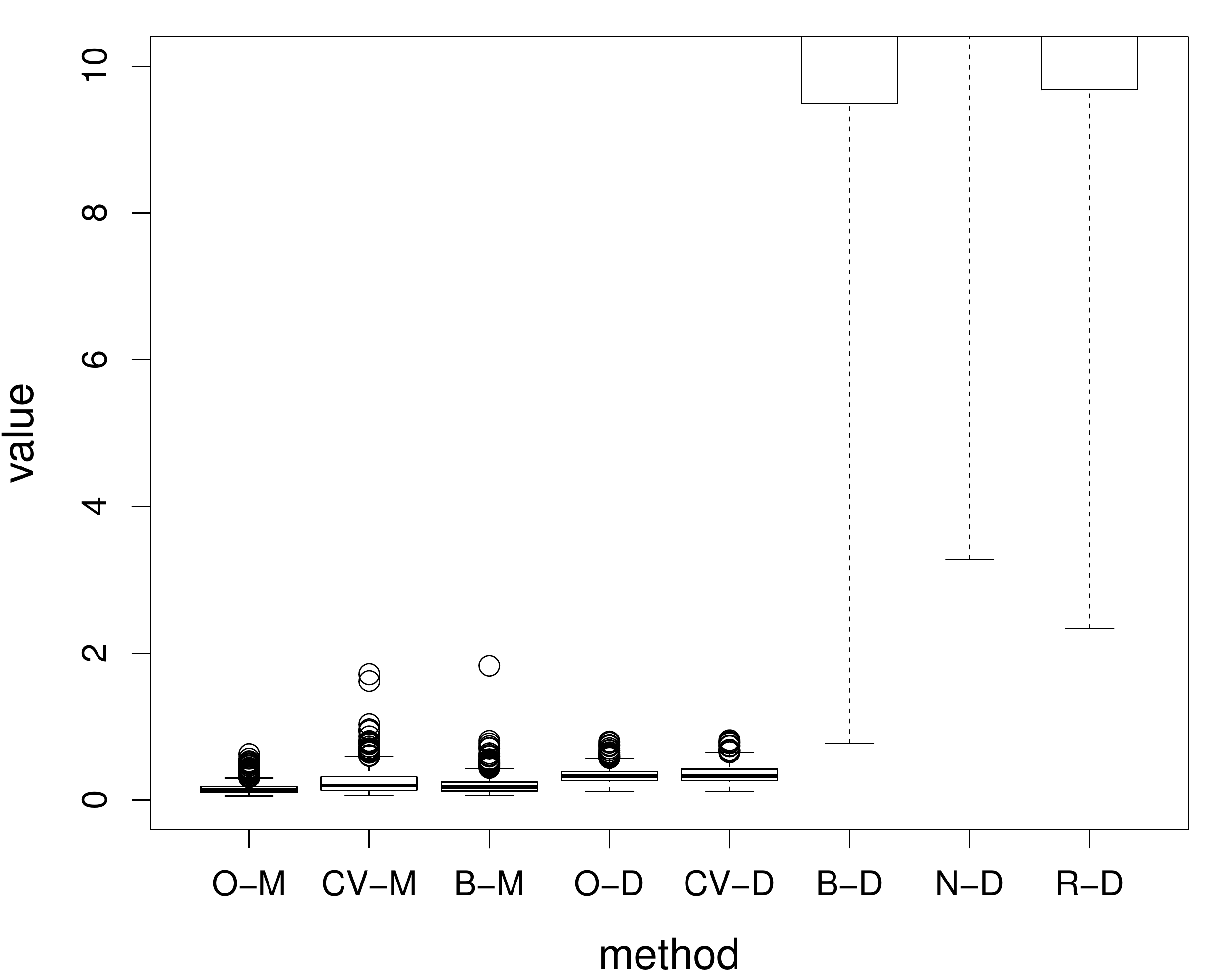} }
	\subfigure[]{ \includegraphics[width=\linewidth]{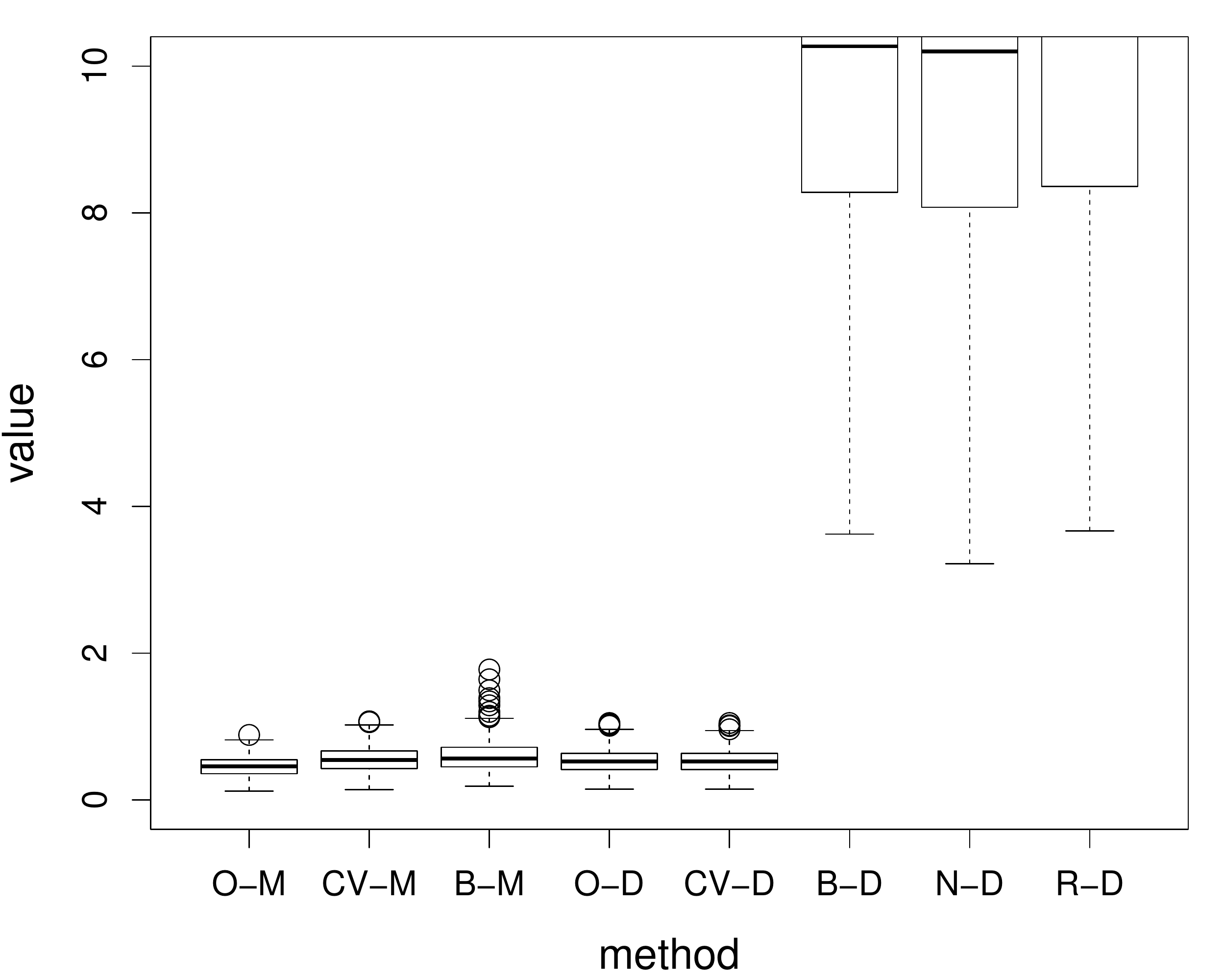} }\\
	\subfigure[]{ \includegraphics[width=\linewidth]{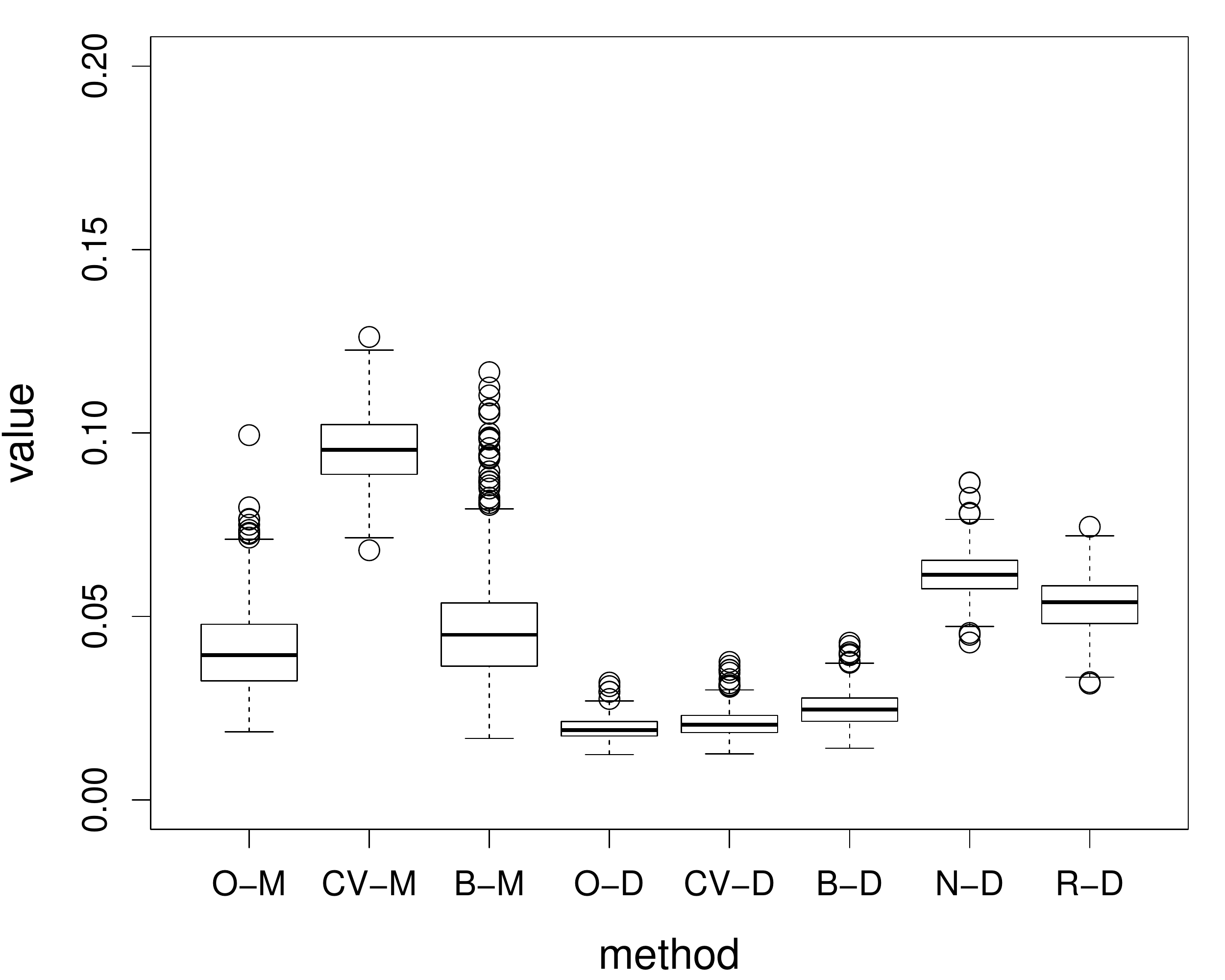} }
	\subfigure[]{ \includegraphics[width=\linewidth]{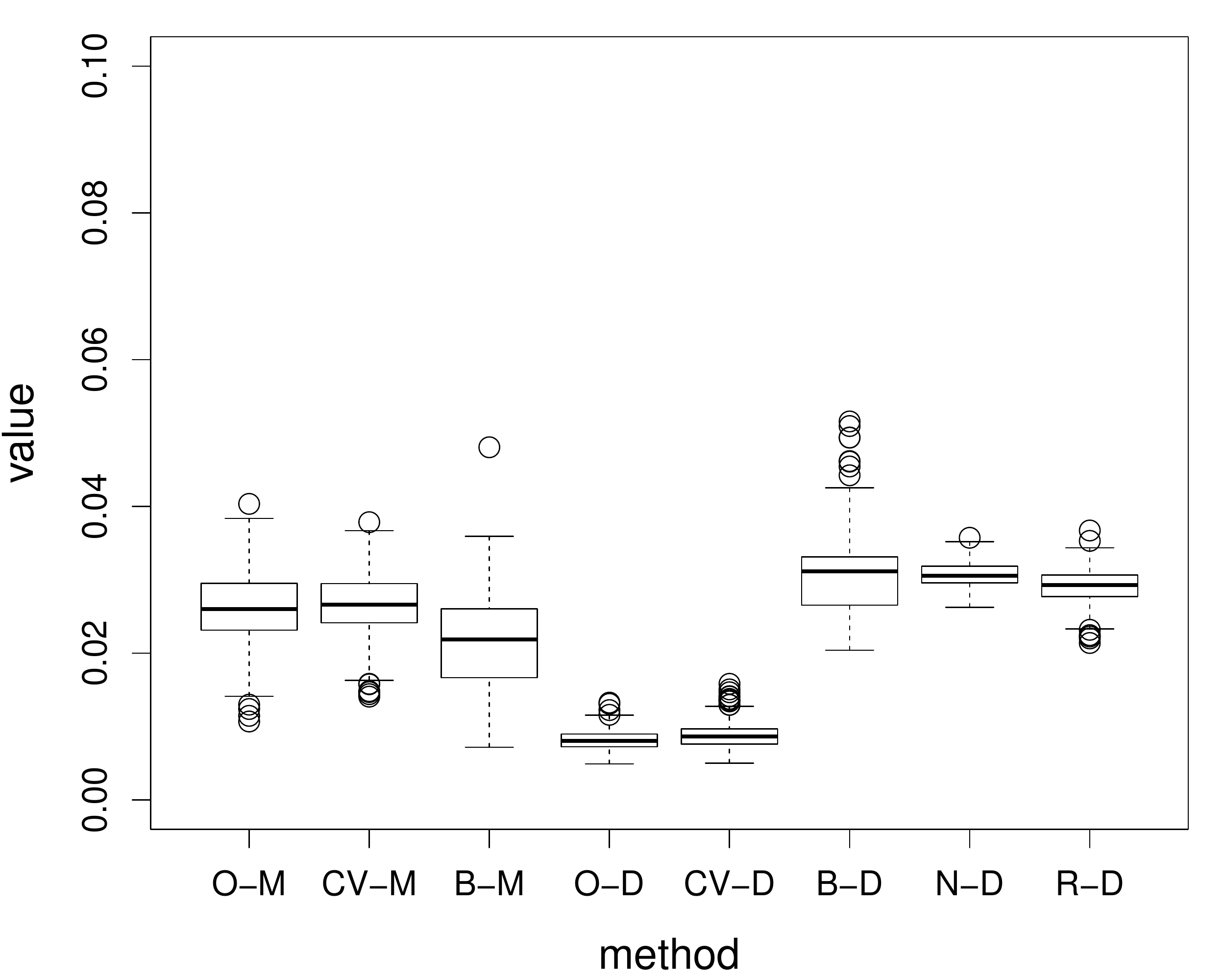} }
	\subfigure[]{ \includegraphics[width=\linewidth]{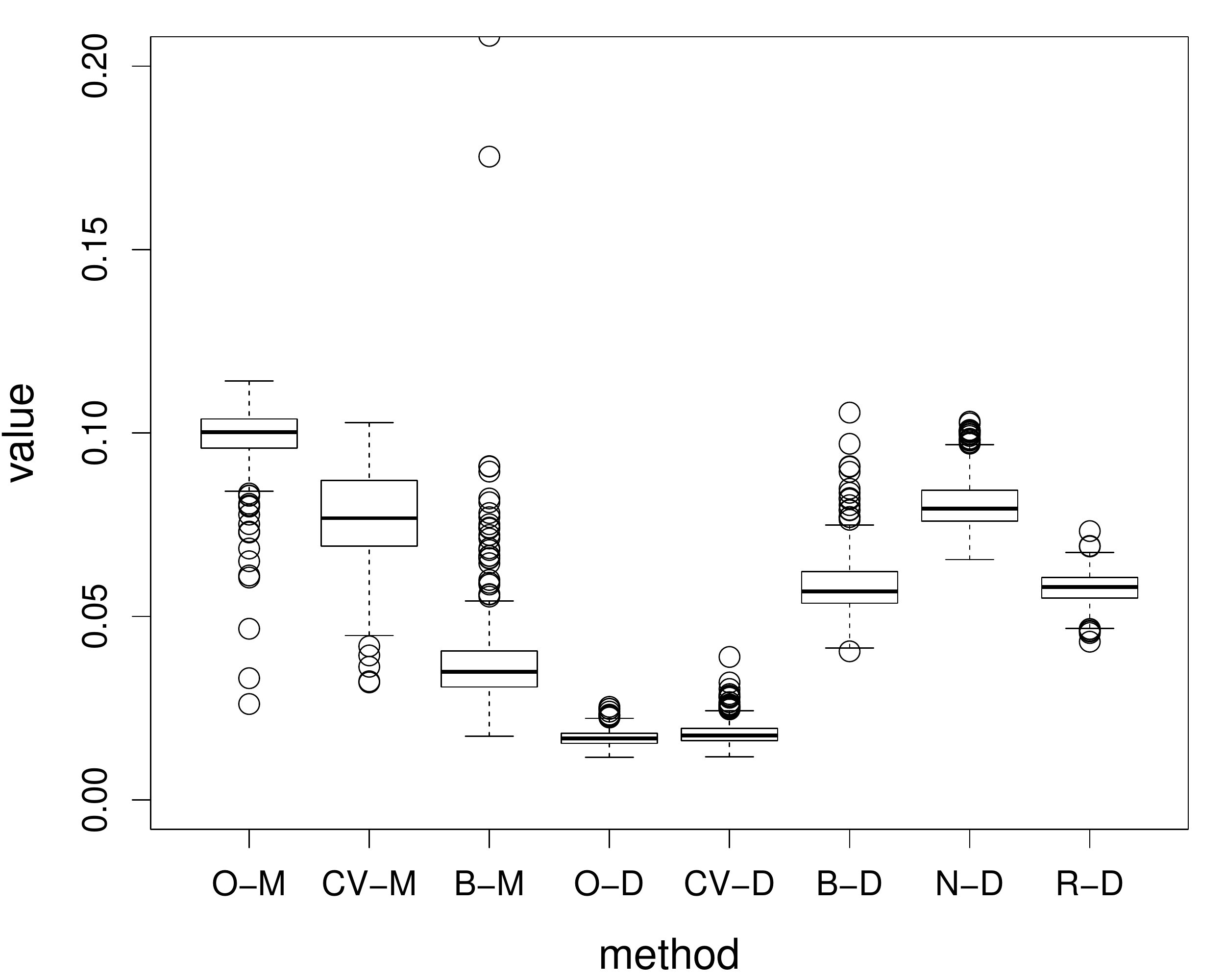} }
	\subfigure[]{ \includegraphics[width=\linewidth]{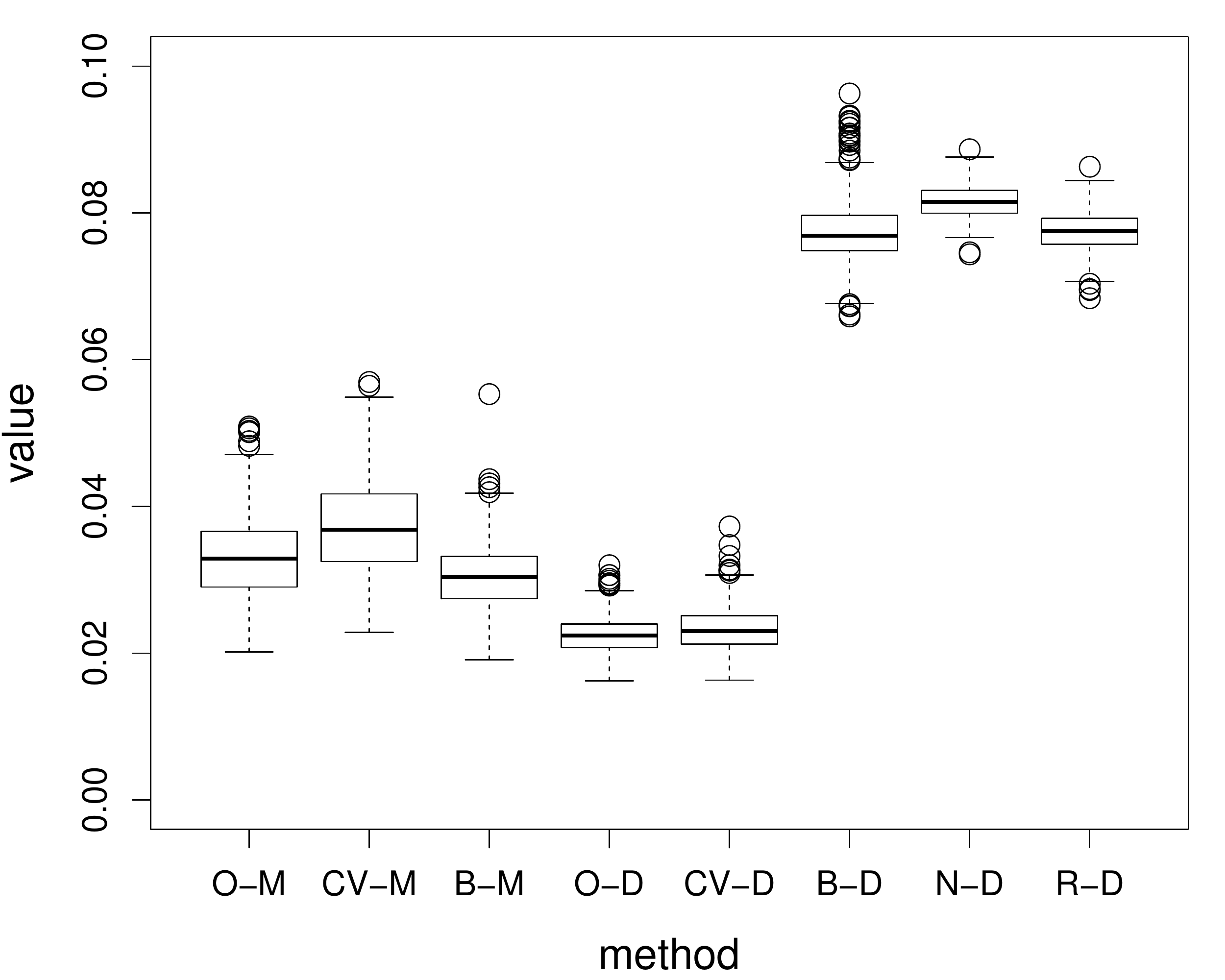} }
	\subfigure[]{ \includegraphics[width=\linewidth]{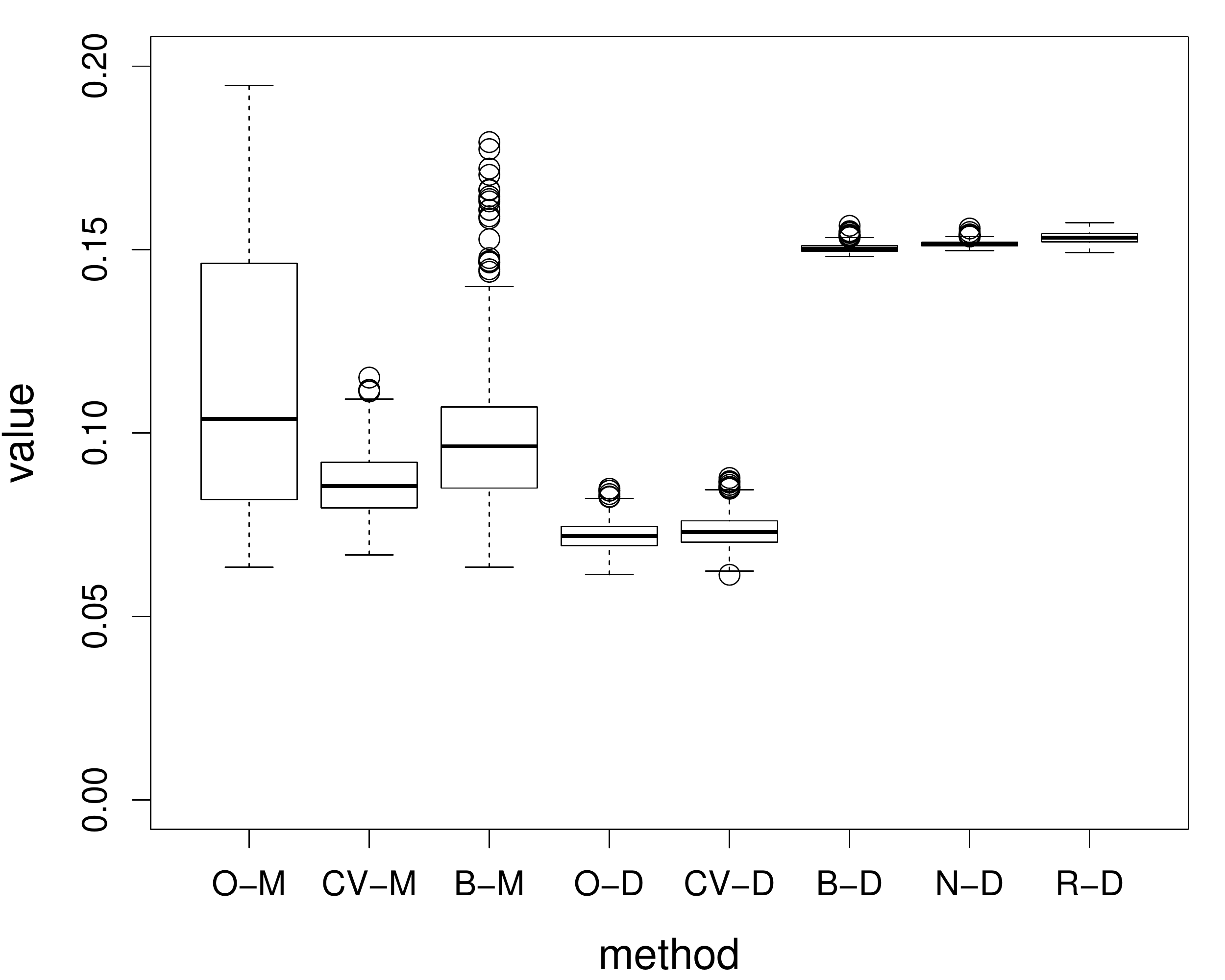} }\\
	\caption{The top row shows boxplots of $\textrm{EISE}_{\hbox {\tiny $M$}}$ under (C1)--(C5) in panels (a)--(e), respectively. The bottom row shows boxplots of $\textrm{EISE}_{\hbox {\tiny $D$}}$ under (C1)--(C5) in panels (f)--(j), respectively. In each panel, the eight adopted $\bh$ corresponding to the eight boxes are the optimal mode-based bandwidths, $\tilde \bh_{\hbox {\tiny $M$}}$ (O-M), the mode-based bandwidths involving CV, $\bh_{\hbox {\tiny $M$}}$ (CV-M), the mode-based bandwidths involving bootstrap, $\bh^*_{\hbox {\tiny $B$}}$ (B-M), the optimal density-based bandwidths, $\tilde \bh_{\hbox {\tiny $D$}}$ (O-D), the density-based bandwidths involving CV, $\bh_{\hbox {\tiny $D$}}$ (CV-D), the density-based bandwidths involving bootstrap, $\bh_{\hbox {\tiny $B$}}$ (B-D), the normal reference, $\bh_{\hbox {\tiny $N$}}$ (N-D), and the regression-based bandwidths, $\bh_{\hbox {\tiny $R$}}$ (R-D).}
	\label{Sim:box}
\end{sidewaysfigure}

Figure~\ref{Sim:box} addresses the second aim by depicting boxplots of $\textrm{EISE}_{\hbox {\tiny $M$}}$ and $\textrm{EISE}_{\hbox {\tiny $D$}}$ evaluated at eight choices of $\bh$. One can see (from the top panels) that the two mode-based methods yield much smaller $\textrm{EISE}_{\hbox {\tiny $M$}}$ than all density-based methods. Between these two methods, the one involving bootstrap (with $\bh=\bh^*_{\hbox {\tiny $B$}}$) produces more variable $\textrm{EISE}_{\hbox {\tiny $M$}}$ than the method involving CV (with $\bh=\bh_{\hbox {\tiny $M$}}$). This is expected due to the much higher variability of $\bh^*_{\hbox {\tiny $B$}}$ than that of $\bh_{\hbox {\tiny $M$}}$ noted earlier. When it comes to density estimation (see the bottom panels), among the four density-based methods, the one involving CV (with $\bh=\bh_{\hbox {\tiny $D$}}$) gives the lowest $\textrm{EISE}_{\hbox {\tiny $D$}}$, which is also expected because of the close agreement between $\bh_{\hbox {\tiny $D$}}$ and $\tilde \bh_{\hbox {\tiny $D$}}$. 

Finally, Table~\ref{Sim:table} presents the MC averages and standard errors of $\textrm{EISE}_{\hbox {\tiny $M$}}$ under the five true model configurations. From there one may gain the perception that using $\bh_{\hbox {\tiny $M$}}$ gives slightly better numerical performance than when $\bh^*_{\hbox {\tiny $B$}}$ is used. When there are many modes as in (C5), the mode-based CV method (leading to $\bh_{\hbox {\tiny $M$}}$) and the density-based CV method (leading to $\bh_{\hbox {\tiny $D$}}$) perform similarly. Additionally, under the most challenging simulation setting, (C3), the two density-based methods that give $\bh_{\hbox {\tiny $R$}}$ and $\bh_{\hbox {\tiny $B$}}$ are surprisingly competitive. 

\subsection{Application to Old Faithful geyser data}
\begin{figure}[p]
	\centering
	\setlength{\linewidth}{12cm}
	\includegraphics[width=\linewidth]{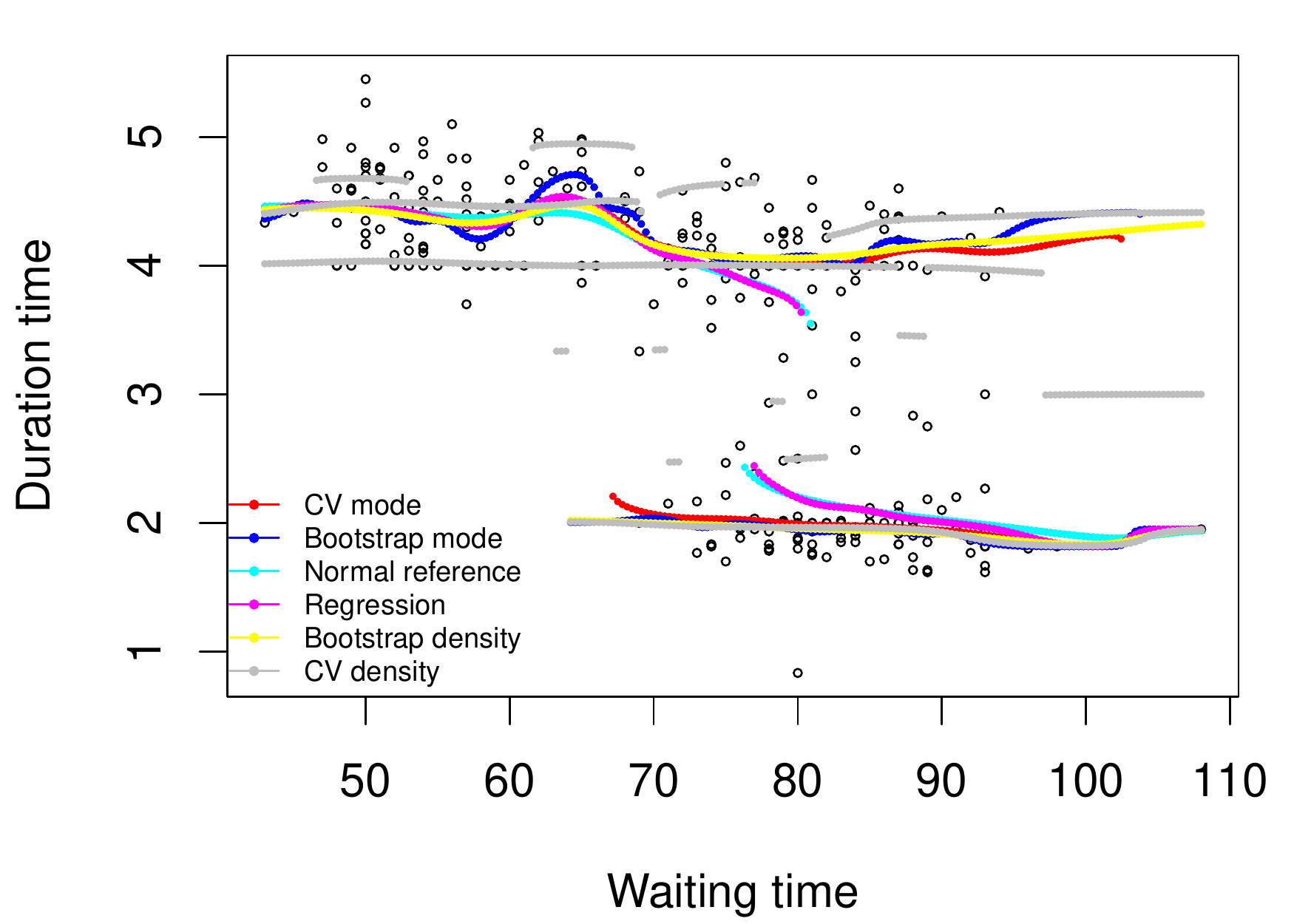}
	\caption{\label{f:geyser} Estimated mode curves resulting from six bandwidth selection methods applied the Old Faithful geyser data (with data points in black circles). The six methods are the mode-based method involving CV (CV mode, red lines), the mode-based method involving bootstrap (Bootstrap mode, blue lines), the normal reference (cyan lines), the regression-based method (pink lines), the density-based method involving bootstrap (Bootstrap density, yellow lines), and the density-based method involving CV (CV density, grey lines).}
\end{figure}

Our empirical comparison of various bandwidth selection methods ends with applying the six considered methods to the Old Faithful geyser data analyzed in \cite{Bashtannyk.Hyndman2001}. This data set consists of 299 observations of the waiting time (in minutes) between eruptions and the duration of the eruption for the Old Faithful geyser in Yellowstone National Park, Wyoming, collected from August 1st to August 15th, 1985. 
Applying the four density-based bandwidth selection methods reviewed in Section~\ref{s:compete} and the two mode-based bandwidth selection methods proposed in Section~\ref{s:cvmode} to this data set yield the following choices of $\bh$: $\bh_{\hbox {\tiny $N$}}=(4.12, 0.87)$, $\bh_{\hbox {\tiny $R$}}=(2.20, 0.87)$, $\bh_{\hbox {\tiny $B$}}=(3.60, 0.40)$, $\bh_{\hbox {\tiny $D$}}=(4.12, 0.09)$, $\bh_{\hbox {\tiny $M$}}=(2.68, 0.60)$, and $\bh^*_{\hbox {\tiny $B$}}=(1.53, 0.37)$. Figure~\ref{f:geyser} presents the data and the six sets of estimated mode curves corresponding to these choices of bandwidths.

As a reminiscence of the overfitting pattern observed in the simulation study, setting $\bh=\bh_{\hbox {\tiny $D$}}$ leads to estimated mode sets that claim far more local modes than the mode estimates from other methods. The estimates resulting from setting $\bh=\bh_{\hbox {\tiny $N$}}$ and  $\bh=\bh_{\hbox {\tiny $R$}}$ are comparable, both less wiggly compared to those resulting from using the mode-based bandwidths,  $\bh_{\hbox {\tiny $M$}}$ and  $\bh^*_{\hbox {\tiny $B$}}$. This may be a sign of underfitting, a pattern repeatedly observed for these two density-based methods in the simulation study. Between the two mode-based bandwidths, the one involving bootstrap, i.e.,  $\bh^*_{\hbox {\tiny $B$}}$, leads to more wiggly estimated mode curves. For this particular data set, the performance of the density-based method involving bootstrap, producing $\bh_{\hbox {\tiny $B$}}$, is similar to that when $\bh_{\hbox {\tiny $M$}}$ is used. 

\section{Discussion}
\label{s:diss}
In this study, we are interested in inferring local modes of $Y$ given $X=x$, and we argue that bandwidth selection methods developed for kernel-based density estimation are not suitable for mode estimation. Even though the four density-based bandwidth selection methods considered in this article only give a small subset of the large collection of existing methods for choosing bandwidths in density estimation, they represent four very different strategies of bandwidth selection in that context; and our numerical studies provide convincing evidence that a bandwidth selection method that performs well in estimating the conditional density typically performs poorly in mode estimation. 

We proposed two bandwidth selection methods tailored for mode estimation, and demonstrated their promising improvement over the density-based methods in estimating conditional modes. The first proposed method is a cross validation procedure for finding $\bh$ that minimizes a CV criterion accounting for the size of a mode set estimate and the distance between this set and the observed response. Due to the difficulty in formulating a proxy for the true mode set at a given observed covariate value, it is unclear if there exists a consistent estimator of $\textrm{ISE}_{\hbox {\tiny $M$}}(\bh)$ given in (\ref{ISE:m}). The CV criterion in (\ref{cv:mode}), $\textrm{CV}_{\hbox {\tiny $M$}}(\bh)$, is constructed with the hope that a large $\textrm{CV}_{\hbox {\tiny $M$}}(\bh)$ usually implies a large $\textrm{ISE}_{\hbox {\tiny $M$}}(\bh)$. Noticing that in the simulation study, under (C1), $\hat h_{2, \hbox {\tiny $M$}}$ tends to be larger than the corresponding optimal choice, $\tilde h_{2, \hbox {\tiny $M$}}$, we believe that $\textrm{CV}_{\hbox {\tiny $M$}}(\bh)$ is an inconsistent estimator of $\textrm{ISE}_{\hbox {\tiny $M$}}(\bh)$. Theoretical properties of this CV criterion deserves more in-depth investigation, which can lead to an improved CV procedure.  

Our second proposed method depends on a mode set estimate, $\hat M^*(\cdot)$, as the byproduct of estimating the conditional density via a finite mixture model. This mode set estimate acts like a proxy for the true mode set in the bootstrap procedure that leads to the selected bandwidths $\bh^*_{\hbox {\tiny $B$}}$. Because estimating a finite mixture model (with unknown number of components) can be subject to high variability, the performance of this bandwidth selection procedure is less stable than the first proposed method. But, with some precautions when fitting a finite mixture model, $\hat M^*(\cdot)$ can be a well-behaved proxy, in which case using it to choose bandwidths via bootstrap often yield satisfactory mode estimates. A question one may raise is why not just use $\hat M^*(x)$ as the ultimate mode estimate instead of implementing the extra bootstrap procedure to select $\bh$ to be used for finding another mode estimate. Indeed, if one wishes to take a parametric or semiparametric route to estimate $M(x)$, one may first estimate the conditional density parametrically or semiparametrically, with finite mixture models as an example, and this (semi)parametric estimate of the conditional density can lead to estimates of local modes. In this study, we estimate $M(x)$ nonparametrically, and we find that the resulting mode estimates using $\bh=\bh^*_{\hbox {\tiny $B$}}$ often improves over $\hat M^*(x)$. 

Relevant works that may provide hints for developing bandwidth selection methods suitable for conditional mode estimation include \cite{chen2015optimal, genovese2016non}, and \cite{chen2016comprehensive}, where the authors considered nonparametric estimation of density ridges of the joint distribution of a multivariate random variable. In these works, the authors used the same bandwidth $h$ for all variables in the joint density and proposed different strategies to choose $h$. In the regression setting where $Y$ and $X$ are treated differently, we observed (in simulation study omitted here) significantly worse mode estimation when assuming $h_1=h_2$ than when this assumption is relaxed. It is possible that methods proposed in these existing works can be revised to allow different bandwidths for different variables, leading to new strategies in the context of conditional mode estimation. Lastly, both our proposed methods struggle more under (C3) compared to the other four cases, where the number of conditional modes varies over $\mathscr{X}$. We believe that this is the situation that calls for variable bandwidths, $\bh(x)$ or $\bh(X_i)$, which is beyond the scope of the current study where all considered methods produce fixed bandwidths $\bh$. One potential direction to follow in order to choose variable bandwidths is to modify the strategies proposed in \cite{comaniciu2001variable}, where the authors also treated all variables symmetrically and chose one variable bandwidth, $h(\bx)$ or $h(\bX_i)$, for all variables.

\bibliographystyle{chicago}

\bibliography{bibliography}

\clearpage 
\thispagestyle{empty}

\begin{table}[p]
	\caption{Monte Carlo averages of $\textrm{EISE}_{\hbox {\tiny $M$}}$ across 500 MC replicates using eight choices of bandwidths. Numbers in parentheses are ($10\times$ standard errors) associated with the averages. Acronym of different methods are the same as those used in Figure~\ref{Sim:box} }
	\label{Sim:table}
	\centering
	{\small
		\begin{tabular}{lcccccccc}
			\hline\noalign{\smallskip}
			& O-M & CV-M & B-M & O-D & CV-D & B-D & N-D & R-D\\
			\hline\noalign{\smallskip}
			(C1) &  0.08   & 0.14   & 0.14   & 1.37   & 1.44   & 0.41  & 0.16   & 0.13\\
			& (0.01) & (0.01) & (0.10) & (0.28) & (0.38) & (0.15) &(0.04) & (0.04)\\
			\hline\noalign{\smallskip}
			(C2) &  0.08   & 0.12   & 0.11   & 0.47   & 0.55   & 0.30  & 0.55   & 0.42\\
			& (0.01) & (0.05) & (0.04) & (0.09) & (0.16) & (0.26) &(0.12) & (0.12)\\
			\hline\noalign{\smallskip}
			(C3) &  0.48   & 1.62   & 2.06   & 3.20   & 3.25   & 1.95  & 2.63   & 1.93\\
			& (0.18) & (0.29) & (0.30) & (0.32) & (0.37) & (0.16) &(0.18) & (0.15)\\		 
			\hline\noalign{\smallskip}
			(C4) &  0.15   & 0.25   & 0.21   & 0.34   & 0.36   & 14.8  & 15.2   & 12.9\\
			& (0.04) & (0.08) & (0.06) & (0.05) & (0.06) & (3.25) &(2.42) & (2.10)\\	
			\hline\noalign{\smallskip}
			(C5) &  0.46   & 0.56   & 0.60   & 0.53   & 0.54   & 10.2  & 10.2   & 10.9\\
			& (0.06) & (0.08) & (0.10) & (0.07) & (0.07) & (1.23) &(1.31) & (1.49)\\	
			\noalign{\smallskip}\hline
		\end{tabular}
	}
\end{table}

\clearpage
\thispagestyle{empty}

\clearpage
\thispagestyle{empty}

\clearpage
\thispagestyle{empty}

\clearpage 
\thispagestyle{empty}

\clearpage 
\thispagestyle{empty}

\clearpage 
\thispagestyle{empty}

\clearpage
\thispagestyle{empty}

\clearpage 
\thispagestyle{empty}

\clearpage 
\thispagestyle{empty}

\clearpage 
\thispagestyle{empty}

\clearpage 
\thispagestyle{empty}

\clearpage
\thispagestyle{empty}
 
\end{document}